%% file: main.tex
\documentclass{jfm}
\usepackage{graphicx}
\usepackage{epstopdf, epsfig}
\usepackage{xcolor}

\journal{Journal of \LaTeX\ Templates}
\newcommand*\diff{\mathop{}\!\mathrm{d}} 
\newcommand{\sm}[1]{{\scriptscriptstyle#1}} 
\newcommand{\xoverbrace}[2][\vphantom{\frac{A}{A}}]{\overbrace{#1#2}} 

\shorttitle{Flame Kernel/Turbulence Interactions under Engine Conditions}
\shortauthor{T. Falkenstein , S. Kang and H. Pitsch}

\title{\textcolor{black}{Analysis of Premixed Flame Kernel/Turbulence Interactions under Engine Conditions \textcolor{black}{based on DNS Data} }}

\author{T. Falkenstein\aff{1},
  S. Kang\aff{2}
 \and H. Pitsch\aff{1}\corresp{\email{office@itv.rwth-aachen.de}}}

\affiliation{\aff{1}Institute for Combustion Technology, RWTH Aachen University, 52056 Aachen, Germany
\aff{2}Department of Mechanical Engineering, Sogang University, Seoul 121-742, Republic of Korea}

\begin{document}

\maketitle

\begin{abstract}
Although the evolution of premixed flames in turbulence has been frequently studied, it is not well understood how small flames interact with large-scale turbulent flow motion. Since this question is of practical importance for the occurrence of cycle-to-cycle variations in spark ignition engines, the objective of the present work is to fundamentally differentiate early flame kernel development from well-established turbulent flame configurations. For this purpose, a DNS database consisting of three flames propagating in homogeneous isotropic turbulence \textcolor{black}{(Falkenstein et al., Combust. Flame, 2019) is considered}. The flames feature different ratios of the initially laminar flame diameter to the integral length scale. To quantify flame kernel development, the time evolution of \textcolor{black}{flame topology and flame front geometry} are analysed in detail. It is shown that \textcolor{black}{some realizations of} the early flame kernel \textcolor{black}{are} substantially influenced by high compressive strain caused by large-scale turbulent flow motion with characteristic length scales greater than the flame kernel size. As a result, the initial spherical kernel topology \textcolor{black}{may} become highly distorted, which \textcolor{black}{is reflected in the stochastic occurrence of excessive curvature variance. Two mechanisms of curvature production resulting from early flame kernel/turbulence interactions are identified by analysis of the mean curvature balance equation. 
Further, it is shown that} the curvature distribution  \textcolor{black}{of small flame kernels becomes} strongly skewed towards positive curvatures, which is contrary to developed turbulent flames.
Hence, the transition of ignition kernels to self-sustaining turbulent flames is very different in nature compared with the development of a statistically planar flame brush. 
\end{abstract}


\section{Introduction}
\label{sec:intro}
\textcolor{black}{The development and control of clean and efficient technical combustion devices is often very challenging due to the potential occurrence of undesirable stochastic phenomena. 
In spark ignition~(SI) engines, efforts to enhance thermal efficiency by lean or exhaust-gas-diluted operation are limited by reduced combustion stability~\citep{Aleiferis04_lean_ccv,Jung17_lean_si_ccv}. In extreme cases, partially burning or misfiring cycles may occur~\citep{Peterson11_exp_misfire} leading to poor efficiency and excessive emissions. At high-load conditions, cycle-to-cycle variations~(CCV) may necessitate significant safety margins to avoid potentially destructive irregular combustion phenomena, such as knock~\citep{Wang17_knock_precs}.} \par
\textcolor{black}{From experimental engine studies, it is well-known that~CCV are correlated with fluctuations in early flame kernel development~\citep{Young81_ccv}. Even under conditions that lead to misfires, spark ignition was found to successfully initiate a flame kernel, which extinguishes during the early growth phase~\citep{Peterson11_exp_misfire}. By using advanced optical diagnostics, prolonged chemical time scales due to exposure of the small flame to locally lean or diluted mixture~\citep{Peterson14_early_flame_exp}, or fluctuations in large-scale flow structures~\citep{Bode17_piv_early_kernel,Zeng18_exp} and kinetic energy~\citep{Buschbeck12_piv_kernel} could be identified as sources for cyclic variability. \citet{Schiffmann17_exp} have shown that the time until the young laminar flame kernel transitions to a turbulent flame correlates with~CCV.} \par 
\textcolor{black}{A particular feature of the combustion process in conventional~SI engines is flame initiation at unfavourably small length scales in terms of resilient flame growth, mainly for three reasons. First, the radius of the young flame kernel is only moderately larger than the flame thickness, which may severely alter the flame structure as well as the reaction/diffusion balance therein. Second, large portions of the initially small burned gas volume may be in contact with colder walls, thus causing significant heat losses. Third, the flame kernel is initially smaller than the energetic flow scales, which may have several implications. As observed in fundamental experiments by~\citet{Akindele82_kernel_flow_interact} and~\citet{Bradley87_kernel_flow_interact}, the young flame kernel can be subject to convective transport attributed to flow structures with larger characteristic length scales than the kernel size, before the onset of noticeable flame wrinkling by turbulence. 
Further, it was hypothesized that the early turbulent flame growth is governed by the high-wavenumber range of the energy spectrum, while lower wavenumbers were thought to be mostly inactive during early flame front/turbulence interactions~\citep{Akindele82_kernel_flow_interact,AbdelGayed84_kernel_flow_interact}. This concept of scale separation based on the characteristic flame kernel size as upper cut-off scale was also introduced into models~\citep{Herweg92_model_valid,Echekki94_kernel_dns_Le_effect}, which were proposed as extensions to previous expressions for turbulent flame brush development~\citep{Scurlock53_planar_flame_devel_theory,Meneveau91_ITNFS_model}. However, the additional effect of flame kernel stretching by hydrodynamic strain mentioned by~\citet{Bradley94_kernel_stretching} has not received much attention and is considered in the present DNS study.} \par
\textcolor{black}{For a quantitative analysis of turbulent premixed flame development, different metrics such as flame area or flame curvature may be considered, which will be briefly introduced hereafter. The global flame front geometry can be characterized by the flame brush thickness or flame surface area~\citep{Peters99_regime_diagr}, which is directly related to the increase in the overall burning rate due to turbulence~\citep{Damkoehler40}. \citet{Candel90_fsd_eq} derived a balance equation for the flame area, here defined as an iso-surface of a scalar~$\vartheta$ propagating with the displacement speed~$\mathrm{s}_{\mathrm{d}}$:}
\begin{equation}  
\frac{1}{{A}_{\vartheta}} \frac{\mathrm{D}_{\vartheta} \left( {A}_{\vartheta} \right)}{\mathrm{D}_{\vartheta}\left(t\right)} =  a_{\mathrm{t}}^{(\vartheta)}
 + \mathrm{s}_{\mathrm{d}}^{(\vartheta)} \kappa^{(\vartheta)}.
\label{eq:one_ov_A_dAdt}
\end{equation}  
\textcolor{black}{According to~(\ref{eq:one_ov_A_dAdt}), flame area may change due to hydrodynamic strain expressed by the tangential strain rate~$a_{\mathrm{t}}$, and the propagation of surface elements with local mean curvature }
\begin{eqnarray}
\kappa^{(\vartheta)} &= & \frac{\partial n_i^{(\vartheta)}}{\partial x_i\;\;}, \label{eq:def_kappa}  \\
n_i^{(\vartheta)} &= & \frac{-1}{\left|\nabla \vartheta \right|}\frac{\partial \vartheta}{\partial x_i} . \label{eq:def_nvec}
\end{eqnarray} 
\textcolor{black}{Note that the sub/superscripts indicating the scalar field will be omitted in most parts of the present  discussion. The total time derivative in a reference frame attached to the propagating scalar iso-surface is defined as:}
\begin{equation}
\frac{\mathrm{D}_{\vartheta} \left( \cdot \right)}{\mathrm{D}_{\vartheta}\left(t\right)} = 
\frac{\partial \left( \cdot \right)}{\partial t} + \left(u_i + \mathrm{s}_{\mathrm{d}}^{(\vartheta)} n_i^{(\vartheta)} \right) \frac{\partial \left( \cdot \right) }{\partial x_i},
\label{eq:sc_tot_d_dt}
\end{equation}
\textcolor{black}{where~$u_i$ denotes the fluid velocity and~$n_i$ is the normal vector pointing into the direction of negative scalar gradient. For $\left( \cdot \right)=\left( \vartheta \right)$, the l.h.s.\ of~(\ref{eq:sc_tot_d_dt}) vanishes, which can be used to determine~$\mathrm{s}_{\mathrm{d}}$.}\par
\textcolor{black}{Flame curvature is a useful quantity to parametrize variations in local flame structure~\citep{Kim07_sc_align}, and to statistically assess the flame front geometry in turbulence. To highlight the connection between total flame area and the first two moments of the curvature probability density function~(PDF), the integral form of (\ref{eq:one_ov_A_dAdt}) can be used to express the evolution of the total flame surface area inside the volumetric domain~$\Omega$ }
\begin{eqnarray}
\frac{1}{\overline{A}_{\Omega}} \frac{\mathrm{D}_{\vartheta} \left( \overline{A}_{\Omega} \right)}{\mathrm{D}_{\vartheta}\left(t\right)} & = &  \left< a_{\mathrm{t}} \right>_{\mathrm{s},\Omega}
 + \left< \mathrm{s}_{\mathrm{d}} \kappa \right>_{\mathrm{s},\Omega}, \label{eq:one_ov_A_dAdt_avg} \\
 ~ & = & \left< a_{\mathrm{t}} \right>_{\mathrm{s},\Omega} 
 + \left< \mathrm{s}_{\mathrm{rn}} \kappa \right>_{\mathrm{s},\Omega}
 - \left< D_{\mathrm{th}}\kappa^2 \right>_{\mathrm{s},\Omega}.
\label{eq:one_ov_A_dAdt_avg_split}
\end{eqnarray} 
\textcolor{black}{Here, $\left<\cdot \right>_{\mathrm{s}}$ denotes generalized surface averaging as proposed by~\citet{Boger98_gen_fsd}, i.e. the averaging operation is defined by computing the surface integral of some quantity over each scalar iso-surface, averaging over all iso-levels, and normalizing by the averaged scalar iso-surface area~${\overline{A}_{\Omega}=\int_{\Omega} \left|\nabla \vartheta \right| \diff V}$. Further, the displacement speed has been decomposed into contributions by reaction/normal-diffusion~($\mathrm{s}_{\mathrm{rn}}$) and tangential diffusion~($-D_{\mathrm{th}}\kappa$) according to~\citet{Echekki99_diff_term_split}, where~$D_{\mathrm{th}}$ is the thermal diffusivity. Hence, the second term in the r.h.s.\ of~(\ref{eq:one_ov_A_dAdt_avg_split}) describes the total area change due to normal-propagation of curved flame regions (kinematic restoration, which could be source or sink for flame area depending on the sign of the curvature),  while the third term corresponds to scalar dissipation~\citep{Peters99_regime_diagr}. The impact of curvature moments on global flame area dynamics can be shown by adapting the decomposition of~$\left< \mathrm{s}_{\mathrm{d}} \kappa \right>_{\mathrm{s},\Omega}$ 
proposed by~\citet{Wang17_jet_dns} to the normal-propagation and scalar dissipation terms in~(\ref{eq:one_ov_A_dAdt_avg_split}):}
\begin{eqnarray}
\left< \mathrm{s}_{\mathrm{rn}} \kappa \right>_{\mathrm{s}} & = & \left< \mathrm{s}_{\mathrm{rn}} \right>_{\mathrm{s}} \left< \kappa \right>_{\mathrm{s}} + 
\left[\left< \mathrm{s}_{\mathrm{rn}} \kappa \right>_{\mathrm{s}} -  \left< \mathrm{s}_{\mathrm{rn}} \right>_{\mathrm{s}} \left< \kappa \right>_{\mathrm{s}} \right], \label{eq:correl_srn_curv} \\ [5pt]
- \left< D_{\mathrm{th}}\kappa^2 \right>_{\mathrm{s}} & = & - \left< D_{\mathrm{th}}\right>_{\mathrm{s}} \left< \kappa^2 \right>_{\mathrm{s}}  - \left[ \left< D_{\mathrm{th}}\kappa^2 \right>_{\mathrm{s}} - \left< D_{\mathrm{th}}\right>_{\mathrm{s}} \left< \kappa^2 \right>_{\mathrm{s}} \right],
\label{eq:correl_D_curv2}
\end{eqnarray} 
\textcolor{black}{where the subscript~$\Omega$ has been omitted for brevity. \citet{Wang17_jet_dns} have shown that the correlation of curvature and displacement speed dominates the overall curvature stretch effect~$\left< \mathrm{s}_{\mathrm{d}} \kappa \right>_{\mathrm{s}}$ in~(\ref{eq:one_ov_A_dAdt_avg}). However, separating the curvature effect on flame area evolution into normal-propagation and scalar dissipation enables a more differentiated understanding.
Although the correlation between curvature and the flame normal propagation velocity in~(\ref{eq:correl_srn_curv}) may only be negligible in the unity-Lewis-number limit and if the flame is fully developed, the contribution of~$\left< \mathrm{s}_{\mathrm{rn}} \kappa \right>_{\mathrm{s}}$ to the global curvature stretch term~$\left< \mathrm{s}_{\mathrm{d}} \kappa \right>_{\mathrm{s}}$ is small under the present conditions (cf. figure~\mbox{S-1}\,(b) in the supp. material). While the scalar dissipation term is of leading order, the correlation between the thermal diffusivity and the squared curvature in~(\ref{eq:correl_D_curv2}) is rather weak and may be neglected (cf. figure~\mbox{S-1}\,(c) in the supp. material). Hence, the following approximations are introduced:}
\begin{eqnarray}
\left< \mathrm{s}_{\mathrm{rn}} \kappa \right>_{\mathrm{s}} & \approx & \left< \mathrm{s}_{\mathrm{rn}} \right>_{\mathrm{s}} \left< \kappa \right>_{\mathrm{s}} , \label{eq:approx_srn_curv} \\ [5pt]
- \left< D_{\mathrm{th}}\kappa^2 \right>_{\mathrm{s}} & \approx & - \left< D_{\mathrm{th}}\right>_{\mathrm{s}} \left< \kappa^2 \right>_{\mathrm{s}} . \label{eq:approx_D_curv2}
\end{eqnarray} 
\textcolor{black}{To relate the evolution of total flame surface area to the surface-weighted curvature PDF, (\ref{eq:approx_srn_curv}) and~(\ref{eq:approx_D_curv2}) are inserted into~(\ref{eq:one_ov_A_dAdt_avg_split}) and rewritten in terms of the first two central curvature moments~$\mu_{\kappa}= \left< \kappa \right>_{\mathrm{s}}$ and~$\sigma_{\kappa}^2 = \left< \kappa^2 \right>_{\mathrm{s}}- \mu_{\kappa}^2$:}
\begin{equation}  
\frac{1}{\overline{A}_{\Omega}} \frac{\mathrm{D}_{\vartheta} \left( \overline{A}_{\Omega} \right)}{\mathrm{D}_{\vartheta}\left(t\right)}  \approx
 \left< a_{\mathrm{t}} \right>_{\mathrm{s}} 
+ \left[ \left< \mathrm{s}_{\mathrm{rn}} \right>_{\mathrm{s}} 
- \left< D_{\mathrm{th}}\right>_{\mathrm{s}}\mu_{\kappa}
\right] \mu_{\kappa}
- \left< D_{\mathrm{th}}\right>_{\mathrm{s}} \sigma_{\kappa}^2
.
\label{eq:one_ov_A_dAdt_avg_mom}
\end{equation}  
\textcolor{black}{It should be noted that the theoretical analysis by~\citet{Peters99_regime_diagr} identified the scalar dissipation term in~(\ref{eq:one_ov_A_dAdt_avg_split}), i.e. the impact of~$\sigma_{\kappa}^2$ on the total flame area (cf.~(\ref{eq:one_ov_A_dAdt_avg_mom})), as the dominant curvature term in the thin reaction zones regime. In empirical studies, the effect of increasing turbulence intensity $u'/\mathrm{s}_{\mathrm{l}}^{\sm{0}}$ on flame wrinkling was quantified by means of flame surface area~\citep{Wenzel00_dns_flame_area} and in terms of~$\sigma_{\kappa}^2$ as the most informative curvature moment in terms of flame front/turbulence interactions~\citep{Shepherd02_curv_pdf_turb,Haq02_curv_pdf_turb,Gashi05_curv_pdf_turb}. } \par
%
\textcolor{black}{In our previous study~\citep{Falkenstein19_kernel_Le1_cnf}, a DNS database of three developing turbulent premixed flames has been established in the limit of unity Lewis number. It was shown that the global heat release rate of small flame kernels exhibits run-to-run variations, which were attributed to stochastic variations in surface area evolution caused by the dissipation term in~(\ref{eq:one_ov_A_dAdt_avg_split}). However, the fundamental role of turbulence in shaping the young flame kernel has not been clarified. 
In this study, the particular nature of flame kernel development will be analysed in terms of the evolution of central curvature moments, which are related to flame area dynamics according to~(\ref{eq:one_ov_A_dAdt_avg_mom}). The occurrence of excessive curvature variance and positive curvature skewness during the early phase of small flame kernel growth will be explained by evaluation of the flame curvature balance equation.  Further, the role of large-scale turbulent flow structures in changing the small flame kernel topology will be shown.} \par
\textcolor{black}{The present manuscript is organized as follows. In~\S\ref{sec:dns_data_base}, a brief summary of the DNS database is given. \textcolor{black}{Characteristic differences in the flame topologies of small flame kernels and developing spherical flames are presented in~\S\ref{sec:flame_topol}. A detailed analysis of flame front geometry evolution in terms of mean curvature dynamics is provided in~\S\ref{sec:flame_wrinkling}.
Length-scale dependent flame front alignment with the flow field is the subject of ~\S\ref{sec:alignment}. }
Finally, the relevance of the present results to actual engines and implications for modelling of early flame kernel development are discussed in~\S\ref{sec:engine_relevance}. }

\section{DNS Database}
\label{sec:dns_data_base}
\textcolor{black}{To isolate the governing physical phenomena stemming from the small initial flame size characteristic for~SI engine combustion, a DNS database consisting of three flame configurations with different geometries and multiple realizations for the most relevant case~\citep{Falkenstein19_kernel_Le1_cnf} is considered. A summary of the parameter variations, flame conditions and the DNS setup is provided in the following sections.} 
\subsection{Flame Geometry Variation}
\label{ssec:flame_var}
A few studies in the literature have proposed regime diagrams for flame kernel/flow interactions derived from two-dimensional DNS data of laminar~\citep{Echekki07_kernel_vortex_map_01,Vasudeo10_kernel_vortex_map_01} or turbulent~\citep{Reddy11_regime_map} flames. Two governing parameters were identified, the ratio of integral length scale (or vortex size) to kernel size, as well as the ratio of turbulent velocity fluctuation (or vortex strength) to burning velocity. While these studies established a qualitative understanding of global flame kernel/flow interactions, quantitative differences between identified regimes and consequences for the developed flames were not reported. In this work, we seek to gain quantitative insight into early flame kernel development under engine conditions by means of three-dimensional DNS with detailed chemistry. \textcolor{black}{Here, we will focus on variations of the ratio of initial flame diameter~$D_{0}$ to the integral length scale~$l_{\mathrm{t}}$, while the ratio of velocity fluctuation to burning velocity~$u'/\mathrm{s}_{\mathrm{l}}^{\sm{0}}$ is kept constant. The main reference case here is the engine flame kernel with~${D_{0}/l_{\mathrm{t}}=0.3}$, which was computed in a constant volume setting. For this case, 
the choice of the initial diameter is not unique since ignition and early flame development cannot be clearly distinguished in real engines. Justification for the selected parameter value will be provided in~\S\ref{sec:engine_relevance}. 
Since the global flame kernel behaviour for~${D_{0}/l_{\mathrm{t}}=0.3}$ depends particularly strongly on the local flow conditions, two kernel realizations have been obtained by igniting the kernel in different locations of the same
initial flow field. The second considered case is a larger flame-kernel dataset that has been designed with~${D_{0}/l_{\mathrm{t}}=2.0}$ such that the initial flame kernel is larger than the integral scale of the turbulence, i.e.\ the initial interaction of the flame kernel with the turbulence is fundamentally different from that of the engine flame kernel. Finally, since both of these flame kernels have mean positive curvature, an 
initially laminar, planar flame front propagating in homogeneous isotropic turbulence is also considered as a well-defined, limiting reference case with~${D_{0}/l_{\mathrm{t}}=\infty}$, which was computed in a constant pressure setting. Here, all flow scales contribute to flame corrugation from the beginning. Additionally, ${D_{0}/l_{\mathrm{t}}}$ does not change over time.
For illustrations of the flames, refer to a previous study by the present authors~\citep{Falkenstein19_kernel_Le1_cnf}.} \par
\subsection{Physical Configuration}
\label{ssec:phys_config}
Mixture state and composition have been chosen to be representative of fully homogeneous, stoichiometric SI engine operation at low load. \textcolor{black}{To enable a better understanding of how the flow alters the young flame kernel, Lewis numbers~(Le) of all species have been artificially set to unity. Flame response in presence of differential diffusion will be investigated in a future study.} Mixture properties of the DNS are summarized in table~\ref{tab:dns_mixture}. \textcolor{black}{The thermodynamic state is given by the pressure~$p^{\left(0\right)} $ and temperature of the unburned gas~$T_{\mathrm{u}}$, which has unity equivalence ratio~$\phi$}.
\begin{table}
  \begin{center}
\def~{\hphantom{0}}
  \begin{tabular}{lcl}
       Mixture   & ~~ &Iso-Octane/Air \\ 
       $p^{\left(0\right)} $ & &$6$ bar  \\
       $T_{\mathrm{u}}$ && $600$ K \\
	   $\phi$ && $1.0$\\
    \textcolor{black}{$\mathrm{s}_{\mathrm{l}}^{\sm{0}}$} & & $0.63$ m/s \\
    \textcolor{black}{$l_{\mathrm{f}}$} & & $71.3$ $\mu \mathrm{m}$ \\
	   $\mathrm{Le}$ && $1.0$ \\
	   \textcolor{black}{Flow Field} && \textcolor{black}{Decaying h.i.t.} \\
	   \textcolor{black}{Combust. Regime} && \textcolor{black}{Thin Rct.\ Zones}\\
  \end{tabular}
  \caption{Flame conditions in the DNS.}
  \label{tab:dns_mixture}
  \end{center}
\end{table}
In order to provide well-defined, but engine-relevant flow configurations, all flames have been computed in decaying homogeneous isotropic turbulence. A comparison between engine and DNS parameters is summarized in table~\ref{tab:dns_params}. 
\textcolor{black}{The turbulent integral length scale and eddy turnover time are denoted by~$l_{\mathrm{t}}$ and~$\tau_{\mathrm{t}}$, and $\eta$ is the Kolmogorov length scale. %
In this study}, the laminar flame thickness~$l_{\mathrm{f}}$ is computed from the maximum temperature gradient, while a turbulent length scale representative for the integral scales is given by $l_{\mathrm{t}} = {{u^{\prime}}^3}/{\overline{\varepsilon}}$, using the mean dissipation rate $\overline{\varepsilon}$. 
Except for the integral length scale of turbulence, which is approximately~2.5 times smaller than in a practical engine, the conditions are realistic for the full duration of the DNS. More details on the flow field initialization and non-reactive DNS results are provided in \textcolor{black}{our previous study~\citep{Falkenstein19_kernel_Le1_cnf}}. \par
\begin{table}
  \begin{center}
\def~{\hphantom{0}}
  \begin{tabular}{lclcl}
       Parameter   &~ & Engine &~ & DNS ($t_{\mathrm{init}} - t_{\mathrm{end}}$) \\ [4pt]
       $\mathrm{Re}_{\mathrm{t}} $&& $100-2390$  && $ 385 - 222$\\[3pt]
	$\frac{u_{\mathrm{rms}}}{\mathrm{s}_{\mathrm{l}}^{\sm{0}}}$&& $2-15$ && $ 7.0 - 3.1 $\\[3pt]
	$\mathrm{Ka} $&& $1-6$ && $13.7 - 4.2$\\[3pt]
	$\mathrm{Da}$ &&$1-100$ && $ 1.4 - 3.6$\\[3pt]
	$\frac{l_{\mathrm{t}}}{\eta}$&& 100 -200 && $87.2 - 57.2$\\[3pt]  
	$\frac{l_{\mathrm{t}}}{l_{\mathrm{f}}}$&& $20 - 147$ && $10.2 - 12.4$\\[3pt]
		$\frac{D_{0}}{l_{\mathrm{t}}}$ &&$<1.0$ && $ 0.3 \;|\; 2.0 \; |\; \infty $\\[3pt]
  \end{tabular}
  \caption{Engine \citep{Heywood94_COMODIA,Heim11_engine_turb} and DNS characteristic numbers.}
  \label{tab:dns_params}
  \end{center}
\end{table}
\textcolor{black}{A summary of the DNS setup is provided in table~\ref{tab:dns_setup}. One important difference between the flame kernel datasets and the planar reference flame is the initialization method. Flame kernels were ignited by a source term in the temperature equation with similar spark duration and size as previously used in literature~\citep{Hesse09_spark_duration,Subramanian09_spark_src,Uranakara17_ign_kernel_3d}. To reliably ignite flame kernels under turbulent conditions, a 40\,\% higher ignition energy than the minimum ignition energy~(MIE) of a laminar flame was used, which results in an early growth phase comparable to engine experiments reported in literature (cf.~\S\ref{sec:engine_relevance}). By contrast, a laminar unstretched flame was imposed into the turbulent flow field as initial condition for the planar-flame DNS to avoid strong dilatation due to the larger burned volume.}
%
\begin{table}
  \begin{center}
\def~{\hphantom{0}}
  \begin{tabular}{lcl}
       \textcolor{black}{Grid Size}   & ~~ & \textcolor{black}{$960^3$} \\ [3pt]
       \textcolor{black}{Domain Size}   & ~~ & \textcolor{black}{$15 \cdot l_{\mathrm{t}}$} \\ [3pt]
       \textcolor{black}{$l_{\mathrm{f}}$/$\Delta x$; $\eta$/$\Delta x$ }   & ~~ & \textcolor{black}{$6$~\citep{Luca19_dns_resolution}; 0.7} \\ [3pt]
       \textcolor{black}{Navier-Stokes Eq.} & ~~ & \textcolor{black}{Low-Mach-Approx.~\citep{Mueller98_lowMach}} \\ [3pt]
       \textcolor{black}{Transport Model} & ~~ &  \textcolor{black}{Curt.-H.~\citep{Hirschfelder54_diff_model}, const.-Le} \\ [3pt]
       \textcolor{black}{Soret Effect} & ~~ & \textcolor{black}{yes} \\ [3pt]
       \textcolor{black}{Chem. Mechanism} & ~~ & \textcolor{black}{26 Spec., based on~\citep{Pitsch96_iso_octane_mech}} \\ [3pt]
       \textcolor{black}{Flame Kernel Init.} & ~~ & \textcolor{black}{Ignition Heat Source} \\ [3pt]
       \textcolor{black}{Sim. Time of Flame-} & ~~ & \textcolor{black}{Engine Kernel I: \,\,($t_{\mathrm{sim.}}=2.3\cdot\tau_{\mathrm{t}}$),}  \\ [3pt]
       \textcolor{black}{Kernel Realizations}  & ~~ & \textcolor{black}{Engine Kernel II: ($t_{\mathrm{sim.}}=1.0\cdot\tau_{\mathrm{t}}$)}\\
  \end{tabular}
  \caption{\textcolor{black}{Computational setup of the DNS.}}
  \label{tab:dns_setup}
  \end{center}
\end{table}

\section{\textcolor{black}{Flame Kernel Topology Evolution}}
\label{sec:flame_topol}
\begin{figure}
\centering
\begin{minipage}[b]{0.45\linewidth}
  \graphicspath{{./data/FLAME_KERNEL_DNS_02/190109_principal_strain_PROG_T_visualize/t_3.005E-05/}}
  \centering
  \makebox[0pt][l]{\quad(a) \textcolor{black}{Engine Kernel~I: $R_{50}=0.4 \cdot l_{\mathrm{t}}$}}\includegraphics[trim={10cm -5cm 40cm -6cm},clip,width=5.5cm]{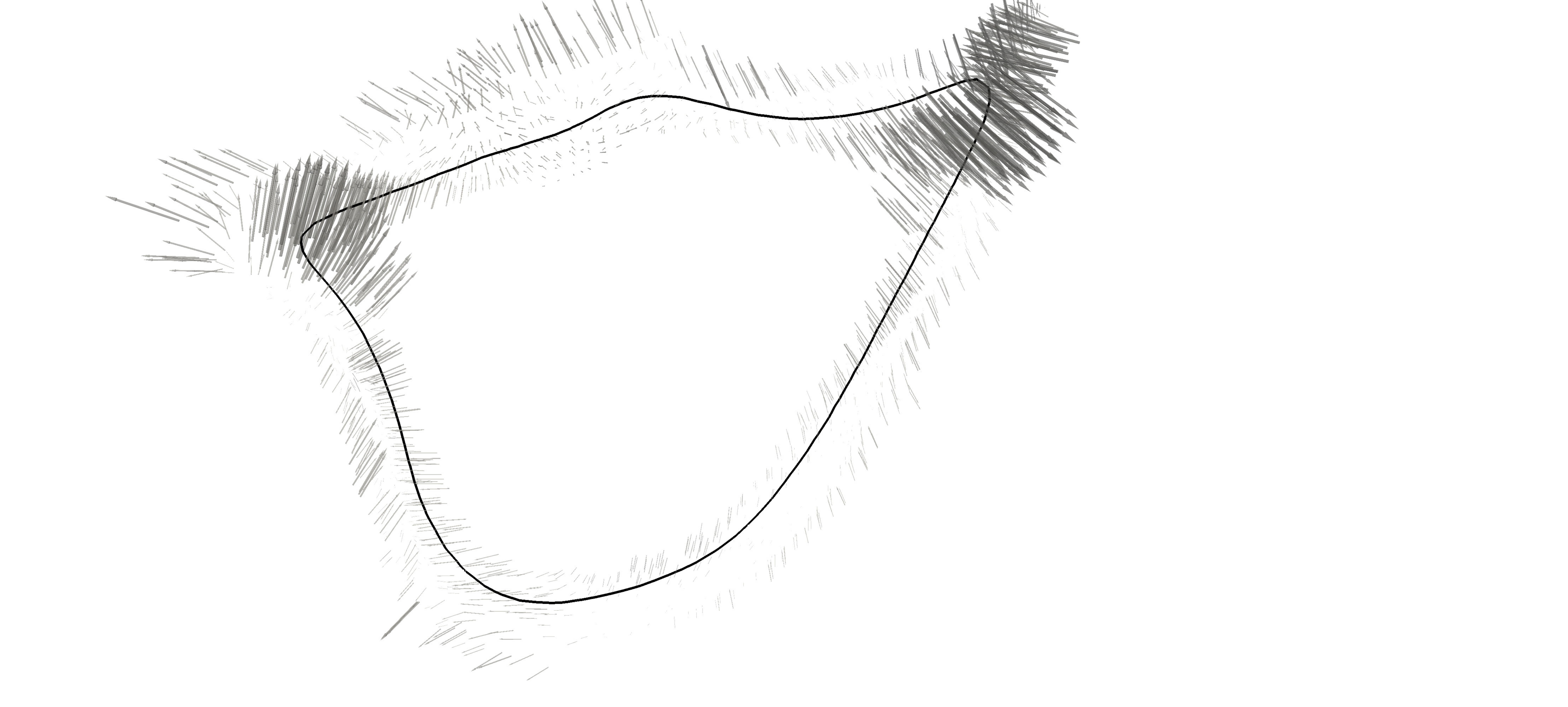}
\end{minipage}
\begin{minipage}[b]{0.45\linewidth}
  \graphicspath{{./data/FLAME_KERNEL_DNS_02/190109_principal_strain_PROG_T_visualize/t_4.505E-05/}}
  \centering
  \makebox[0pt][l]{\quad(b) \textcolor{black}{Engine Kernel~I: $R_{50}=0.5 \cdot l_{\mathrm{t}}$}}\includegraphics[trim={10cm -5cm 40cm -6cm},clip,width=5.5cm]{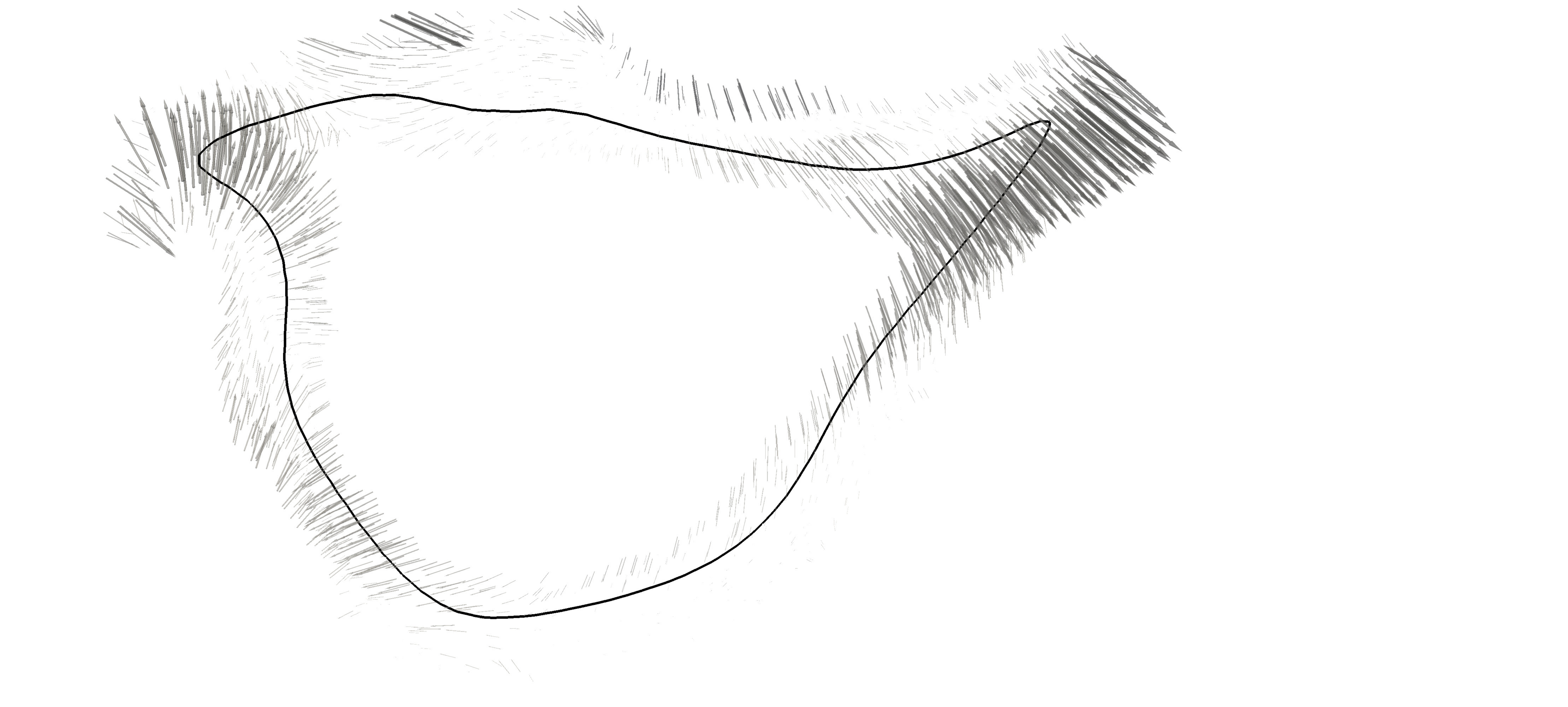}
\end{minipage}
\begin{minipage}[b]{0.45\linewidth}
  \graphicspath{{./data/FLAME_KERNEL_DNS_03/181002_alignment_curv_strain_progT/t_3.006E-05_filt/}}
  \centering
  \makebox[0pt][l]{\quad(c) \textcolor{black}{Large Kernel: $R_{50}=1.0 \cdot l_{\mathrm{t}}$}}\includegraphics[trim={15cm -5cm 35cm -6cm},clip,width=5.5cm]{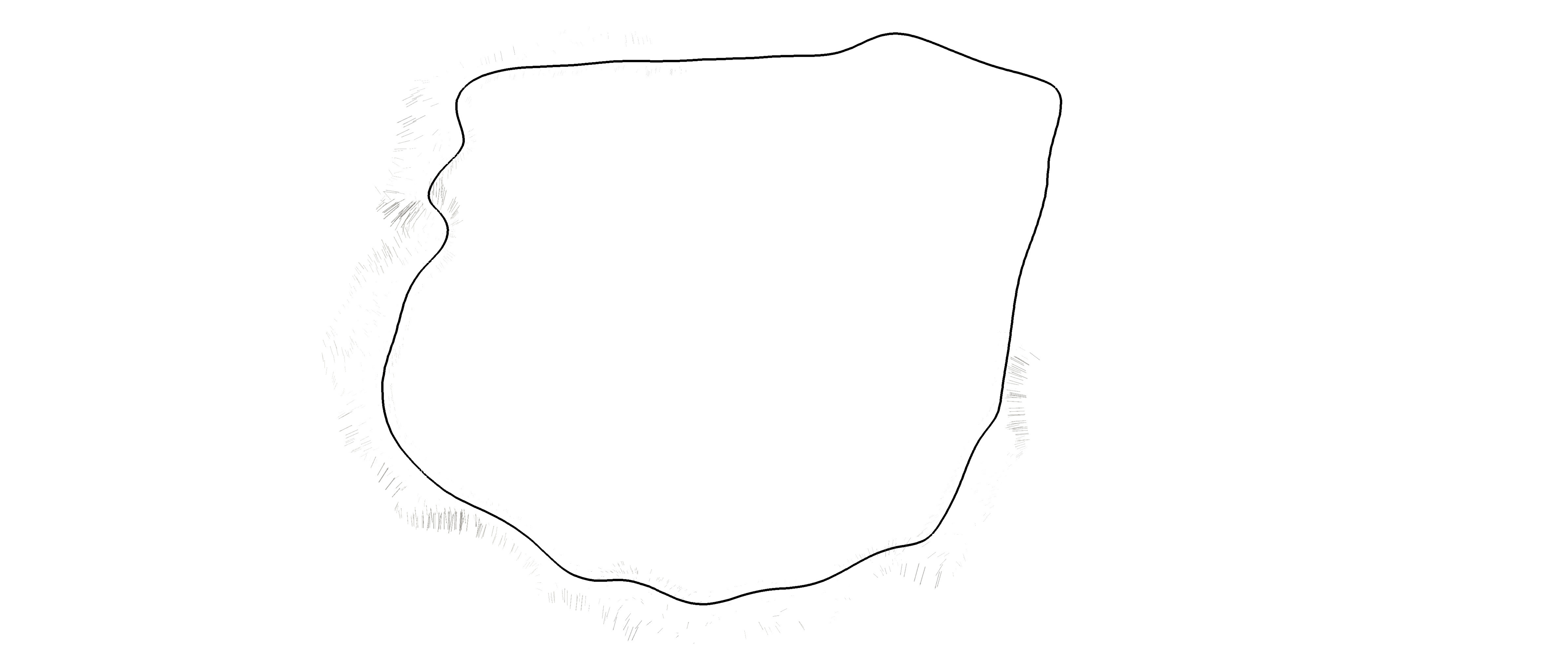}
\end{minipage}
\begin{minipage}[b]{0.45\linewidth}
  \graphicspath{{./data/FLAME_KERNEL_DNS_03/181002_alignment_curv_strain_progT/t_4.501E-05_filt/}}
  \centering
  \makebox[0pt][l]{\quad(d) \textcolor{black}{Large Kernel: $R_{50}=1.1 \cdot l_{\mathrm{t}}$}}\includegraphics[trim={15cm -5cm 35cm -6cm},clip,width=5.5cm]{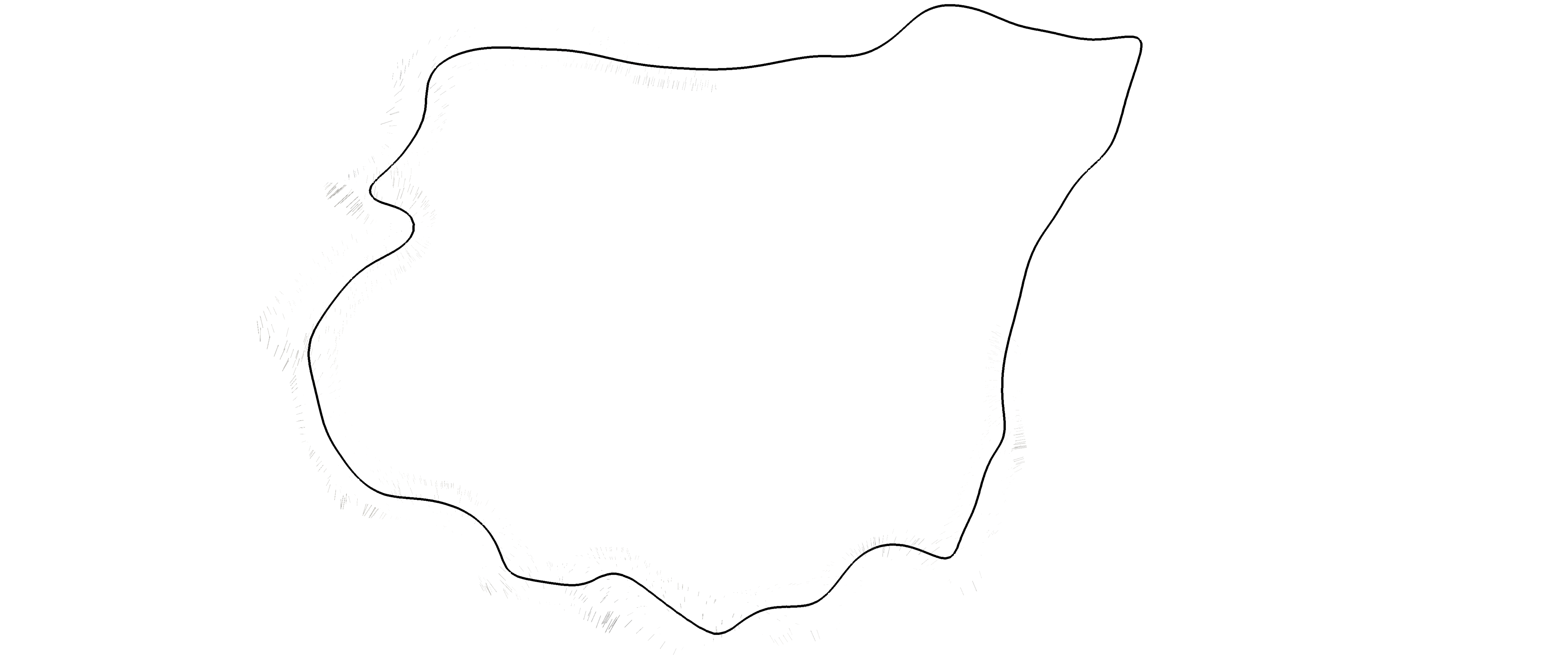}
\end{minipage}
\caption{Temperature iso-contours and eigenvectors of the most compressive strain \textcolor{black}{of Engine flame kernel I}~(a,b), \textcolor{black}{and of the large flame kernel plotted at reduced scale~(c,d) at ($t=0.17\cdot\tau_{\mathrm{t}}$) and ($t=0.25\cdot\tau_{\mathrm{t}}$), respectively}.}
\label{fig:kernel_contour_strain_topol}
\end{figure}

In fully developed turbulent flames, the flame brush thickness is proportional to the integral length scale~\citep{Peters92_jfm}. However, during early flame kernel development in engines, the size of the flame kernel is typically smaller or on the same order of the energy-containing flow scales. \textcolor{black}{Accordingly, some models for early flame kernel development are based on the idea that only flow scales smaller than the flame size contribute to early flame kernel area production through surface wrinkling~\citep{Herweg92_model_valid,Echekki94_kernel_dns_Le_effect}.} \textcolor{black}{In figure~\ref{fig:kernel_contour_strain_topol}, flame front segments of one engine-relevant flame kernel realization and of the large flame kernel are shown at two time instants during early flame development. The comparison of both volumetric flame topologies indicates that the engine-relevant flame kernel significantly deviates from a wrinkled spherical flame, which will be quantified in section~\ref{ssec:kernel_deform}. While the flame surface of the large flame kernel becomes successively wrinkled by turbulence, the small kernel exhibits only moderate deviations from the flame-mean curvature in most parts, but features some regions with excessive positive curvature. Hence, the curvature distribution as a measure for the flame front geometry can be expected to differ substantially between both flames, which will be investigated in~\S\ref{sec:flame_wrinkling}. 
Additionally, figures~\ref{fig:kernel_contour_strain_topol}\,(a) and~(b) suggest that the flame front normal orientation of the engine-relevant kernel is preferentially aligned with the most compressive principal strain vectors and that high-strain regions tend to form flat nibs. To identify the role of large-scale flow structures in shaping the young flame, the length-scale dependence of flame front/strain alignment statistics will be analysed in~\S\ref{sec:alignment}. }
\subsection{Early Flame Kernel Deformation}
\label{ssec:kernel_deform}
\textcolor{black}{Figure~\ref{fig:kernel_contour_strain_topol} shows that the topology of the engine kernel is quite different from that of the larger kernel. This will be discussed based on quantitative measures in this section.} Qualitative regime diagrams which characterize flame kernel/flow interactions based on a velocity- and a length scale ratio~\citep{Echekki07_kernel_vortex_map_01,Vasudeo10_kernel_vortex_map_01,Reddy11_regime_map} distinguish between kernel wrinkling and kernel breakthrough or breakup behaviour. In the latter regime, high straining flow motion leads to a global distortion of the small flame and eventually generates spatially separated regions of burned mixture, possibly resulting in local extinction. 
In the present engine-relevant DNS datasets, the flame kernels maintain a coherent flame surface at all times. Still, the early flame/turbulence interaction significantly differs from the large flame kernel, as will be shown below. \par
\begin{figure}
\centering
\includegraphics[trim={0cm 3cm 2cm 2cm},clip,width=7cm]{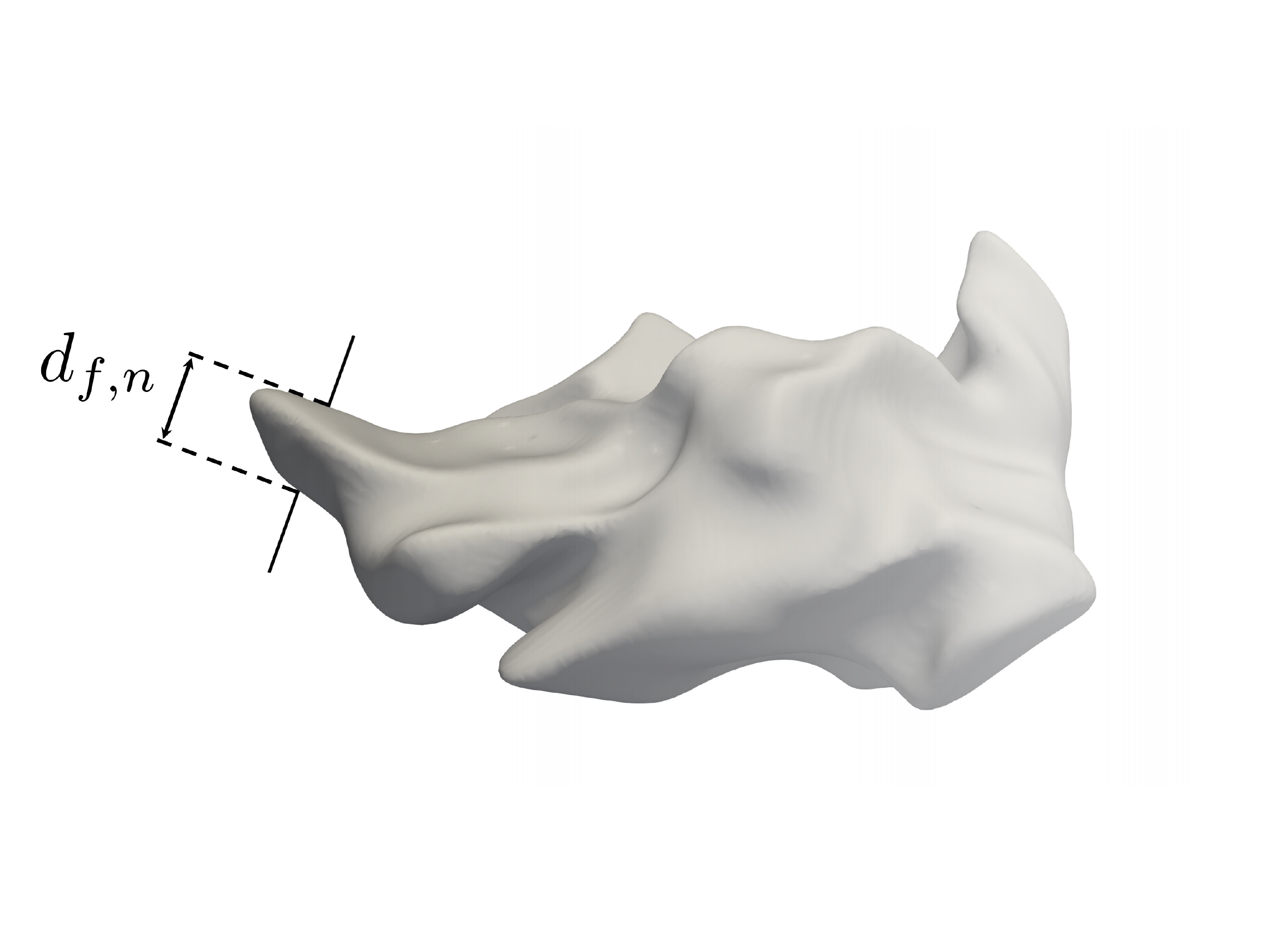}
\caption{Definition of the burned region thickness~$d_{f,n}$. }
\label{fig:sketch_Lstl_def}
\end{figure}

To track flame topology changes, \citet{Echekki07_kernel_vortex_map_01} used flame length as a parameter in a study on two-dimensional laminar vortex/flame kernel interactions. In cases which were attributed to the kernel breakthrough regime, interactions on the product side between the leading and the trailing edges of the highly distorted flame fronts were identified. 
In this work, the local thickness~$d_{f,n}$ of the burned-gas region \textcolor{black}{enclosed by a temperature iso-surface} is used to parameterize the \textcolor{black}{volumetric flame topology}. \textcolor{black}{At every surface point, $d_{f,n}$ is determined by launching a ray into the opposite direction of the local normal vector and computing the distance to the closest intersection with the iso-surface on the opposite side of the flame. Hence, $d_{f,n}$ characterizes the distance between the leading and the trailing edge of two flame segments on either side of the flame kernel.} 
An illustration is provided in figure~\ref{fig:sketch_Lstl_def}.
\graphicspath{{./data/FLAME_KERNEL_DNS_02/180916_local_geom_lseg/visualize}}
 \textcolor{black}{To obtain a flame topology measure that takes into account the flame kernel size and is not affected by the choice of the temperature iso-surface, $d_{f,n}$ has been normalized by the equivalent diameter~$D_{\mathrm{v}}$ of a sphere with the same burned volume as enclosed by the iso-surface. } \par
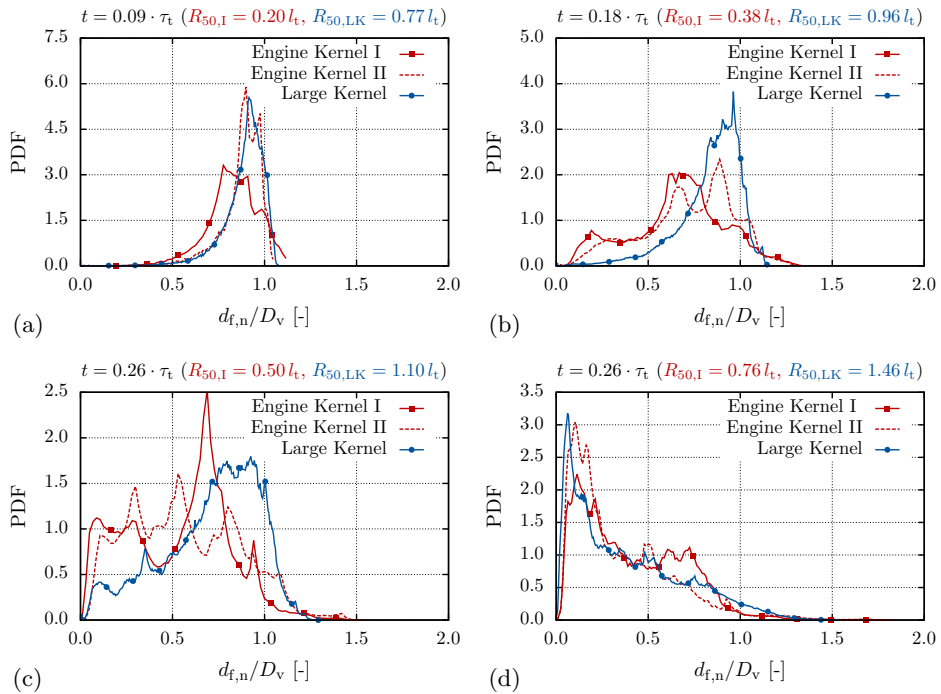
\begin{figure}
\centering
\graphicspath{{./data/FLAME_KERNEL_DNS_02/180916_local_geom_lseg_T_1700_to_2000K/t1p5e-5/}}
\begin{minipage}[b]{0.45\linewidth}
  \centering
  \makebox[0pt][l]{\quad(a)}\scalebox{0.7}{\input{./data/FLAME_KERNEL_DNS_02/180916_local_geom_lseg_T_1700_to_2000K/t1p5e-5/Lstl_PDF_normDv_0p10Tau_02_LOC01_03_JFM.tex}}
\end{minipage}
\graphicspath{{./data/FLAME_KERNEL_DNS_02/180916_local_geom_lseg_T_1700_to_2000K/t3p0e-5/}}
\begin{minipage}[b]{0.45\linewidth}
  \centering
  \makebox[0pt][l]{\quad(b)}\scalebox{0.7}{\input{./data/FLAME_KERNEL_DNS_02/180916_local_geom_lseg_T_1700_to_2000K/t3p0e-5/Lstl_PDF_norm_Dv_0p18Tau_02_LOC01_03_JFM.tex}}
\end{minipage}
\graphicspath{{./data/FLAME_KERNEL_DNS_02/180916_local_geom_lseg_T_1700_to_2000K/t4p5e-5/}}
\begin{minipage}[b]{0.45\linewidth}
  \centering
  \makebox[0pt][l]{\quad(c)}\scalebox{0.7}{\input{./data/FLAME_KERNEL_DNS_02/180916_local_geom_lseg_T_1700_to_2000K/t4p5e-5/Lstl_PDF_norm_Dv_0p26Tau_02_LOC01_03_JFM.tex}}
\end{minipage}
\graphicspath{{./data/FLAME_KERNEL_DNS_02/180916_local_geom_lseg_T_1700_to_2000K/t9p0e-5/}}
\begin{minipage}[b]{0.45\linewidth}
  \centering
  \makebox[0pt][l]{\quad(d)}\scalebox{0.7}{\input{./data/FLAME_KERNEL_DNS_02/180916_local_geom_lseg_T_1700_to_2000K/t9p0e-5/Lstl_PDF_norm_Dv_0p53Tau_02_LOC01_03_JFM.tex}}
\end{minipage}
\caption{PDFs of normalized burned region thickness at different times. \textcolor{black}{$D_{\mathrm{v}}$ is the diameter of a sphere with the same burned volume as the actual flame.} (a) $t=0.09\cdot\tau_{\mathrm{t}}$, (b) $t=0.18\cdot\tau_{\mathrm{t}}$, (c) $t=0.26\cdot\tau_{\mathrm{t}}$ and (d) $t=0.53\cdot\tau_{\mathrm{t}}$.}
\label{fig:kernel_02_03_Lstl}
\end{figure}

\textcolor{black}{The~PDFs of~${d_{f,n}/D_{\mathrm{v}}}$ shown in figure~\ref{fig:kernel_02_03_Lstl} have been computed from~13 temperature iso-surfaces across the reaction zone of each flame and} are considered as a measure for the overall flame shape. For a perfect sphere, one would obtain a delta function at unity. 
Close to the flame kernel breakthrough regime, a significant portion of the flame is expected to feature small distances between the leading and the trailing edges. 
For investigations on kernel extinction, normalization by the laminar flame thickness might be more suitable. \par
As can be seen in the figure, the geometry of \textcolor{black}{all three} flame kernels is initially determined by the ignition heat source. However, the PDFs develop very differently. While the large flame kernel maintains its overall shape during early flame growth, the small kernels quickly develop very thin geometric patterns \textcolor{black}{with higher statistical significance}. 
This is clearly visible during ($t=0.18..0.26\cdot \tau_{\mathrm{t}}$), when the overall \textcolor{black}{multi}modal distributions of ${d_{f,n}/D_{\mathrm{v}}}$ indicate distorted engine flame kernel topologies, \textcolor{black}{which significantly deviate from spherical shape}. By contrast, wrinkling of the large flame kernel by turbulent eddies slowly develops thin flame structures \textcolor{black}{(`fingers')}. This rather gradual process is indicated by the long monotonous tail of the PDF at ($t=0.18\cdot \tau_{\mathrm{t}}$). 
\textcolor{black}{Flame front corrugation by turbulence eventually causes high probability of thin structures, while distances~$d_{f,n}$ on the order of the equivalent flame diameter~$D_{\mathrm{v}}$ become rather unlikely.} In fact, the PDFs of \textcolor{black}{all three} flame kernels approach a similar distribution at ($t=0.53\cdot \tau_{\mathrm{t}}$). This indicates that the different nature of early turbulent flame kernel development depending on the length scale ratio~$D_{0}/l_{\mathrm{t}}$ is restricted to the very initial phase. \textcolor{black}{With respect to differences between both engine-relevant flame kernel realizations, Engine Kernel~I is initially subject to stronger deformation (cf.\ figure~\ref{fig:kernel_02_03_Lstl}\,(a)) caused by large-scale turbulent flow structures, as will be shown in \S~\ref{sec:alignment}. While turbulence rapidly generates small geometric length scales in similar portions of the flame kernels during ($t=0.18..0.26\cdot \tau_{\mathrm{t}}$), noticeable differences between both realizations are shown at larger~${d_{f,n}/D_{\mathrm{v}}}$. The gradual disappearance of local maxima on the order of at least the equivalent flame radius may be interpreted as a measure for the transition to a fully developed turbulent flame.} \par



\section{\textcolor{black}{Flame Front Geometry Evolution}}
\label{sec:flame_wrinkling}
\textcolor{black}{In the previous section, it was shown that the topology of the early flame kernel has a specific signature owing to its small size compared with the large turbulent scales, which can have a significant impact on the early transition to a fully turbulent flame. Since the global heat release rate is coupled to the evolution of total flame area, which is in turn related to the first two central moments of the curvature distribution (cf.~(\ref{eq:one_ov_A_dAdt_avg_mom})), the dynamics of flame curvature will be investigated hereafter. In section~\ref{sec:curv_var}, the dependence of flame wrinkling on~$D_{0}/l_{\mathrm{t}}$ will be assessed based on the curvature variance evolution. Run-to-run variations in the early flame front geometry of engine-relevant flame kernels will be attributed to flame/turbulence interactions by analysis of the mean curvature balance equation in section~\ref{sec:curv_trsp}. 
Further, the development of flat nibs with high positive curvature at the tips shown for the engine-relevant flame kernel configuration (cf.\ figure~\ref{fig:kernel_contour_strain_topol}) suggests a characteristic evolution of the third central moment (skewness) of the curvature distribution, which will be considered as a marker for characteristic flame kernel/turbulence interactions in section~\ref{ssec:mean_curv_skew}.}
\subsection{\textcolor{black}{Variance of the Mean Curvature Distribution}}
\label{sec:curv_var}
\begin{figure}
  \centering
  \scalebox{0.7}{\input{./data/FLAME_KERNEL_DNS_02/181003_curv_strain_kin_restor_diss_divg_skew/flame_kernel_curvVar_02_LOC01_03_P02_JFM_HALF.tex }}
\caption{ \textcolor{black}{Evolution of the curvature variance for all flames. The upper x-axis indicates the size of Engine Kernel~I.}}
\label{fig:curvVar_K02_K03_P02}
\end{figure}
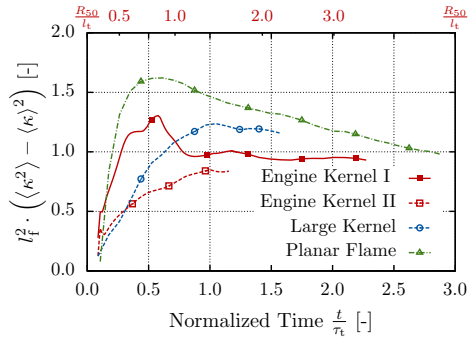

\textcolor{black}{In the following, the flame front geometry evolution of all flames will be quantified by the respective curvature variance histories, which are shown in figure~\ref{fig:curvVar_K02_K03_P02}. Note that the upper x-axis indicates the size of Engine Kernel~I. }
%
%
\textcolor{black}{If the transition from a quasi-laminar flame kernel to a fully developed turbulent flame was governed by the high-wavenumber range of the energy spectrum characterized by an upper cut-off scale on the order of the flame diameter, one would expect the rate of variance formation to increase with~$D_{0}/l_{\mathrm{t}}$. By comparing the results from the planar flame, the large flame kernel and Engine Kernel~II, this assumption seems to hold. However, Engine Kernel~I develops curvature variance significantly faster than the large flame kernel and noticeably exceeds the variance level of Engine Kernel~II during the early development phase.} \par
%
\graphicspath{{./data/FLAME_KERNEL_DNS_02/181003_curv_strain_kin_restor_diss_divg_skew/}}
\textcolor{black}{Since the governing mechanisms of curvature production and dissipation vary in positively and negatively curved flame zones~\citep{Dopazo18_curvEq}, both regions will be distinguished in the following analysis. For brevity, only the curvature variance conditioned on the sign of the curvature (cf.~(\mbox{S-2.1}) in the supp. material) will be discussed here, while conditionally averaged curvatures are depicted in figure~\mbox{S-3} in the supplementary material. As shown in figure~\ref{fig:curvVar_minus_plus_02_03_P02}\,(a), the conditional variance evaluated in negatively curved regions follows similar trends as discussed for the unconditional variance. By contrast, the conditional curvature variance evaluated in positively curved regions of Engine Kernel~I even exceeds the maximum reached for the planar flame  (cf.\ figure~\ref{fig:curvVar_minus_plus_02_03_P02}\,(b)). This excess variance quickly decays to similar levels as observed in the planar flame and the large flame kernel during the time interval ($0.6\tau_{\mathrm{t}}<t<0.8\tau_{\mathrm{t}}$). Simultaneously, there is a local minimum in the conditional curvature variance in negatively curved regions of Engine Kernel~I, which then recovers the variance level of Engine Kernel~II. Hence, excessive curvature variance occurs in both positively and negatively curved regions. Note that the respective local maxima develop at different times.} \textcolor{black}{More insights into these dynamics will be provided in section~\ref{sec:curv_trsp} by analysis of the mean curvature transport equation. } 
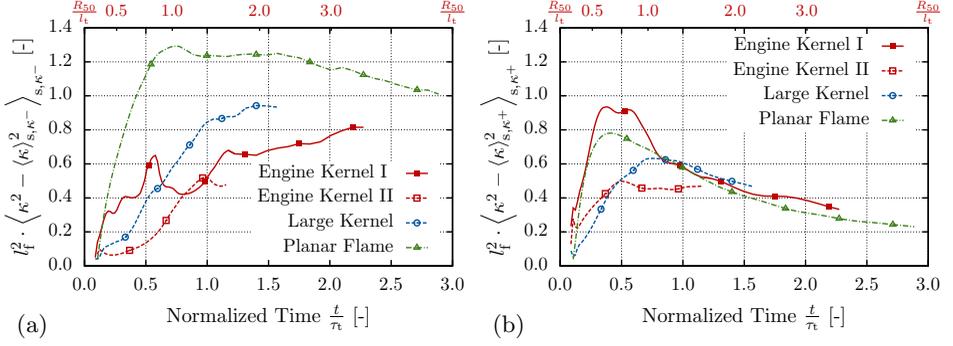
\begin{figure}
 \centering
\graphicspath{{./data/FLAME_KERNEL_DNS_02/181003_curv_strain_kin_restor_diss_divg_skew_190712/}}
\begin{minipage}[b]{0.45\linewidth}
  \centering
  \makebox[0pt][l]{\quad(a)}\scalebox{0.7}{\input{./data/FLAME_KERNEL_DNS_02/181003_curv_strain_kin_restor_diss_divg_skew_190712/flame_kernel_curvVar_minus_02_LOC01_03_P02_JFM_HALF.tex}}
\end{minipage}
\graphicspath{{./data/FLAME_KERNEL_DNS_02/181003_curv_strain_kin_restor_diss_divg_skew_190712/}}
\begin{minipage}[b]{0.45\linewidth}
  \centering
  \makebox[0pt][l]{\quad(b)}\scalebox{0.7}{\input{./data/FLAME_KERNEL_DNS_02/181003_curv_strain_kin_restor_diss_divg_skew_190712/flame_kernel_curvVar_plus_02_LOC01_03_P02_JFM_HALF.tex}}
\end{minipage}
\caption{\textcolor{black}{Evolution of the conditional curvature variance in regions of negative~(a) and positive~(b) curvature for all flames. The upper x-axes indicate the size of Engine Kernel~I.}}
\label{fig:curvVar_minus_plus_02_03_P02}
\end{figure}
\textcolor{black}{It should be noted that according to~(\ref{eq:one_ov_A_dAdt_avg_mom}), the afore-discussed decay of excessive curvature variance is responsible for a plateau in total flame area rate-of-change observed for Engine Kernel~I~\citep{Falkenstein19_kernel_Le1_cnf}, similar to some realizations of previous two- \citep{Thevenin02_dns_2d_3d} and three-dimensional flame kernel DNS~\citep{Thevenin05_kernel_dns_3d}.} 
\subsection{\textcolor{black}{Mean Curvature Transport}}
\label{sec:curv_trsp}
We start from the mean curvature transport formulation for scalar iso-surfaces propagating with non-uniform displacement speed, as recently derived by~\citet{Dopazo18_curvEq}. 
\textcolor{black}{Advection and the reaction-diffusion imbalance result in a total velocity of every scalar iso-surface element, which is given by}
\begin{equation}
u_i^{(T)} = u_i + \mathrm{s}_{\mathrm{d}}^{(T)} n^{(T)}_i.
\label{eq:u_surfT}
\end{equation}
\textcolor{black}{For the definition of the displacement speed $\mathrm{s}_{\mathrm{d}}^{(T)}$, refer to~(\ref{eq:sc_tot_d_dt}). The superscript denotes the scalar field used for analysis, which is the temperature in this case. }
In order to follow the physical analysis of curvature dynamics by~\citet{Pope88_surf_turb} more closely, the equation is slightly reformulated as (cf. appx.~\ref{appCurv} for details):

\begin{eqnarray}
\frac{\mathrm{D}_{T} \left( \kappa^{(T)} \right)}{\mathrm{D}_{T}\left(t\right)} = \kappa^{(T)} a_{\mathrm{n}}^{(T)} - 2 S_{ij}^{(T)} \frac{\p n^{(T)}_j}{\p x_i} 
- \left[ \frac{\p^2 u_j^{(T)}}{\p x_i^2} n_j 
- \frac{\p^2 u_{\mathrm{n}}^{(T)}}{\p x_{\mathrm{n}}^2} \right].
\label{eq:curv_eq}
\end{eqnarray}
The total~\textcolor{black}{(kinematic)} strain rate tensor is given by
\begin{equation}
S_{ij}^{(T)} = \frac{1}{2} \left(\frac{\p u_i^{(T)}}{\p x_j} + \frac{\p u_j^{(T)}}{\p x_i} \right),
\label{eq:Sij_surfT}
\end{equation}
and the total normal strain rate is
\begin{equation}
a_{\mathrm{n}}^{(T)} = n^{(T)}_i \frac{\p u_i^{(T)}}{\p x_j} n^{(T)}_j = \frac{\p u_{\mathrm{n}}^{(T)}}{\p x_{\mathrm{n}}}.
\label{eq:aN_surfT}
\end{equation}
For brevity, the superscript will be omitted  below. In order to separate effects related to the velocity field (superscript $u$) from those related to flame propagation (superscript $\mathrm{p}$), (\ref{eq:curv_eq}) can be expanded as
\begin{eqnarray}
\frac{\mathrm{D}_{T} \left( \kappa \right)}{\mathrm{D}_{T}\left(t\right)} &=&
\xoverbrace{\kappa a_{\mathrm{n}}}^{ \mathfrak{T}_{\mathrm{n}}^{u} }  + 
\xoverbrace{\kappa \frac{\p \mathrm{s}_{\mathrm{d}}}{\p x_{\mathrm{n}}}}^{ \mathfrak{T}_{\mathrm{n}}^{\mathrm{p}} }
\;\xoverbrace{- 2 S_{ij} \frac{\p n_j}{\p x_i}}^{ \mathfrak{T}_{\mathrm{t}}^{u} }
\;\xoverbrace{- 2 S_{ij}^\mathrm{p} \frac{\p n_j}{\p x_i}}^{ \mathfrak{T}_{\mathrm{t}}^{\mathrm{p}} }  \nonumber \\
&& \underbrace{ - \left[ \frac{\p^2 u_j}{\p x_i^2} n_j - \frac{\p^2 u_{\mathrm{n}}}{\p x_{\mathrm{n}}^2} \right]}_{ \mathfrak{T}_{\mathrm{b}}^{u} }
\; \underbrace{- \left[ \frac{\p^2 \left(\mathrm{s}_{\mathrm{d}} n_j \right)}{\p x_i^2} n_j - \frac{\p^2 \mathrm{s}_{\mathrm{d}} }{\p x_{\mathrm{n}}^2} \right]}_{ \mathfrak{T}_{\mathrm{b}}^{\mathrm{p}} },
\label{eq:curv_eq_all_terms}
\end{eqnarray}
where the flame propagation strain tensor is defined as
\begin{equation}
S_{ij}^\mathrm{p} = \frac{1}{2} \left(\frac{\p \left(\mathrm{s}_{\mathrm{d}} n_i \right) }{\p x_j} + \frac{\p \left( \mathrm{s}_{\mathrm{d}} n_j \right)}{\p x_i} \right).
\label{eq:Sp_ij}
\end{equation}
According to~\citet{Pope88_surf_turb}, the evolution of curvature is determined by a balance between bending and straining of the scalar iso-surface. In the r.h.s.\ of~(\ref{eq:curv_eq_all_terms}), three main effects can be identified among the flow-field and flame-propagation-related terms. Effects of normal strain (subscript n) due to non-uniformities of the flow and propagation velocities along the normal direction of an already curved scalar iso-surface are accounted for by~$\mathfrak{T}_{\mathrm{n}}^{u}$ and~$\mathfrak{T}_{\mathrm{n}}^{\mathrm{p}}$. Additionally, the geometry of a surface with non-zero curvature may be altered by flow and propagation velocity gradients in the tangential plane, which is quantified by the tangential strain terms (subscript t)~$\mathfrak{T}_{\mathrm{t}}^{u}$ and~$\mathfrak{T}_{\mathrm{t}}^{\mathrm{p}}$. 
The third effect is added by the terms~$\mathfrak{T}_{\mathrm{b}}^{u}$ and~$\mathfrak{T}_{\mathrm{b}}^{\mathrm{p}}$, which consider second derivatives of the flow and propagation velocities in the iso-surface-tangential directions and cause surface bending (subscript b). Among these three phenomena, bending is the only mechanism which can cause an initially planar iso-surface geometry to become wrinkled. \textcolor{black}{Note that flame kernels are initialized with (possibly large) positive curvature, i.e.\ the~$\mathfrak{T}_{\mathrm{n}}^{\ast}$ and~$\mathfrak{T}_{\mathrm{t}}^{\ast}$ terms may significantly contribute to the curvature balance from  the beginning.} \par
For the analysis of flame curvature dynamics, all terms in the r.h.s.\ of~(\ref{eq:curv_eq_all_terms}) have been integrated over the flame structure, but split into contributions from regions with positive and negative curvature \textcolor{black}{(cf.~(\mbox{S-2.1}) in the supp. material)}.  
A statistical analysis on curvature dynamics in a jet flame, but using the curvature equation formulation proposed by~\citet{Dopazo18_curvEq} has been recently carried out by~\citet{Cifuentes18_curv_eq_kempf}.
Note that in the formulation~(\ref{eq:curv_eq_all_terms}), the two bending effects are related to the terms considered by~\citet{Dopazo18_curvEq} as ($\mathfrak{T}_{\mathrm{b}}^{u} = T_2+T_4+T_5$) and ($\mathfrak{T}_{\mathrm{b}}^{\mathrm{p}}=T_7+T_9+T_{10}$). As shown by~\citet{Cifuentes18_curv_eq_kempf}, the terms $T_2$ and $T_4$, as well as $T_7$ and $T_9$ mostly balance each other when conditioned upon the reaction progress variable. \par 
\textcolor{black}{The following discussion is focussed on the distinctive flame curvature distribution evolution of Engine Kernel~I, characterized by the temporary occurrence of high variance (cf.\ figure~\ref{fig:curvVar_K02_K03_P02}). 
It will be shown that due to the small initial kernel diameter and correspondingly high positive curvature, strong tangential strain may immediately amplify the curvature magnitude through stretching~($\mathfrak{T}_{\mathrm{t}}^{u}$). By contrast, rapid production of large negative curvature magnitudes by tangential derivatives in flame displacement speed~($\mathfrak{T}_{\mathrm{t}}^{\mathrm{p}}$) relies on the initial formation of negatively curved flame regions by second derivatives of the velocity field through bending~($\mathfrak{T}_{\mathrm{b}}^{u}$).} \par
\begin{figure}
\centering
\graphicspath{{./data/FLAME_KERNEL_DNS_02/181002_curvEq_Ti_diss_skew_mp_curv_mp_progT/}}
\begin{minipage}[b]{0.45\linewidth}
  \centering
  \makebox[0pt][l]{\quad(a)}\scalebox{0.7}{\input{./data/FLAME_KERNEL_DNS_02/181002_curvEq_Ti_diss_skew_mp_curv_mp_progT/flame_kernel_curvEq_flow_flame_net_terms_prodMINUS_02_LOC01_JFM_HALF.tex}}
\end{minipage}
\graphicspath{{./data/FLAME_KERNEL_DNS_02/181002_curvEq_Ti_diss_skew_mp_curv_mp_progT/}}
\begin{minipage}[b]{0.45\linewidth}
  \centering
  \makebox[0pt][l]{\quad(b)}\scalebox{0.7}{\input{./data/FLAME_KERNEL_DNS_02/181002_curvEq_Ti_diss_skew_mp_curv_mp_progT/flame_kernel_curvEq_flow_flame_net_terms_prodPLUS_02_LOC01_JFM_HALF.tex}}
\end{minipage}
\graphicspath{{./data/FLAME_KERNEL_DNS_02/181002_curvEq_Ti_diss_skew_mp_curv_mp_progT/}}
\begin{minipage}[b]{0.45\linewidth}
  \centering
  \makebox[0pt][l]{\quad(c)}\scalebox{0.7}{\input{./data/FLAME_KERNEL_DNS_02/181002_curvEq_Ti_diss_skew_mp_curv_mp_progT/flame_kernel_curvEq_flow_terms_prodMINUS_Bending_1p2TAU_02_LOC01_JFM_HALF.tex}}
\end{minipage}
\graphicspath{{./data/FLAME_KERNEL_DNS_02/181002_curvEq_Ti_diss_skew_mp_curv_mp_progT/}}
\begin{minipage}[b]{0.45\linewidth}
  \centering
  \makebox[0pt][l]{\quad(d)}\scalebox{0.7}{\input{./data/FLAME_KERNEL_DNS_02/181002_curvEq_Ti_diss_skew_mp_curv_mp_progT/flame_kernel_curvEq_flow_terms_prodPLUS_Bending_1p2TAU_02_LOC01_JFM_HALF.tex}}
\end{minipage}
\graphicspath{{./data/FLAME_KERNEL_DNS_02/181002_curvEq_Ti_a_diss_skew_mp_curv_mp_progT/}}
\begin{minipage}[b]{0.45\linewidth}
  \centering
  \makebox[0pt][l]{\quad(e)}\scalebox{0.7}{\input{./data/FLAME_KERNEL_DNS_02/181002_curvEq_Ti_a_diss_skew_mp_curv_mp_progT/flame_kernel_curvEq_flame_terms_prodMINUS_Bending_1p2TAU_02_LOC01_JFM_HALF.tex}}
\end{minipage}
\graphicspath{{./data/FLAME_KERNEL_DNS_02/181002_curvEq_Ti_a_diss_skew_mp_curv_mp_progT/}}
\begin{minipage}[b]{0.45\linewidth}
  \centering
  \makebox[0pt][l]{\quad(f)}\scalebox{0.7}{\input{./data/FLAME_KERNEL_DNS_02/181002_curvEq_Ti_a_diss_skew_mp_curv_mp_progT/flame_kernel_curvEq_flame_terms_prodPLUS_Bending_1p2TAU_02_LOC01_JFM_HALF.tex}}
\end{minipage}
\caption{\textcolor{black}{Net production terms of mean curvature (cf.~r.h.s.\ in (\ref{eq:curv_eq_all_terms})) for all flame kernels~(a,b). Production of negative curvature by velocity field~(c) and flame propagation~(e) and production of positive curvature by velocity field~(d) and flame propagation~(f) for Engine Kernel~I (solid lines) and Engine Kernel~II (dashed lines). The upper x-axes indicate the size of Engine Kernel~I.}} 
\label{fig:kernel_DcurvDt_terms_bending_zoom}
\end{figure}
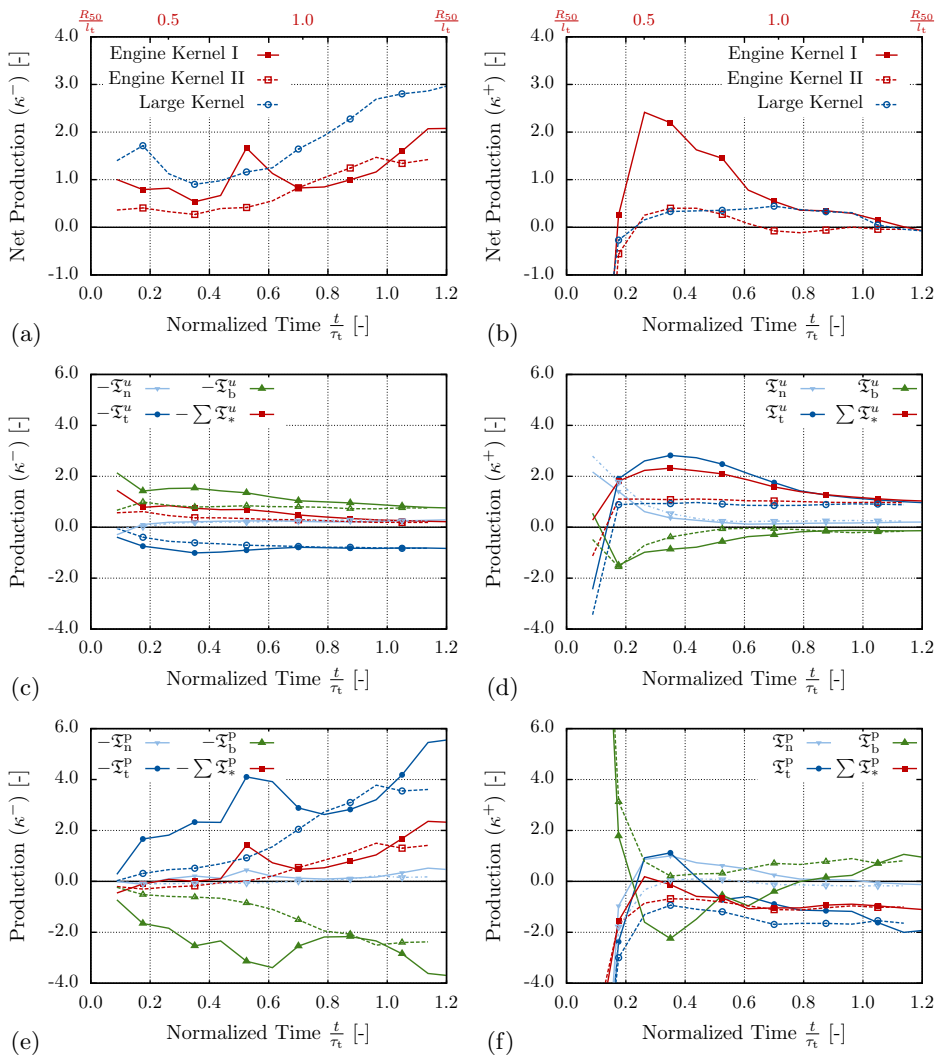

\textcolor{black}{In figures~\ref{fig:kernel_DcurvDt_terms_bending_zoom}\,(a) and~(b), the histories of net curvature production in negatively and positively curved flame regions are shown for all three flame kernels. Note that the sign convention in the figures is such that positive terms describe production of curvature, i.e. the lines show the evolution of $\left(\sum \mathrm{sgn}\left(\kappa\right)\cdot \mathfrak{T}_{\ast}^{\ast}\right)$. According to the local maxima in conditional curvature variance of Engine Kernel~I (cf.\ figure~\ref{fig:curvVar_minus_plus_02_03_P02}), one would expect initially excessive production of positive curvature, and slightly delayed but pronounced production of negative curvature. This is confirmed in figures~\ref{fig:kernel_DcurvDt_terms_bending_zoom}\,(b) and~(a) by comparison to the other two flame kernels. Further, it is shown that the engine-relevant flame kernel configuration exhibits strong run-to-run variations in curvature production. The physical mechanisms leading to the temporary occurrence of substantial curvature production differ between positively and negatively curved flame regions, as will be described hereafter.} \par
\textcolor{black}{In figures~\ref{fig:kernel_DcurvDt_terms_bending_zoom}\,(d) and~(f), the production of positive curvature by flow terms $\mathfrak{T}_{\ast}^u$ and flame propagation terms $\mathfrak{T}_{\ast}^{\mathrm{p}}$ is depicted for both engine-relevant flame kernels. For (${t>1.0\cdot\tau_{\mathrm{t}}}$), only minor differences between flame kernel realizations persist and the present results are in good qualitative agreement with previous studies~\citep{Cifuentes18_curv_eq_kempf,Alqallaf19_kernel_curv_eq}. However, noticeable deviations between both flame kernels exist during the early development phase. From figure~\ref{fig:kernel_DcurvDt_terms_bending_zoom}\,(d) it can be concluded that the main reason for stronger initial production of positive curvature in Engine Kernel~I is the temporarily increased flow tangential strain term ($\mathfrak{T}_{\mathrm{t}}^{u}$). It should be noted that this term is particularly effective, if the flame front features high positive curvature (cf.\ figure~\mbox{S-4}\,(a) in the supp. material), 
which is intrinsic to the engine-relevant flame kernel configuration. Further, also the dissipative net effect of flame propagation on positively curved flame regions (cf.\ figure~\ref{fig:kernel_DcurvDt_terms_bending_zoom}\,(f)) is weaker in Engine Kernel~I compared to~II during (${0.2\tau_{\mathrm{t}}<t<0.5\tau_{\mathrm{t}}}$). Note that the straining/bending balance due to flame propagation is reversed during this phase, which is a direct consequence of increased probability of high positive curvatures. As shown in previous studies~\citep{Cifuentes18_curv_eq_kempf,Alqallaf19_kernel_curv_eq}, the curvature-conditioned means of $\mathfrak{T}_{\mathrm{t}}^{\mathrm{p}}$ and $\mathfrak{T}_{\mathrm{b}}^{\mathrm{p}}$ exhibit a saddle point at ${\kappa = 0}$, while both terms substantially deviate from zero only at elevated curvatures (cf.\ figures~\mbox{S-4}\,(c) and~(d) in the supp. material). 
Hence, the temporary occurrence of extreme curvatures may change the respective overall straining/bending effect in sign. Additionally, the dissipative effect of flame propagation on positively curved regions in Engine Kernel~I is weakened by the increased level of  $\mathfrak{T}_{\mathrm{n}}^{\mathrm{p}}$ during early flame kernel development.} \par
\textcolor{black}{Although the development of excessive curvature variance in negatively curved regions of Engine Kernel~I is less pronounced compared to positively curved zones (cf.\ figure~\ref{fig:curvVar_minus_plus_02_03_P02}), noticeable differences in negative curvature production exist between both flame kernel realizations, as shown in figures~\ref{fig:kernel_DcurvDt_terms_bending_zoom}\,(c) and~(e). At late times, the production of negative curvature is governed by the flame propagation term~$\mathfrak{T}_{\mathrm{t}}^{\mathrm{p}}$. Since the curvature-conditioned mean of this term is small at curvature radii well above the laminar flame thickness (cf.\ figure~\mbox{S-4}\,(c) in the supp. material), the initial generation of negatively curved flame kernel regions may be accelerated by a different mechanism. As shown in figure~\ref{fig:kernel_DcurvDt_terms_bending_zoom}\,(c), bending caused by second derivatives of the fluid velocity ($\mathfrak{T}_{\mathrm{b}}^{u}$) is initially the dominant process, which is strongest in Engine Kernel~I. Hence, the flow field is responsible for the formation of some negatively curved flame kernel regions, where the curvature magnitude is amplified by the flame propagation term~$\mathfrak{T}_{\mathrm{t}}^{\mathrm{p}}$. Note that the combination of both effects seems to require a certain lag time, which explains the delayed local maximum in conditional variance in figure~\ref{fig:curvVar_minus_plus_02_03_P02}\,(a).} \par
\textcolor{black}{To summarize, the analysis of mean flame curvature transport terms has shown that Engine Kernel~I temporarily exhibits high curvature variance due to two primary effects caused by early flame kernel/flow interactions. First, high tangential strain produces excessive positive curvatures. This effect is favoured by the initially large positive curvature intrinsic to the engine-relevant flame kernel configuration, but requires certain flow conditions at the ignition location, which may vary strongly between different kernel realizations. Second, turbulent eddies cause flame front bending through second derivatives of the velocity field, which leads to initial formation of negative curvature. Moderate negative curvature magnitudes are further amplified by flame propagation, which leads to cusp formation. In the next section, it will be shown that the initial amplification of positive curvature accompanied by inhibited formation of negative curvature results in a positively skewed curvature distribution of small flame kernels, which is not observed in the other flame configurations. }

\subsection{Skewness of the Mean Curvature Distribution}
\label{ssec:mean_curv_skew}
It is well known that in both laminar~\citep{Zeldovich66_lam_cusp} and \textcolor{black}{weakly} turbulent~\citep{Karlovitz51_turb} premixed flames, propagation of corrugated fronts causes cusp formation by Huygens' principle leading to skewness towards negative curvatures\textcolor{black}{~\citep{Shephered92_curv_symm_turb,Shephered02_curv_turb}}. In \textcolor{black}{highly} turbulent flows, two additional mechanisms of \textcolor{black}{negatively curved} cusp formation resulting from flow-induced flame collisions were proposed by~\citet{Poludnenko11_cusp_turb}. \textcolor{black}{Since the curvature skewness contains the signatures of flame propagation, thermal expansion, and turbulent flame wrinkling, ~\citet{Creta16_lam_turb_skew} proposed the skewness of the curvature PDF as a marker for the occurrence of Darrieus-Landau instabilities in weakly turbulent flames. Hereafter,} \textcolor{black}{curvature skewness will be shown to contain the signature of characteristic flame kernel/turbulence interactions.}\par
\begin{figure}
\centering
\graphicspath{{./data/FLAME_KERNEL_DNS_02/181003_curv_strain_kin_restor_diss_divg_skew/}}
\begin{minipage}[b]{0.45\linewidth}
\scalebox{0.7}{\input{./data/FLAME_KERNEL_DNS_02/181003_curv_strain_kin_restor_diss_divg_skew/flame_kernel_curvSkew_02_LOC01_03_P02_JFM_HALF.tex}}
\end{minipage}
\caption{\textcolor{black}{Skewness of the surface-weighted curvature distribution as function of time.}}
\label{fig:curv_skew_02_03_P02}
\end{figure}
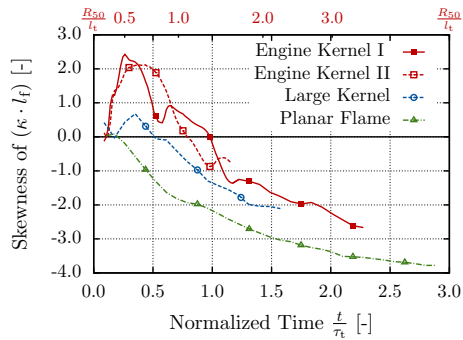

 
\textcolor{black}{In figure~\ref{fig:curv_skew_02_03_P02}, the skewness of the surface-weighted curvature distribution is depicted for each flame as a function of time. Towards the end of the simulations, the curvature distributions of all flames are indeed negatively skewed.} This behaviour can be explained by the dominance of the flame propagation terms in the mean curvature balance equation, particularly $\mathfrak{T}_{\mathrm{t}}^{\mathrm{p}}$, which produce negative curvature (cf.\ figure~\ref{fig:kernel_DcurvDt_terms_bending_zoom}\,(e)).
\textcolor{black}{With respect to flame kernel development, recall that the analysis in section~\ref{sec:curv_trsp} has shown that the initial amplification of high positive curvature by~$\mathfrak{T}_{\mathrm{t}}^{u}$ is a feature of the engine-relevant flame kernel configuration due to flame initiation with small diameter. According to figures~\ref{fig:kernel_contour_strain_topol}\,(a) and~(b), excessive positive curvatures are generated in regions with high compressive strain, which is consistent with the negative sign of~$\mathfrak{T}_{\mathrm{t}}^{u}$ and may be further investigated by considerations in a flame-tangential, local coordinate system~\citep{Cifuentes18_curv_eq_kempf}. In figure~\mbox{S-5} of the supplementary material, preferential alignment of the principal axes corresponding to the most positive curvature and the most compressive strain is demonstrated for both engine-relevant flame kernels. The resulting amplification of positive curvature during the early kernel development phase induces a long tail on the positive side of the curvature distribution. This effect has been observed in both engine-relevant flame kernel realizations, but to different extent. Regarding negative curvature, recall that the initial formation is caused by bending due to turbulent eddies, which induce non-zero second derivatives of the flow velocity. In figure~\ref{fig:kernel_DcurvDt_terms_bending_zoom}\,(a) it was shown that the net production of negatively curved flame regions is weaker for the engine-relevant flame kernels compared to the large kernel. Hence, the flow scales responsible for the bending effect may not be able to effectively alter the flame surface of kernels with small size. The combination of delayed production of negatively curved and strong production of positively curved zones causes the curvature distribution of both Engine Kernel~I and~II to become noticeably skewed towards positive curvature, as shown in figure~\ref{fig:curv_skew_02_03_P02}. Note that the delayed production of negative curvature is particularly evident for Engine Kernel~II, while Engine Kernel~I initially exhibits very strong production of positive curvature (cf.\ figures~\ref{fig:kernel_DcurvDt_terms_bending_zoom}\,(a) and~(b)). } 
In section~4 of the supplementary material, it is shown that the occurrence of strong positive skewness is not just an artefact of the present engine-relevant flame kernel realizations, but unique to the small flame kernel configuration. \textcolor{black}{Hence, the curvature skewness contains information on the early flame kernel topology evolution caused by the delayed formation of negatively as opposed to positively curved regions, and most prominently, on the effect of high compressive strain leading to excessive positive curvatures. The latter effect will be related to large-scale turbulent flow motion in \S\ref{sec:alignment}.} \par
\section{\textcolor{black}{Length-Scale-Dependent Flame Front/Flow Field Alignment}}
\label{sec:alignment}
\textcolor{black}{Alignment statistics have been used by several researchers to quantify the evolution of non-reacting~\citep{Ashurst87_sc_align} and reacting~\citep{Kim07_sc_align,Chakraborty07_sc_align} scalar fields. Local alignment between scalar gradients and strain principal axes determines normal and tangential strain rates, which in turn appear in the transport equations of scalar gradient magnitude or scalar dissipation rate, iso-surface area~(cf.~(\ref{eq:one_ov_A_dAdt})) and curvature~(cf.~(\ref{eq:curv_eq_all_terms})), among others. Non-reacting scalar gradients tend to align with the most compressive strain~\citep{Ashurst87_sc_align}, while in premixed flames, predominance of heat release induces preferential alignment with the most extensive strain as Damk{\"o}hler numbers are increased~\citep{Chakraborty07_sc_align}.} 
\textcolor{black}{In this work, alignment statistics for all principal strain components have been computed and related to data from literature (cf.\ figure~\mbox{S-7} in the supp. material), while the following discussion will be limited to the most compressive principal strain for brevity. Evidence will be provided for the dominant role of large-scale turbulent flow structures in changing the volumetric flame topology of early, engine-relevant flame kernels. } \par
Recall that in section~\ref{ssec:kernel_deform} it has been shown that significant differences exist in the very early \textcolor{black}{topological} development of the small and large flame kernels, \textcolor{black}{which is reflected in the evolution of curvature skewness as discussed in section~\ref{ssec:mean_curv_skew}}. Since the only difference between the two flame kernel configurations is the length scale ratio $D_{0}/l_{\mathrm{t}}$, it is hypothesized that the nature of early flame/turbulence interactions is significantly different in both cases. In particular, it is assumed that the engine-relevant flame kernel is initially governed by large-scale flow structures that change the overall flame topology, while the large flame kernel is subject to flame front wrinkling by flow scales which are mostly smaller than the flame size. Hence, the objective of the present analysis is to quantify the length-scale dependence of flame front orientation with respect to the velocity field. To achieve this, \textcolor{black}{small-scale velocity fluctuations were removed from the flow field by applying a low-pass filter. If early flame kernel/turbulence interactions were governed by large-scale flow structures, preferential alignment of the flame front normal vectors with the principal axes of the strain tensor evaluated from the filtered as opposed to the unfiltered velocity field would be expected. Since the focus of this analysis is on length scales of at least the same order as the flame kernel size, the filtered velocity field has been obtained by applying a box filter of size ($\Delta = 0.5 \cdot l_{\mathrm{t}}$), which is approximately equal to a characteristic radius of the small flame kernels at ($t=0.25\cdot\tau_{\mathrm{t}}$). }
This analysis is to some extent similar to a study on scale-dependence of enstrophy production by~\citet{Buxton11_enthrophy_prod}. In that study, statistics of angles between vorticity and strain tensor eigenvectors were calculated from velocity fields filtered at different scales and compared to analyse scale dependence. \par

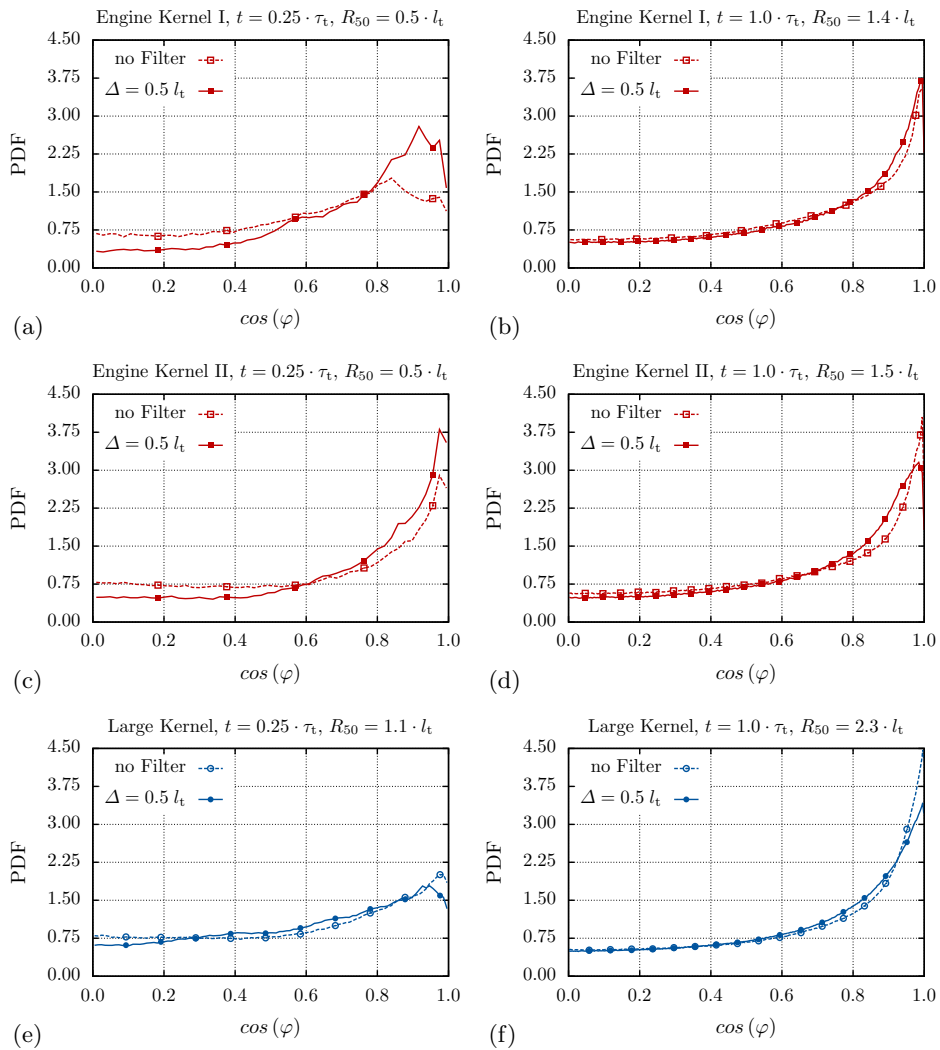
\begin{figure}
\centering
\graphicspath{{./data/FLAME_KERNEL_DNS_02/181021_alignment_curv_strain_progT_filt33/}}
\begin{minipage}[b]{0.45\linewidth}
  \centering
  \makebox[0pt][l]{\quad(a)}\scalebox{0.7}{\input{./data/FLAME_KERNEL_DNS_02/181021_alignment_curv_strain_progT_filt33/flame_kernel_nvec_ComprStrain_align_filt33_0p25TAU_02_JFM_HALF.tex}}
\end{minipage}
\graphicspath{{./data/FLAME_KERNEL_DNS_02/181021_alignment_curv_strain_progT_filt33/}}
\begin{minipage}[b]{0.45\linewidth}
  \centering
  \makebox[0pt][l]{\quad(b)}\scalebox{0.7}{\input{./data/FLAME_KERNEL_DNS_02/181021_alignment_curv_strain_progT_filt33/flame_kernel_nvec_ComprStrain_align_filt33_1p00TAU_02_JFM_HALF.tex}}
\end{minipage}
%
\graphicspath{{./data/FLAME_KERNEL_DNS_02_LOC1/181021_alignment_curv_strain_progT_filt33/}}
\begin{minipage}[b]{0.45\linewidth}
  \centering
  \makebox[0pt][l]{\quad(c)}\scalebox{0.7}{\input{./data/FLAME_KERNEL_DNS_02_LOC1/181021_alignment_curv_strain_progT_filt33/flame_kernel_nvec_ComprStrain_align_filt33_0p25TAU_02_LOC1_JFM_HALF.tex}}
\end{minipage}
\graphicspath{{./data/FLAME_KERNEL_DNS_02_LOC1/181021_alignment_curv_strain_progT_filt33/}}
\begin{minipage}[b]{0.45\linewidth}
  \centering
  \makebox[0pt][l]{\quad(d)}\scalebox{0.7}{\input{./data/FLAME_KERNEL_DNS_02_LOC1/181021_alignment_curv_strain_progT_filt33/flame_kernel_nvec_ComprStrain_align_filt33_1p00TAU_02_LOC1_JFM_HALF.tex}}
\end{minipage}
\graphicspath{{./data/FLAME_KERNEL_DNS_03/181021_alignment_curv_strain_progT_filt33/}}
\begin{minipage}[b]{0.45\linewidth}
  \centering
  \makebox[0pt][l]{\quad(e)}\scalebox{0.7}{\input{./data/FLAME_KERNEL_DNS_03/181021_alignment_curv_strain_progT_filt33/flame_kernel_nvec_ComprStrain_align_filt33_0p25TAU_03_JFM_HALF.tex}}
\end{minipage}
\graphicspath{{./data/FLAME_KERNEL_DNS_03/181021_alignment_curv_strain_progT_filt33/}}
\begin{minipage}[b]{0.45\linewidth}
  \centering
  \makebox[0pt][l]{\quad(f)}\scalebox{0.7}{\input{./data/FLAME_KERNEL_DNS_03/181021_alignment_curv_strain_progT_filt33/flame_kernel_nvec_ComprStrain_align_filt33_1p00TAU_03_JFM_HALF.tex}}
\end{minipage}
\caption{\textcolor{black}{Alignment of the flame front normal vectors with the principal axes of the most compressive strain for Engine Kernel~I~((a), (b)),  Engine Kernel~II~((c), (d)) and the large flame kernel~((e), (f)) at ($t=0.25\cdot \tau_{\mathrm{t}}$) and  ($t=1.0\cdot \tau_{\mathrm{t}}$).}}
\label{fig:kernel_02_03_alignment_filter}
\end{figure}

In figure~\ref{fig:kernel_02_03_alignment_filter}, \textcolor{black}{PDFs of the cosines of the angle between the flame normal vectors and the most compressive principal strain axes are plotted for all flame kernels at ($t=0.25\cdot\tau_{\mathrm{t}}$) and ($t=1.0\cdot\tau_{\mathrm{t}}$). Note that normal vectors have been computed from temperature gradients throughout the flame structure, without conditioning on a particular iso-surface. In the r.h.s.\ of figure~\ref{fig:kernel_02_03_alignment_filter} it is shown that flame normals eventually align with the most compressive strain in all three flame kernels. Further, removing small flow scales from the velocity field by applying a low-pass filter only has a moderate effect on the alignment statistics at ($t=1.0\cdot\tau_{\mathrm{t}}$). 
By contrast, the preferential  flame front alignment with the large flow scales for Engine Kernel~I is much more pronounced at ($t=0.25\cdot\tau_{\mathrm{t}}$), as shown in figure~\ref{fig:kernel_02_03_alignment_filter}\,(a). In line with the initially high positive curvature production by strain (cf.\ figure~\ref{fig:kernel_DcurvDt_terms_bending_zoom}\,(d)) and the strong deformation of the overall flame topology (cf.\ figure~\ref{fig:kernel_02_03_Lstl}), the flame front alignment statistics of Engine Kernel~I exhibit the strongest length-scale dependence among all three flame kernels. Although preferential alignment with the large flow scales is also observed in Engine Kernel~II at this early time instant, the PDF is much less affected by the low-pass filtering than in case of Engine Kernel~I. Note that the length-scale dependence may vary significantly in magnitude over time, but the discussed qualitative differences between both small flame kernel realizations were found throughout the early development phase. In contrast to the engine-relevant flame kernels, the large flame kernel does not exhibit preferential flame front alignment with large-scale flow structures at both shown times, as expected. } \par

To conclude the analysis of early flame kernel\textcolor{black}{/turbulence interactions}, it has been shown that the small flame kernel is subject to strong compressive strain caused by turbulent eddies that are at least as large as the flame radius. 
As a result, the young flame becomes strongly distorted, though a coherent kernel structure is maintained under the present conditions. 
\textcolor{black}{ Since these effects are highly dependent on the local flow conditions, significant run-to-run variations in curvature variance may occur, which affect the evolution of total flame area through~(\ref{eq:one_ov_A_dAdt_avg_mom}).} 
The observed large-scale-dominated early flame/turbulence interactions rapidly generate burned regions of flat shape, which feature high positive curvatures caused by tangential strain. Consequently, the curvature PDF of the small flame kernels is inversely skewed for ($t\lesssim 1.0\cdot \tau_{\mathrm{t}}$). In other words, the early flame kernel is subject to local strain caused by turbulent length scales larger than the flame size, which leads to a production of small curvature radii. This process is in some sense similar to scalar turbulence with $\mathrm{Sc}>1$, where small scalar structures (large gradients) in a range between the Kolmogorov and the Batchelor scale ($\eta_{\mathrm{B}}<l<\eta$) are produced by strain at larger length scales ($l>\eta$)~\citep{Batchelor59}. Different from this statistical effect in scalar mixing, the dominance of flame curvature production by strain due to local, large-scale flow motion is reduced as the flame grows in size and flame/turbulence interaction approaches statistical behaviour.

\section{\textcolor{black}{Relevance and Implications for Engine Combustion and Modelling}}
\label{sec:engine_relevance}
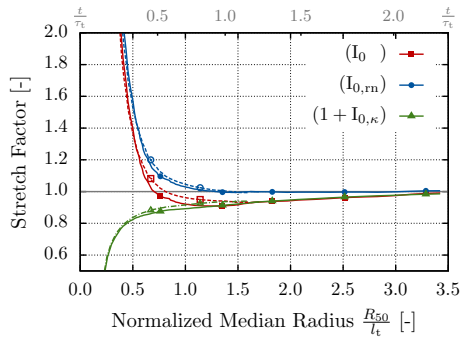
\begin{figure}
  \centering
  \scalebox{0.7}{\input{./data/FLAME_KERNEL_DNS_02/181003_curv_strain_kin_restor_diss_divg_skew/kernel_dns_mean_I0_Sd_1d_sizeT_02_LOC01_JFM_HALF.tex}}
\caption{Flame displacement speed deviation from a laminar unstretched flame as function of flame kernel size \textcolor{black}{for Engine Kernel~I (solid lines) and Engine Kernel~II (dashed lines)}.}
\label{fig:kernel_02_mean_I0_R50}
\end{figure}

\textcolor{black}{From laser-optical engine experiments it was found that even when misfires occur, a small flame kernel is initiated by spark ignition~\citep{Peterson11_exp_misfire}, which rapidly expands and interacts with the flow. However, ignition kernels may be affected by turbulence even before the critical radius for self-sustaining flame propagation is reached. The effect of high-temperature ignition pocket stretching by turbulence was identified by~\citet{Castela16_spark_dns} in a DNS study on plasma-assisted combustion using a sophisticated ignition model. In a recent DNS study, \citet{Uranakara17_ign_kernel_3d} observed global flame kernel extinction within less than one laminar flame time. Hence, interactions of the early ignition or flame kernel with the flow field can be expected to play an important role, even when the young flame is still affected by the spark. 
In the present work, the ignition heat source is meant to represent the macroscopic spark effect, which accelerates the early flame kernel growth as shown in figure~\ref{fig:kernel_02_mean_I0_R50} by the the global stretch factor (${\overline{\mathrm{I}}_0=\overline{\mathrm{I}}_{0,\mathrm{rn}} + \overline{\mathrm{I}}_{0,\kappa}}$) plotted as function of flame kernel radius. The time evolution of all three stretch factor components has been discussed in our previous study~\citep{Falkenstein19_kernel_Le1_cnf}. Here, it should be noted that flame propagation is very consistent with experimental observations by~\citet{Herweg92_model_valid}, who found that the minimum burning velocity of the young flame occurs at a flame radius of approximately one integral length scale. In the present study, interactions of large-scale turbulent flow structures with the flame kernel topology were identified as a characteristic mechanism of early flame kernel development, which occurs while the flame is still affected by the initial ignition energy supply~(${\overline{\mathrm{I}}_0 < 2}$). Possible interactions between severe flame kernel distortion and the development of a self-sustaining flame structure from a high-temperature ignition pocket are aspects suggested for future investigations on kernel extinction. } \par
\graphicspath{{./data/FLAME_KERNEL_DNS_02/181003_curv_strain_kin_restor_diss_divg_skew/}}
The results of the present study have several practical implications, since one of the key challenges in SI engine development is the reduction of cycle-to-cycle variations~\citep{Young81_ccv,Fansler15_ccv}, which are thought to be strongly dependent on early flame kernel development. More recently, simulations of stochastic phenomena in engines have become feasible by using the Large-Eddy-Simulation~(LES) technique~\citep{Rutland11_LES_review,Hasse16_LES}. However, computation of a series of engine cycles introduces the need for modelling spark ignition and early flame kernel development, since the associated time and length scales are very small. As a consequence, the ability of a simulation framework to predict cycle-to-cycle variations is particularly dependent on the accuracy of the early flame kernel model. Under the conditions investigated in the present work, \textcolor{black}{run-to-run variations} were identified which may need to be captured by predictive models. \textcolor{black}{It has been shown that} large-scale turbulent flow structures \textcolor{black}{can} lead to strong topology changes of the young flame, which may substantially alter early kernel growth due to preferential diffusion effects present in common transportation fuel/air mixtures ($\mathrm{Le}_{\mathrm{eff}}\approx 2$), or to cause global extinction by flame kernel breakup in extreme cases. This effect of large-scale flow structures is in contrast to the assumption of kernel wrinkling by a successively wider range of turbulent length scales, which implies that initially only small flow scales are important~\citep{Herweg92_model_valid,Echekki94_kernel_dns_Le_effect}. In LES, information on large flow structures is naturally available since these are resolved on the computational grid. In order to capture interactions with the small flame, Eulerian flame kernel models were proposed by~\citet{Richard07_spark_model} and~\citet{Colin11_spark_model}. However, these models can only reproduce early distortion of the flame as observed in the present study, if the young kernel shape is resolved on the computational grid when the flame diameter is smaller than the integral length scale ($\mathcal{O}(2~\mathrm{mm})$ in engines). Since the flame kernel grows rapidly in size, this can be achieved by temporary mesh refinement as long as ($t\lesssim0.5\cdot\tau_{\mathrm{t}}$). For rigorous model testing, a DNS database of several flame realizations computed up to ($t=1.0\cdot\tau_{\mathrm{t}}$) would be sufficient according to the present results. \par
\section{Conclusions}
\label{sec:conclusions}
\textcolor{black}{Since cycle-to-cycle variations in SI engines are correlated with the very early combustion phase, a better understanding of the governing effects during turbulent flame kernel development is desirable. In this work, characteristic topological changes intrinsic to the engine-relevant flame kernel configuration have been identified and attributed to flame interactions with large-scale turbulent flow structures, which cause noticeable run-to-run variations.}
\textcolor{black}{Specifically, analyses of flame front geometry evolution in three flame configurations with different~$D_{0}/l_{\mathrm{t}}$ have shown the following:}
\begin{itemize}
\item \textcolor{black}{The global curvature variance during flame development does not follow a clear trend 
\makebox[18pt][l]{} for small $D_{0}/l_{\mathrm{t}}$, but varies significantly between different flame kernel realizations.}
\item \textcolor{black}{Although sometimes used as modelling assumption, the development of curvature 
\makebox[18pt][l]{} variance or flame wrinkling does not seem to be restricted by an upper cut-off scale 
\makebox[18pt][l]{} on the order of the kernel size.}
\item \textcolor{black}{In some realizations of the engine-relevant flame kernel configuration, excessive
\makebox[21pt][l]{}curvature variance is produced during the initial kernel development phase.}
\item \textcolor{black}{The decay of this excess variance causes a plateau in net flame surface area
\makebox[21pt][l]{}production~\citep{Falkenstein19_kernel_Le1_cnf}, which can be explained by~(\ref{eq:one_ov_A_dAdt_avg_mom}).}
\end{itemize}
\textcolor{black}{The physical origin of the stochastic occurrence of high curvature variance has been further investigated based on the mean curvature transport equation, which enabled the following conclusions:}
\begin{itemize}
\item \textcolor{black}{Curvature evolution in both positively and negatively curved flame regions is
\makebox[21pt][l]{}substantially influenced by the local flow conditions, which leads to pronounced
\makebox[21pt][l]{}run-to-run variations.}
\item \textcolor{black}{The small initial kernel diameter and correspondingly high positive curvature
\makebox[21pt][l]{}intrinsic to the engine-relevant flame kernel configuration favours the immediate
\makebox[21pt][l]{}amplification of positive curvature by tangential strain.}
\item \textcolor{black}{This initial curvature production by flow straining may lead to a long tail on the
\makebox[18pt][l]{} positive side of the curvature distribution.}
\item \textcolor{black}{By contrast, production of large negative curvature magnitudes by tangential
\makebox[21pt][l]{}derivatives in flame displacement speed relies on the initial formation of negatively
\makebox[21pt][l]{}curved flame regions by turbulent eddies, which induce non-zero second derivatives 
\makebox[21pt][l]{}of the fluid velocity (bending).  Compared to other flame configurations, this effect
\makebox[21pt][l]{}is less effective in case of small flame kernels, possibly due to the small flame
\makebox[21pt][l]{}diameter respective of the relevant flow scales. }
\item Consequently, the curvature distribution of small flame kernels becomes strongly
\makebox[21pt][l]{}skewed towards positive curvatures, which is the opposite of what is observed for
\makebox[21pt][l]{}developed flames.
\end{itemize}
\textcolor{black}{Further, the impact of flame kernel initiation with smaller diameter than the energy-containing turbulent length scales has been examined, which is a characteristic feature of SI engine combustion. The analysis of length-scale dependent flame front alignment with the flow field has shown the following:}
\begin{itemize}
\item At early times ($t\approx 0.25\cdot \tau_{\mathrm{t}}$), flame normal vectors are preferentially aligned with
\makebox[21pt][l]{}the most compressive principal axis of the strain tensor computed from a low-pass-
\makebox[21pt][l]{}filtered velocity field, where flow scales up to the kernel size have been removed.
\item Consequently, the small flame kernel is subject to compressive strain caused by
\makebox[21pt][l]{}turbulent eddies that are at least as large as the flame radius\textcolor{black}{, i.e.\ the mean flame
\makebox[21pt][l]{}front curvature radius and the governing flow scales are of similar order.}
\item \textcolor{black}{This characteristic flame kernel/turbulence interaction has a strong impact on the
\makebox[21pt][l]{}volumetric flame topology and flame front geometry, and distinguishes early flame
\makebox[21pt][l]{}kernel development from the evolution of a statistically planar flame brush.}
\end{itemize}
It is hypothesized that the demonstrated flame kernel distortion by large-scale turbulent flow structures is a major contributor to cycle-to-cycle variations in engines and may even result in global extinction under extreme conditions. 
In the presence of preferential diffusion effects, flame response to high stretch caused by strain and topology changes might be severe, and will be investigated in the future. \par
In the present work, only two flame realizations were considered. Hence, conclusions on the impact of fluctuations in early flame kernel growth on heat release at later times or on cycle-to-cycle variations in engines could not be drawn. Still, characteristic observations from the small-flame-kernel datasets were confirmed in local regions of the other flames to ensure the validity of the conclusions on early flame kernel/flow interactions. \par

\vspace{\baselineskip}
The authors from RWTH Aachen University gratefully acknowledge partial funding by Honda R\&D and by the Deutsche Forschungsgemeinschaft (DFG, German Research Foundation) under Germany's Excellence Strategy - Exzellenzcluster 2186 `The Fuel Science Center' ID: 390919832.\par
The authors gratefully acknowledge the Gauss Centre for Supercomputing e.V. (www.gauss-centre.eu) for funding this project by providing computing time on the GCS Supercomputer Super-MUC at Leibniz Supercomputing Centre (LRZ, www.lrz.de).\par
Data analyses were performed with computing resources granted by RWTH Aachen University under project thes0373. \par
S.K. gratefully acknowledges financial support from the National Research Foundation of Korea (NRF) grant by the Korea government (MSIP) (No. 2017R1A2B3008273). \par

\appendix
%
\section{}
\subsection{Derivation of the Mean Curvature Transport Equation}
\label{appCurv}
In order to derive the form of the mean curvature evolution equation used in this study (cf.~(\ref{eq:curv_eq})), we formulate equation~(38) from the study of~\citet{Dopazo18_curvEq} in the present notation:
\begin{eqnarray}
\frac{\mathrm{D}_{\vartheta} \left( \kappa^{(\vartheta)} \right)}{\mathrm{D}_{\vartheta}\left(t\right)} = \kappa^{(\vartheta)} a_{\mathrm{n}}^{(\vartheta)} + \frac{\p a_{\mathrm{n}}^{(\vartheta)}}{\p x_{\mathrm{n}}} - 2 S_{ij}^{(\vartheta)} \frac{\p n^{(\vartheta)}_j}{\p x_i} - \frac{\p S_{ij}^{(\vartheta)}}{\p x_i} n^{(\vartheta)}_j + \frac{\p W_{ij}^{(\vartheta)}}{\p x_i} n^{(\vartheta)}_j,
\label{eq:curv_eq_dopazo}
\end{eqnarray}
where $\vartheta$ refers to the scalar field used to define iso-surfaces. The velocity of an iso-surface element is given by:
\begin{equation}
u_i^{(\vartheta)} = u_i + \mathrm{s}_{\mathrm{d}}^{(\vartheta)} n^{(\vartheta)}_i.
\label{eq:u_surf}
\end{equation}
For the definitions of the normal vector $n^{(\vartheta)}_i$ and displacement speed $\mathrm{s}_{\mathrm{d}}^{(\vartheta)}$, refer to~(\ref{eq:def_nvec}) and~(\ref{eq:sc_tot_d_dt}). The scalar iso-surface described in the moving reference frame is subject to the total normal strain rate,
\begin{equation}
a_{\mathrm{n}}^{(\vartheta)} = n^{(\vartheta)}_i \frac{\p u_i^{(\vartheta)}}{\p x_j} n^{(\vartheta)}_j = \frac{\p u_{\mathrm{n}}^{(\vartheta)}}{\p x_{\mathrm{n}}},
\label{eq:aN_surf}
\end{equation}
and the total strain rate and rotation rate tensors are defined as: 
\begin{eqnarray}
S_{ij}^{(\vartheta)} = \frac{1}{2} \left(\frac{\p u_i^{(\vartheta)}}{\p x_j} + \frac{\p u_j^{(\vartheta)}}{\p x_i} \right), \\
W_{ij}^{(\vartheta)} = \frac{1}{2} \left(\frac{\p u_i^{(\vartheta)}}{\p x_j} - \frac{\p u_j^{(\vartheta)}}{\p x_i} \right).
\label{eq:Sij_Wij_surf}
\end{eqnarray}
To rewrite~(\ref{eq:curv_eq_dopazo}) in terms of second derivatives of the total velocity, we use
\begin{equation}
-S_{ij}^{(\vartheta)} + W_{ij}^{(\vartheta)} = - \frac{\p u_j^{(\vartheta)}}{\p x_i}
\label{eq:curv_eq_merge_T4T5}
\end{equation}
to combine the last two terms in the r.h.s.\ of~(\ref{eq:curv_eq_dopazo}). The second term in the r.h.s.\ is modified by using the normal strain definition (cf.~(\ref{eq:aN_surf})) to obtain:
\begin{eqnarray}
\frac{\mathrm{D}_{\vartheta} \left( \kappa^{(\vartheta)} \right)}{\mathrm{D}_{\vartheta}\left(t\right)} = \kappa^{(\vartheta)} a_{\mathrm{n}}^{(\vartheta)} - 2 S_{ij}^{(\vartheta)} \frac{\p n^{(\vartheta)}_j}{\p x_i} 
- \left[ \frac{\p^2 u_j^{(\vartheta)}}{\p x_i^2} n^{(\vartheta)}_j 
- \frac{\p^2 u_{\mathrm{n}}^{(\vartheta)}}{\p x_{\mathrm{n}}^2} \right],
\label{eq:curv_eq_app}
\end{eqnarray}
which is equivalent to~(\ref{eq:curv_eq}). The first two terms in the r.h.s.\ describe the effects of normal and tangential strain on curvature evolution, respectively. The last two terms in brackets are responsible for iso-surface bending by second derivatives of the total velocity in the tangential plane~\citep{Pope88_surf_turb}.

\bibliographystyle{jfm}
\bibliography{references/flame_kernel_01.bib,references/les4ice16.bib}

\end{document}

%% file: data/FLAME_KERNEL_DNS_02/180916_local_geom_lseg_T_1700_to_2000K/t1p5e-5/Lstl_PDF_normDv_0p10Tau_02_LOC01_03_JFM.tex
\begingroup
  \makeatletter
  \providecommand\color[2][]{%
    \GenericError{(gnuplot) \space\space\space\@spaces}{%
      Package color not loaded in conjunction with
      terminal option `colourtext'%
    }{See the gnuplot documentation for explanation.%
    }{Either use 'blacktext' in gnuplot or load the package
      color.sty in LaTeX.}%
    \renewcommand\color[2][]{}%
  }%
  \providecommand\includegraphics[2][]{%
    \GenericError{(gnuplot) \space\space\space\@spaces}{%
      Package graphicx or graphics not loaded%
    }{See the gnuplot documentation for explanation.%
    }{The gnuplot epslatex terminal needs graphicx.sty or graphics.sty.}%
    \renewcommand\includegraphics[2][]{}%
  }%
  \providecommand\rotatebox[2]{#2}%
  \@ifundefined{ifGPcolor}{%
    \newif\ifGPcolor
    \GPcolortrue
  }{}%
  \@ifundefined{ifGPblacktext}{%
    \newif\ifGPblacktext
    \GPblacktexttrue
  }{}%
  \let\gplgaddtomacro\g@addto@macro
  \gdef\gplbacktext{}%
  \gdef\gplfronttext{}%
  \makeatother
  \ifGPblacktext
    \def\colorrgb#1{}%
    \def\colorgray#1{}%
  \else
    \ifGPcolor
      \def\colorrgb#1{\color[rgb]{#1}}%
      \def\colorgray#1{\color[gray]{#1}}%
      \expandafter\def\csname LTw\endcsname{\color{white}}%
      \expandafter\def\csname LTb\endcsname{\color{black}}%
      \expandafter\def\csname LTa\endcsname{\color{black}}%
      \expandafter\def\csname LT0\endcsname{\color[rgb]{1,0,0}}%
      \expandafter\def\csname LT1\endcsname{\color[rgb]{0,1,0}}%
      \expandafter\def\csname LT2\endcsname{\color[rgb]{0,0,1}}%
      \expandafter\def\csname LT3\endcsname{\color[rgb]{1,0,1}}%
      \expandafter\def\csname LT4\endcsname{\color[rgb]{0,1,1}}%
      \expandafter\def\csname LT5\endcsname{\color[rgb]{1,1,0}}%
      \expandafter\def\csname LT6\endcsname{\color[rgb]{0,0,0}}%
      \expandafter\def\csname LT7\endcsname{\color[rgb]{1,0.3,0}}%
      \expandafter\def\csname LT8\endcsname{\color[rgb]{0.5,0.5,0.5}}%
    \else
      \def\colorrgb#1{\color{black}}%
      \def\colorgray#1{\color[gray]{#1}}%
      \expandafter\def\csname LTw\endcsname{\color{white}}%
      \expandafter\def\csname LTb\endcsname{\color{black}}%
      \expandafter\def\csname LTa\endcsname{\color{black}}%
      \expandafter\def\csname LT0\endcsname{\color{black}}%
      \expandafter\def\csname LT1\endcsname{\color{black}}%
      \expandafter\def\csname LT2\endcsname{\color{black}}%
      \expandafter\def\csname LT3\endcsname{\color{black}}%
      \expandafter\def\csname LT4\endcsname{\color{black}}%
      \expandafter\def\csname LT5\endcsname{\color{black}}%
      \expandafter\def\csname LT6\endcsname{\color{black}}%
      \expandafter\def\csname LT7\endcsname{\color{black}}%
      \expandafter\def\csname LT8\endcsname{\color{black}}%
    \fi
  \fi
  \setlength{\unitlength}{0.0500bp}%
  \begin{picture}(5328.00,3684.00)%
    \gplgaddtomacro\gplbacktext{%
      \csname LTb\endcsname%
      \put(858,704){\makebox(0,0)[r]{\strut{}\normalsize 0.0}}%
      \csname LTb\endcsname%
      \put(858,1190){\makebox(0,0)[r]{\strut{}\normalsize 1.5}}%
      \csname LTb\endcsname%
      \put(858,1676){\makebox(0,0)[r]{\strut{}\normalsize 3.0}}%
      \csname LTb\endcsname%
      \put(858,2161){\makebox(0,0)[r]{\strut{}\normalsize 4.5}}%
      \csname LTb\endcsname%
      \put(858,2647){\makebox(0,0)[r]{\strut{}\normalsize 6.0}}%
      \csname LTb\endcsname%
      \put(858,3133){\makebox(0,0)[r]{\strut{}\normalsize 7.5}}%
      \csname LTb\endcsname%
      \put(990,484){\makebox(0,0){\strut{}\normalsize 0.0}}%
      \csname LTb\endcsname%
      \put(1975,484){\makebox(0,0){\strut{}\normalsize 0.5}}%
      \csname LTb\endcsname%
      \put(2961,484){\makebox(0,0){\strut{}\normalsize 1.0}}%
      \csname LTb\endcsname%
      \put(3946,484){\makebox(0,0){\strut{}\normalsize 1.5}}%
      \csname LTb\endcsname%
      \put(4931,484){\makebox(0,0){\strut{}\normalsize 2.0}}%
      \put(352,1918){\rotatebox{-270}{\makebox(0,0){\strut{}\large PDF}}}%
      \put(2960,154){\makebox(0,0){\strut{}\large ${d_{\mathrm{f,n}}}/{D_{\mathrm{v}}}$ [-]}}%
      \put(2960,3353){\makebox(0,0){\strut{}\normalsize  $t=0.09\cdot \tau_{\mathrm{t}}$ (\color[HTML]{C00000}$R_{50,\mathrm{I}}=0.20 \, l_{\mathrm{t}}$, \color[HTML]{00549F}$R_{50,\mathrm{LK}}=0.77 \, l_{\mathrm{t}}$\color[HTML]{000000})}}%
    }%
    \gplgaddtomacro\gplfronttext{%
      \csname LTb\endcsname%
      \put(4274,2960){\makebox(0,0)[r]{\strut{}\normalsize Engine Kernel I\,\,}}%
      \csname LTb\endcsname%
      \put(4274,2740){\makebox(0,0)[r]{\strut{}\normalsize Engine Kernel II}}%
      \csname LTb\endcsname%
      \put(4274,2520){\makebox(0,0)[r]{\strut{}\normalsize Large Kernel}}%
    }%
    \gplbacktext
    \put(0,0){\includegraphics{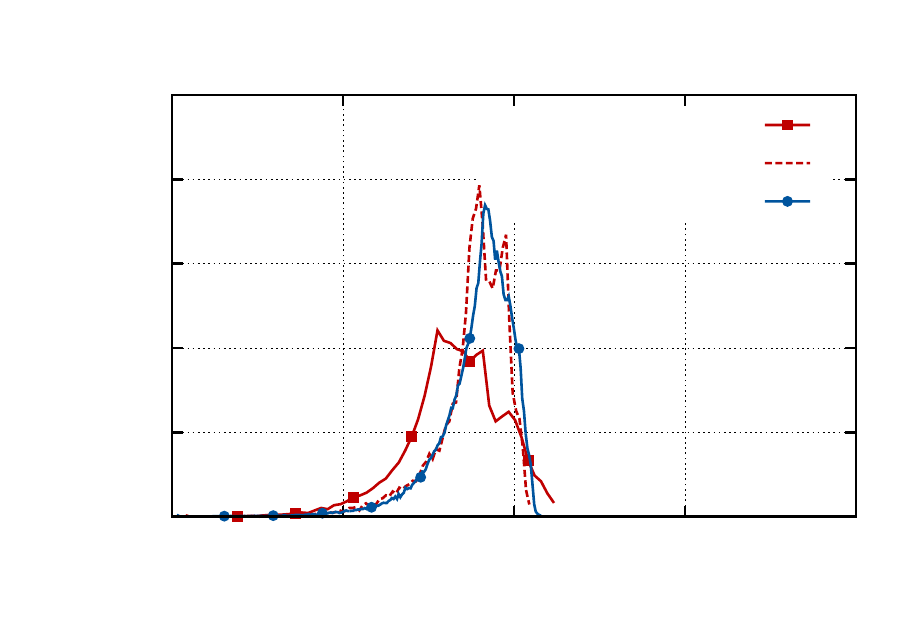}}%
    \gplfronttext
  \end{picture}%
\endgroup

%% file: data/FLAME_KERNEL_DNS_02/180916_local_geom_lseg_T_1700_to_2000K/t3p0e-5/Lstl_PDF_norm_Dv_0p18Tau_02_LOC01_03_JFM.tex
\begingroup
  \makeatletter
  \providecommand\color[2][]{%
    \GenericError{(gnuplot) \space\space\space\@spaces}{%
      Package color not loaded in conjunction with
      terminal option `colourtext'%
    }{See the gnuplot documentation for explanation.%
    }{Either use 'blacktext' in gnuplot or load the package
      color.sty in LaTeX.}%
    \renewcommand\color[2][]{}%
  }%
  \providecommand\includegraphics[2][]{%
    \GenericError{(gnuplot) \space\space\space\@spaces}{%
      Package graphicx or graphics not loaded%
    }{See the gnuplot documentation for explanation.%
    }{The gnuplot epslatex terminal needs graphicx.sty or graphics.sty.}%
    \renewcommand\includegraphics[2][]{}%
  }%
  \providecommand\rotatebox[2]{#2}%
  \@ifundefined{ifGPcolor}{%
    \newif\ifGPcolor
    \GPcolortrue
  }{}%
  \@ifundefined{ifGPblacktext}{%
    \newif\ifGPblacktext
    \GPblacktexttrue
  }{}%
  \let\gplgaddtomacro\g@addto@macro
  \gdef\gplbacktext{}%
  \gdef\gplfronttext{}%
  \makeatother
  \ifGPblacktext
    \def\colorrgb#1{}%
    \def\colorgray#1{}%
  \else
    \ifGPcolor
      \def\colorrgb#1{\color[rgb]{#1}}%
      \def\colorgray#1{\color[gray]{#1}}%
      \expandafter\def\csname LTw\endcsname{\color{white}}%
      \expandafter\def\csname LTb\endcsname{\color{black}}%
      \expandafter\def\csname LTa\endcsname{\color{black}}%
      \expandafter\def\csname LT0\endcsname{\color[rgb]{1,0,0}}%
      \expandafter\def\csname LT1\endcsname{\color[rgb]{0,1,0}}%
      \expandafter\def\csname LT2\endcsname{\color[rgb]{0,0,1}}%
      \expandafter\def\csname LT3\endcsname{\color[rgb]{1,0,1}}%
      \expandafter\def\csname LT4\endcsname{\color[rgb]{0,1,1}}%
      \expandafter\def\csname LT5\endcsname{\color[rgb]{1,1,0}}%
      \expandafter\def\csname LT6\endcsname{\color[rgb]{0,0,0}}%
      \expandafter\def\csname LT7\endcsname{\color[rgb]{1,0.3,0}}%
      \expandafter\def\csname LT8\endcsname{\color[rgb]{0.5,0.5,0.5}}%
    \else
      \def\colorrgb#1{\color{black}}%
      \def\colorgray#1{\color[gray]{#1}}%
      \expandafter\def\csname LTw\endcsname{\color{white}}%
      \expandafter\def\csname LTb\endcsname{\color{black}}%
      \expandafter\def\csname LTa\endcsname{\color{black}}%
      \expandafter\def\csname LT0\endcsname{\color{black}}%
      \expandafter\def\csname LT1\endcsname{\color{black}}%
      \expandafter\def\csname LT2\endcsname{\color{black}}%
      \expandafter\def\csname LT3\endcsname{\color{black}}%
      \expandafter\def\csname LT4\endcsname{\color{black}}%
      \expandafter\def\csname LT5\endcsname{\color{black}}%
      \expandafter\def\csname LT6\endcsname{\color{black}}%
      \expandafter\def\csname LT7\endcsname{\color{black}}%
      \expandafter\def\csname LT8\endcsname{\color{black}}%
    \fi
  \fi
  \setlength{\unitlength}{0.0500bp}%
  \begin{picture}(5328.00,3684.00)%
    \gplgaddtomacro\gplbacktext{%
      \csname LTb\endcsname%
      \put(858,704){\makebox(0,0)[r]{\strut{}\normalsize 0.0}}%
      \csname LTb\endcsname%
      \put(858,1190){\makebox(0,0)[r]{\strut{}\normalsize 1.0}}%
      \csname LTb\endcsname%
      \put(858,1676){\makebox(0,0)[r]{\strut{}\normalsize 2.0}}%
      \csname LTb\endcsname%
      \put(858,2161){\makebox(0,0)[r]{\strut{}\normalsize 3.0}}%
      \csname LTb\endcsname%
      \put(858,2647){\makebox(0,0)[r]{\strut{}\normalsize 4.0}}%
      \csname LTb\endcsname%
      \put(858,3133){\makebox(0,0)[r]{\strut{}\normalsize 5.0}}%
      \csname LTb\endcsname%
      \put(990,484){\makebox(0,0){\strut{}\normalsize 0.0}}%
      \csname LTb\endcsname%
      \put(1975,484){\makebox(0,0){\strut{}\normalsize 0.5}}%
      \csname LTb\endcsname%
      \put(2961,484){\makebox(0,0){\strut{}\normalsize 1.0}}%
      \csname LTb\endcsname%
      \put(3946,484){\makebox(0,0){\strut{}\normalsize 1.5}}%
      \csname LTb\endcsname%
      \put(4931,484){\makebox(0,0){\strut{}\normalsize 2.0}}%
      \put(352,1918){\rotatebox{-270}{\makebox(0,0){\strut{}\large PDF}}}%
      \put(2960,154){\makebox(0,0){\strut{}\large ${d_{\mathrm{f,n}}}/{D_{\mathrm{v}}}$ [-]}}%
      \put(2960,3353){\makebox(0,0){\strut{}\normalsize  $t=0.18\cdot \tau_{\mathrm{t}}$ (\color[HTML]{C00000}$R_{50,\mathrm{I}}=0.38 \, l_{\mathrm{t}}$, \color[HTML]{00549F}$R_{50,\mathrm{LK}}=0.96 \, l_{\mathrm{t}}$\color[HTML]{000000})}}%
    }%
    \gplgaddtomacro\gplfronttext{%
      \csname LTb\endcsname%
      \put(4274,2960){\makebox(0,0)[r]{\strut{}\normalsize Engine Kernel I\,\,}}%
      \csname LTb\endcsname%
      \put(4274,2740){\makebox(0,0)[r]{\strut{}\normalsize Engine Kernel II}}%
      \csname LTb\endcsname%
      \put(4274,2520){\makebox(0,0)[r]{\strut{}\normalsize Large Kernel}}%
    }%
    \gplbacktext
    \put(0,0){\includegraphics{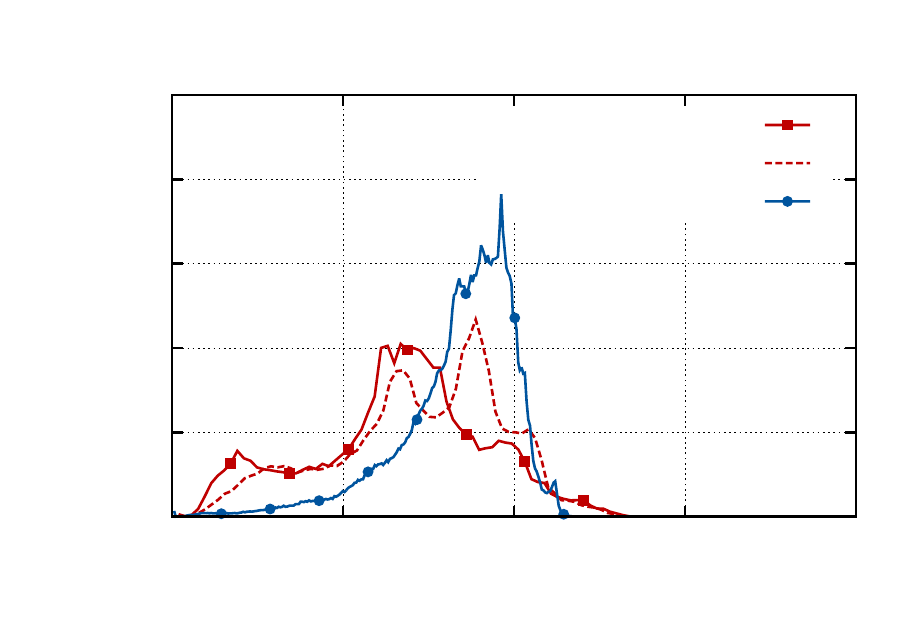}}%
    \gplfronttext
  \end{picture}%
\endgroup

%% file: data/FLAME_KERNEL_DNS_02/180916_local_geom_lseg_T_1700_to_2000K/t4p5e-5/Lstl_PDF_norm_Dv_0p26Tau_02_LOC01_03_JFM.tex
\begingroup
  \makeatletter
  \providecommand\color[2][]{%
    \GenericError{(gnuplot) \space\space\space\@spaces}{%
      Package color not loaded in conjunction with
      terminal option `colourtext'%
    }{See the gnuplot documentation for explanation.%
    }{Either use 'blacktext' in gnuplot or load the package
      color.sty in LaTeX.}%
    \renewcommand\color[2][]{}%
  }%
  \providecommand\includegraphics[2][]{%
    \GenericError{(gnuplot) \space\space\space\@spaces}{%
      Package graphicx or graphics not loaded%
    }{See the gnuplot documentation for explanation.%
    }{The gnuplot epslatex terminal needs graphicx.sty or graphics.sty.}%
    \renewcommand\includegraphics[2][]{}%
  }%
  \providecommand\rotatebox[2]{#2}%
  \@ifundefined{ifGPcolor}{%
    \newif\ifGPcolor
    \GPcolortrue
  }{}%
  \@ifundefined{ifGPblacktext}{%
    \newif\ifGPblacktext
    \GPblacktexttrue
  }{}%
  \let\gplgaddtomacro\g@addto@macro
  \gdef\gplbacktext{}%
  \gdef\gplfronttext{}%
  \makeatother
  \ifGPblacktext
    \def\colorrgb#1{}%
    \def\colorgray#1{}%
  \else
    \ifGPcolor
      \def\colorrgb#1{\color[rgb]{#1}}%
      \def\colorgray#1{\color[gray]{#1}}%
      \expandafter\def\csname LTw\endcsname{\color{white}}%
      \expandafter\def\csname LTb\endcsname{\color{black}}%
      \expandafter\def\csname LTa\endcsname{\color{black}}%
      \expandafter\def\csname LT0\endcsname{\color[rgb]{1,0,0}}%
      \expandafter\def\csname LT1\endcsname{\color[rgb]{0,1,0}}%
      \expandafter\def\csname LT2\endcsname{\color[rgb]{0,0,1}}%
      \expandafter\def\csname LT3\endcsname{\color[rgb]{1,0,1}}%
      \expandafter\def\csname LT4\endcsname{\color[rgb]{0,1,1}}%
      \expandafter\def\csname LT5\endcsname{\color[rgb]{1,1,0}}%
      \expandafter\def\csname LT6\endcsname{\color[rgb]{0,0,0}}%
      \expandafter\def\csname LT7\endcsname{\color[rgb]{1,0.3,0}}%
      \expandafter\def\csname LT8\endcsname{\color[rgb]{0.5,0.5,0.5}}%
    \else
      \def\colorrgb#1{\color{black}}%
      \def\colorgray#1{\color[gray]{#1}}%
      \expandafter\def\csname LTw\endcsname{\color{white}}%
      \expandafter\def\csname LTb\endcsname{\color{black}}%
      \expandafter\def\csname LTa\endcsname{\color{black}}%
      \expandafter\def\csname LT0\endcsname{\color{black}}%
      \expandafter\def\csname LT1\endcsname{\color{black}}%
      \expandafter\def\csname LT2\endcsname{\color{black}}%
      \expandafter\def\csname LT3\endcsname{\color{black}}%
      \expandafter\def\csname LT4\endcsname{\color{black}}%
      \expandafter\def\csname LT5\endcsname{\color{black}}%
      \expandafter\def\csname LT6\endcsname{\color{black}}%
      \expandafter\def\csname LT7\endcsname{\color{black}}%
      \expandafter\def\csname LT8\endcsname{\color{black}}%
    \fi
  \fi
  \setlength{\unitlength}{0.0500bp}%
  \begin{picture}(5328.00,3684.00)%
    \gplgaddtomacro\gplbacktext{%
      \csname LTb\endcsname%
      \put(858,704){\makebox(0,0)[r]{\strut{}\normalsize 0.0}}%
      \csname LTb\endcsname%
      \put(858,1190){\makebox(0,0)[r]{\strut{}\normalsize 0.5}}%
      \csname LTb\endcsname%
      \put(858,1676){\makebox(0,0)[r]{\strut{}\normalsize 1.0}}%
      \csname LTb\endcsname%
      \put(858,2161){\makebox(0,0)[r]{\strut{}\normalsize 1.5}}%
      \csname LTb\endcsname%
      \put(858,2647){\makebox(0,0)[r]{\strut{}\normalsize 2.0}}%
      \csname LTb\endcsname%
      \put(858,3133){\makebox(0,0)[r]{\strut{}\normalsize 2.5}}%
      \csname LTb\endcsname%
      \put(990,484){\makebox(0,0){\strut{}\normalsize 0.0}}%
      \csname LTb\endcsname%
      \put(1975,484){\makebox(0,0){\strut{}\normalsize 0.5}}%
      \csname LTb\endcsname%
      \put(2961,484){\makebox(0,0){\strut{}\normalsize 1.0}}%
      \csname LTb\endcsname%
      \put(3946,484){\makebox(0,0){\strut{}\normalsize 1.5}}%
      \csname LTb\endcsname%
      \put(4931,484){\makebox(0,0){\strut{}\normalsize 2.0}}%
      \put(352,1918){\rotatebox{-270}{\makebox(0,0){\strut{}\large PDF}}}%
      \put(2960,154){\makebox(0,0){\strut{}\large ${d_{\mathrm{f,n}}}/{D_{\mathrm{v}}}$ [-]}}%
      \put(2960,3353){\makebox(0,0){\strut{}\normalsize  $t=0.26\cdot \tau_{\mathrm{t}}$ (\color[HTML]{C00000}$R_{50,\mathrm{I}}=0.50 \, l_{\mathrm{t}}$, \color[HTML]{00549F}$R_{50,\mathrm{LK}}=1.10 \, l_{\mathrm{t}}$\color[HTML]{000000})}}%
    }%
    \gplgaddtomacro\gplfronttext{%
      \csname LTb\endcsname%
      \put(4274,2960){\makebox(0,0)[r]{\strut{}\normalsize Engine Kernel I\,\,}}%
      \csname LTb\endcsname%
      \put(4274,2740){\makebox(0,0)[r]{\strut{}\normalsize Engine Kernel II}}%
      \csname LTb\endcsname%
      \put(4274,2520){\makebox(0,0)[r]{\strut{}\normalsize Large Kernel}}%
    }%
    \gplbacktext
    \put(0,0){\includegraphics{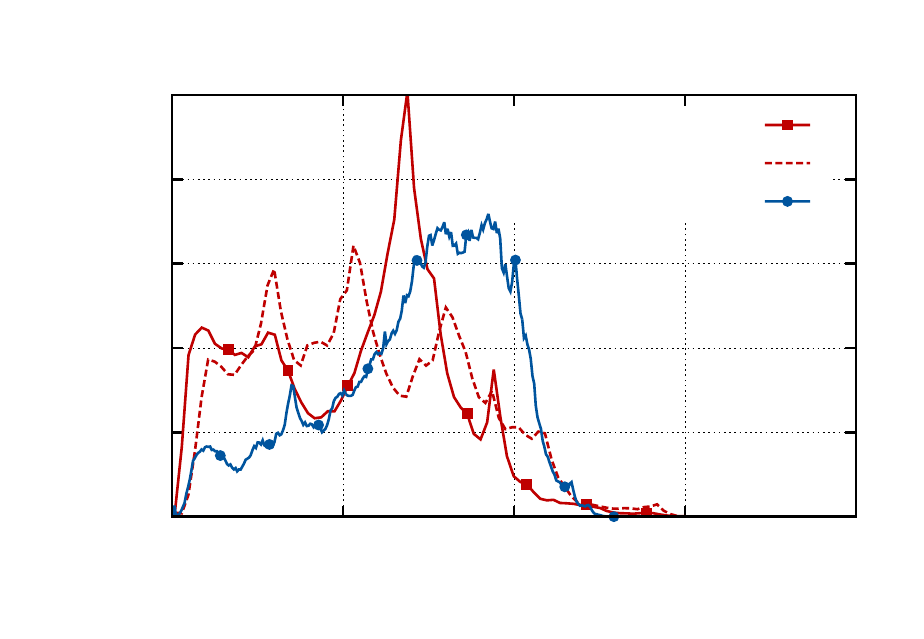}}%
    \gplfronttext
  \end{picture}%
\endgroup

%% file: data/FLAME_KERNEL_DNS_02/180916_local_geom_lseg_T_1700_to_2000K/t9p0e-5/Lstl_PDF_norm_Dv_0p53Tau_02_LOC01_03_JFM.tex
\begingroup
  \makeatletter
  \providecommand\color[2][]{%
    \GenericError{(gnuplot) \space\space\space\@spaces}{%
      Package color not loaded in conjunction with
      terminal option `colourtext'%
    }{See the gnuplot documentation for explanation.%
    }{Either use 'blacktext' in gnuplot or load the package
      color.sty in LaTeX.}%
    \renewcommand\color[2][]{}%
  }%
  \providecommand\includegraphics[2][]{%
    \GenericError{(gnuplot) \space\space\space\@spaces}{%
      Package graphicx or graphics not loaded%
    }{See the gnuplot documentation for explanation.%
    }{The gnuplot epslatex terminal needs graphicx.sty or graphics.sty.}%
    \renewcommand\includegraphics[2][]{}%
  }%
  \providecommand\rotatebox[2]{#2}%
  \@ifundefined{ifGPcolor}{%
    \newif\ifGPcolor
    \GPcolortrue
  }{}%
  \@ifundefined{ifGPblacktext}{%
    \newif\ifGPblacktext
    \GPblacktexttrue
  }{}%
  \let\gplgaddtomacro\g@addto@macro
  \gdef\gplbacktext{}%
  \gdef\gplfronttext{}%
  \makeatother
  \ifGPblacktext
    \def\colorrgb#1{}%
    \def\colorgray#1{}%
  \else
    \ifGPcolor
      \def\colorrgb#1{\color[rgb]{#1}}%
      \def\colorgray#1{\color[gray]{#1}}%
      \expandafter\def\csname LTw\endcsname{\color{white}}%
      \expandafter\def\csname LTb\endcsname{\color{black}}%
      \expandafter\def\csname LTa\endcsname{\color{black}}%
      \expandafter\def\csname LT0\endcsname{\color[rgb]{1,0,0}}%
      \expandafter\def\csname LT1\endcsname{\color[rgb]{0,1,0}}%
      \expandafter\def\csname LT2\endcsname{\color[rgb]{0,0,1}}%
      \expandafter\def\csname LT3\endcsname{\color[rgb]{1,0,1}}%
      \expandafter\def\csname LT4\endcsname{\color[rgb]{0,1,1}}%
      \expandafter\def\csname LT5\endcsname{\color[rgb]{1,1,0}}%
      \expandafter\def\csname LT6\endcsname{\color[rgb]{0,0,0}}%
      \expandafter\def\csname LT7\endcsname{\color[rgb]{1,0.3,0}}%
      \expandafter\def\csname LT8\endcsname{\color[rgb]{0.5,0.5,0.5}}%
    \else
      \def\colorrgb#1{\color{black}}%
      \def\colorgray#1{\color[gray]{#1}}%
      \expandafter\def\csname LTw\endcsname{\color{white}}%
      \expandafter\def\csname LTb\endcsname{\color{black}}%
      \expandafter\def\csname LTa\endcsname{\color{black}}%
      \expandafter\def\csname LT0\endcsname{\color{black}}%
      \expandafter\def\csname LT1\endcsname{\color{black}}%
      \expandafter\def\csname LT2\endcsname{\color{black}}%
      \expandafter\def\csname LT3\endcsname{\color{black}}%
      \expandafter\def\csname LT4\endcsname{\color{black}}%
      \expandafter\def\csname LT5\endcsname{\color{black}}%
      \expandafter\def\csname LT6\endcsname{\color{black}}%
      \expandafter\def\csname LT7\endcsname{\color{black}}%
      \expandafter\def\csname LT8\endcsname{\color{black}}%
    \fi
  \fi
  \setlength{\unitlength}{0.0500bp}%
  \begin{picture}(5328.00,3684.00)%
    \gplgaddtomacro\gplbacktext{%
      \csname LTb\endcsname%
      \put(858,704){\makebox(0,0)[r]{\strut{}\normalsize 0.0}}%
      \csname LTb\endcsname%
      \put(858,1051){\makebox(0,0)[r]{\strut{}\normalsize 0.5}}%
      \csname LTb\endcsname%
      \put(858,1398){\makebox(0,0)[r]{\strut{}\normalsize 1.0}}%
      \csname LTb\endcsname%
      \put(858,1745){\makebox(0,0)[r]{\strut{}\normalsize 1.5}}%
      \csname LTb\endcsname%
      \put(858,2092){\makebox(0,0)[r]{\strut{}\normalsize 2.0}}%
      \csname LTb\endcsname%
      \put(858,2439){\makebox(0,0)[r]{\strut{}\normalsize 2.5}}%
      \csname LTb\endcsname%
      \put(858,2786){\makebox(0,0)[r]{\strut{}\normalsize 3.0}}%
      \csname LTb\endcsname%
      \put(858,3133){\makebox(0,0)[r]{\strut{}\normalsize 3.5}}%
      \csname LTb\endcsname%
      \put(990,484){\makebox(0,0){\strut{}\normalsize 0.0}}%
      \csname LTb\endcsname%
      \put(1975,484){\makebox(0,0){\strut{}\normalsize 0.5}}%
      \csname LTb\endcsname%
      \put(2961,484){\makebox(0,0){\strut{}\normalsize 1.0}}%
      \csname LTb\endcsname%
      \put(3946,484){\makebox(0,0){\strut{}\normalsize 1.5}}%
      \csname LTb\endcsname%
      \put(4931,484){\makebox(0,0){\strut{}\normalsize 2.0}}%
      \put(352,1918){\rotatebox{-270}{\makebox(0,0){\strut{}\large PDF}}}%
      \put(2960,154){\makebox(0,0){\strut{}\large ${d_{\mathrm{f,n}}}/{D_{\mathrm{v}}}$ [-]}}%
      \put(2960,3353){\makebox(0,0){\strut{}\normalsize  $t=0.26\cdot \tau_{\mathrm{t}}$ (\color[HTML]{C00000}$R_{50,\mathrm{I}}=0.76 \, l_{\mathrm{t}}$, \color[HTML]{00549F}$R_{50,\mathrm{LK}}=1.46 \, l_{\mathrm{t}}$\color[HTML]{000000})}}%
    }%
    \gplgaddtomacro\gplfronttext{%
      \csname LTb\endcsname%
      \put(4274,2960){\makebox(0,0)[r]{\strut{}\normalsize Engine Kernel I\,\,}}%
      \csname LTb\endcsname%
      \put(4274,2740){\makebox(0,0)[r]{\strut{}\normalsize Engine Kernel II}}%
      \csname LTb\endcsname%
      \put(4274,2520){\makebox(0,0)[r]{\strut{}\normalsize Large Kernel}}%
    }%
    \gplbacktext
    \put(0,0){\includegraphics{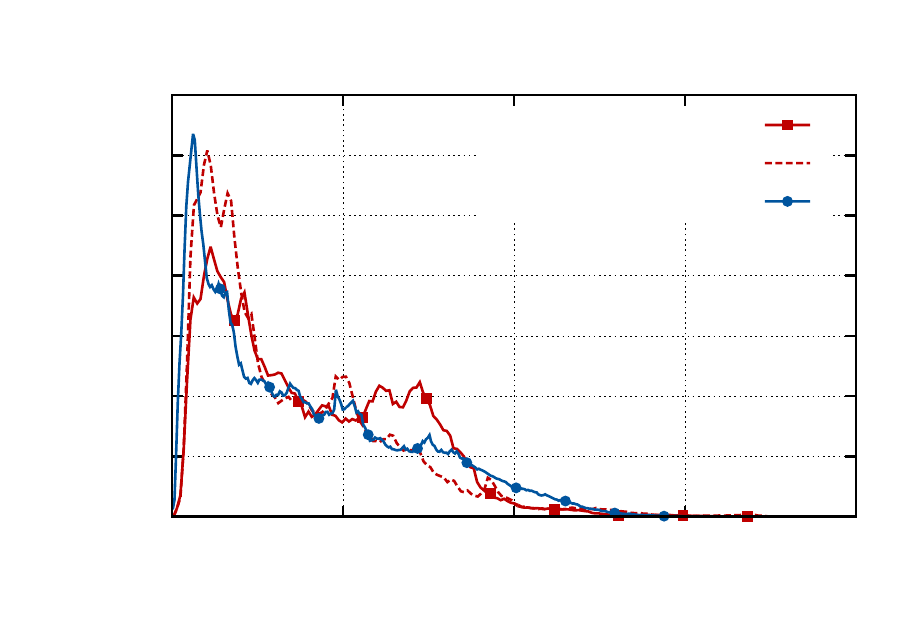}}%
    \gplfronttext
  \end{picture}%
\endgroup

%% file: data/FLAME_KERNEL_DNS_02/181003_curv_strain_kin_restor_diss_divg_skew/flame_kernel_curvVar_02_LOC01_03_P02_JFM_HALF.tex
\begingroup
  \makeatletter
  \providecommand\color[2][]{%
    \GenericError{(gnuplot) \space\space\space\@spaces}{%
      Package color not loaded in conjunction with
      terminal option `colourtext'%
    }{See the gnuplot documentation for explanation.%
    }{Either use 'blacktext' in gnuplot or load the package
      color.sty in LaTeX.}%
    \renewcommand\color[2][]{}%
  }%
  \providecommand\includegraphics[2][]{%
    \GenericError{(gnuplot) \space\space\space\@spaces}{%
      Package graphicx or graphics not loaded%
    }{See the gnuplot documentation for explanation.%
    }{The gnuplot epslatex terminal needs graphicx.sty or graphics.sty.}%
    \renewcommand\includegraphics[2][]{}%
  }%
  \providecommand\rotatebox[2]{#2}%
  \@ifundefined{ifGPcolor}{%
    \newif\ifGPcolor
    \GPcolortrue
  }{}%
  \@ifundefined{ifGPblacktext}{%
    \newif\ifGPblacktext
    \GPblacktexttrue
  }{}%
  \let\gplgaddtomacro\g@addto@macro
  \gdef\gplbacktext{}%
  \gdef\gplfronttext{}%
  \makeatother
  \ifGPblacktext
    \def\colorrgb#1{}%
    \def\colorgray#1{}%
  \else
    \ifGPcolor
      \def\colorrgb#1{\color[rgb]{#1}}%
      \def\colorgray#1{\color[gray]{#1}}%
      \expandafter\def\csname LTw\endcsname{\color{white}}%
      \expandafter\def\csname LTb\endcsname{\color{black}}%
      \expandafter\def\csname LTa\endcsname{\color{black}}%
      \expandafter\def\csname LT0\endcsname{\color[rgb]{1,0,0}}%
      \expandafter\def\csname LT1\endcsname{\color[rgb]{0,1,0}}%
      \expandafter\def\csname LT2\endcsname{\color[rgb]{0,0,1}}%
      \expandafter\def\csname LT3\endcsname{\color[rgb]{1,0,1}}%
      \expandafter\def\csname LT4\endcsname{\color[rgb]{0,1,1}}%
      \expandafter\def\csname LT5\endcsname{\color[rgb]{1,1,0}}%
      \expandafter\def\csname LT6\endcsname{\color[rgb]{0,0,0}}%
      \expandafter\def\csname LT7\endcsname{\color[rgb]{1,0.3,0}}%
      \expandafter\def\csname LT8\endcsname{\color[rgb]{0.5,0.5,0.5}}%
    \else
      \def\colorrgb#1{\color{black}}%
      \def\colorgray#1{\color[gray]{#1}}%
      \expandafter\def\csname LTw\endcsname{\color{white}}%
      \expandafter\def\csname LTb\endcsname{\color{black}}%
      \expandafter\def\csname LTa\endcsname{\color{black}}%
      \expandafter\def\csname LT0\endcsname{\color{black}}%
      \expandafter\def\csname LT1\endcsname{\color{black}}%
      \expandafter\def\csname LT2\endcsname{\color{black}}%
      \expandafter\def\csname LT3\endcsname{\color{black}}%
      \expandafter\def\csname LT4\endcsname{\color{black}}%
      \expandafter\def\csname LT5\endcsname{\color{black}}%
      \expandafter\def\csname LT6\endcsname{\color{black}}%
      \expandafter\def\csname LT7\endcsname{\color{black}}%
      \expandafter\def\csname LT8\endcsname{\color{black}}%
    \fi
  \fi
  \setlength{\unitlength}{0.0500bp}%
  \begin{picture}(5328.00,3684.00)%
    \gplgaddtomacro\gplbacktext{%
      \csname LTb\endcsname%
      \put(858,704){\makebox(0,0)[r]{\strut{}\normalsize 0.0}}%
      \csname LTb\endcsname%
      \put(858,1339){\makebox(0,0)[r]{\strut{}\normalsize 0.5}}%
      \csname LTb\endcsname%
      \put(858,1974){\makebox(0,0)[r]{\strut{}\normalsize 1.0}}%
      \csname LTb\endcsname%
      \put(858,2608){\makebox(0,0)[r]{\strut{}\normalsize 1.5}}%
      \csname LTb\endcsname%
      \put(858,3243){\makebox(0,0)[r]{\strut{}\normalsize 2.0}}%
      \csname LTb\endcsname%
      \put(990,484){\makebox(0,0){\strut{}\normalsize 0.0}}%
      \csname LTb\endcsname%
      \put(1647,484){\makebox(0,0){\strut{}\normalsize 0.5}}%
      \csname LTb\endcsname%
      \put(2304,484){\makebox(0,0){\strut{}\normalsize 1.0}}%
      \csname LTb\endcsname%
      \put(2961,484){\makebox(0,0){\strut{}\normalsize 1.5}}%
      \csname LTb\endcsname%
      \put(3617,484){\makebox(0,0){\strut{}\normalsize 2.0}}%
      \csname LTb\endcsname%
      \put(4274,484){\makebox(0,0){\strut{}\normalsize 2.5}}%
      \csname LTb\endcsname%
      \put(4931,484){\makebox(0,0){\strut{}\normalsize 3.0}}%
      \colorrgb{0.75,0.00,0.00}%
      \put(990,3408){\makebox(0,0){\strut{}\color[HTML]{C00000}$\frac{R_{50}}{l_{\mathrm{t}}}$}}%
      \colorrgb{0.75,0.00,0.00}%
      \put(1334,3408){\makebox(0,0){\strut{}\color[HTML]{C00000}0.5}}%
      \colorrgb{0.75,0.00,0.00}%
      \put(1930,3408){\makebox(0,0){\strut{}\color[HTML]{C00000}1.0}}%
      \colorrgb{0.75,0.00,0.00}%
      \put(2865,3408){\makebox(0,0){\strut{}\color[HTML]{C00000}2.0}}%
      \colorrgb{0.75,0.00,0.00}%
      \put(3632,3408){\makebox(0,0){\strut{}\color[HTML]{C00000}3.0}}%
      \colorrgb{0.75,0.00,0.00}%
      \put(4931,3408){\makebox(0,0){\strut{}\color[HTML]{C00000}$\frac{R_{50}}{l_{\mathrm{t}}}$}}%
      \csname LTb\endcsname%
      \put(352,1973){\rotatebox{-270}{\makebox(0,0){\strut{}\large  $l_{\mathrm{f}}^2 \cdot \left( \left< \kappa^2 \right> - \left< \kappa \right>^2 \right) $ [-]}}}%
      \put(2960,154){\makebox(0,0){\strut{}\large Normalized Time $\frac{t}{\tau_{\mathrm{t}}}$ [-]}}%
    }%
    \gplgaddtomacro\gplfronttext{%
      \csname LTb\endcsname%
      \put(4301,1691){\makebox(0,0)[r]{\strut{}\normalsize Engine Kernel I\,\,}}%
      \csname LTb\endcsname%
      \put(4301,1427){\makebox(0,0)[r]{\strut{}\normalsize Engine Kernel II}}%
      \csname LTb\endcsname%
      \put(4301,1163){\makebox(0,0)[r]{\strut{}\normalsize Large Kernel}}%
      \csname LTb\endcsname%
      \put(4301,899){\makebox(0,0)[r]{\strut{}\normalsize Planar Flame}}%
    }%
    \gplbacktext
    \put(0,0){\includegraphics{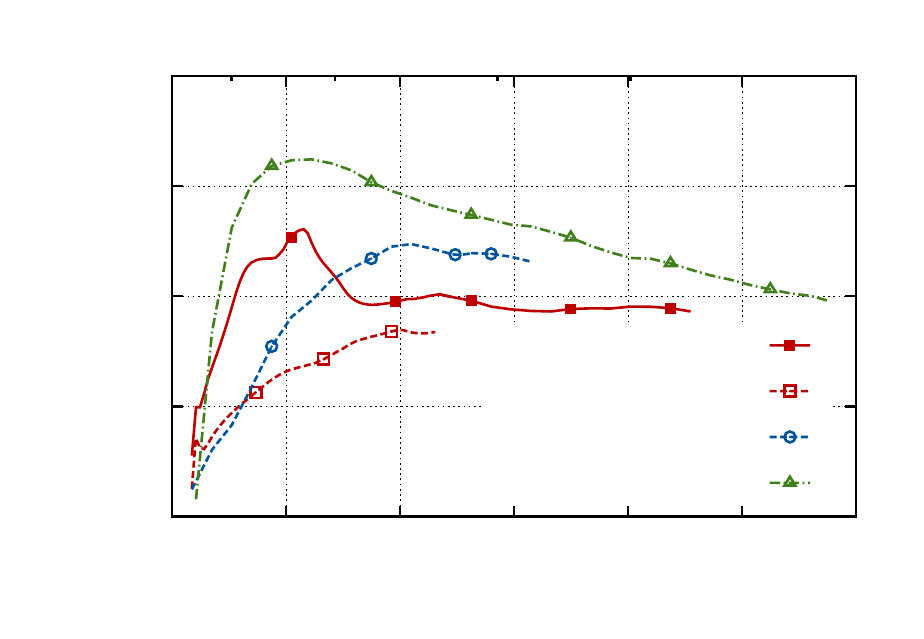}}%
    \gplfronttext
  \end{picture}%
\endgroup

%% file: data/FLAME_KERNEL_DNS_02/181003_curv_strain_kin_restor_diss_divg_skew_190712/flame_kernel_curvVar_minus_02_LOC01_03_P02_JFM_HALF.tex
\begingroup
  \makeatletter
  \providecommand\color[2][]{%
    \GenericError{(gnuplot) \space\space\space\@spaces}{%
      Package color not loaded in conjunction with
      terminal option `colourtext'%
    }{See the gnuplot documentation for explanation.%
    }{Either use 'blacktext' in gnuplot or load the package
      color.sty in LaTeX.}%
    \renewcommand\color[2][]{}%
  }%
  \providecommand\includegraphics[2][]{%
    \GenericError{(gnuplot) \space\space\space\@spaces}{%
      Package graphicx or graphics not loaded%
    }{See the gnuplot documentation for explanation.%
    }{The gnuplot epslatex terminal needs graphicx.sty or graphics.sty.}%
    \renewcommand\includegraphics[2][]{}%
  }%
  \providecommand\rotatebox[2]{#2}%
  \@ifundefined{ifGPcolor}{%
    \newif\ifGPcolor
    \GPcolortrue
  }{}%
  \@ifundefined{ifGPblacktext}{%
    \newif\ifGPblacktext
    \GPblacktexttrue
  }{}%
  \let\gplgaddtomacro\g@addto@macro
  \gdef\gplbacktext{}%
  \gdef\gplfronttext{}%
  \makeatother
  \ifGPblacktext
    \def\colorrgb#1{}%
    \def\colorgray#1{}%
  \else
    \ifGPcolor
      \def\colorrgb#1{\color[rgb]{#1}}%
      \def\colorgray#1{\color[gray]{#1}}%
      \expandafter\def\csname LTw\endcsname{\color{white}}%
      \expandafter\def\csname LTb\endcsname{\color{black}}%
      \expandafter\def\csname LTa\endcsname{\color{black}}%
      \expandafter\def\csname LT0\endcsname{\color[rgb]{1,0,0}}%
      \expandafter\def\csname LT1\endcsname{\color[rgb]{0,1,0}}%
      \expandafter\def\csname LT2\endcsname{\color[rgb]{0,0,1}}%
      \expandafter\def\csname LT3\endcsname{\color[rgb]{1,0,1}}%
      \expandafter\def\csname LT4\endcsname{\color[rgb]{0,1,1}}%
      \expandafter\def\csname LT5\endcsname{\color[rgb]{1,1,0}}%
      \expandafter\def\csname LT6\endcsname{\color[rgb]{0,0,0}}%
      \expandafter\def\csname LT7\endcsname{\color[rgb]{1,0.3,0}}%
      \expandafter\def\csname LT8\endcsname{\color[rgb]{0.5,0.5,0.5}}%
    \else
      \def\colorrgb#1{\color{black}}%
      \def\colorgray#1{\color[gray]{#1}}%
      \expandafter\def\csname LTw\endcsname{\color{white}}%
      \expandafter\def\csname LTb\endcsname{\color{black}}%
      \expandafter\def\csname LTa\endcsname{\color{black}}%
      \expandafter\def\csname LT0\endcsname{\color{black}}%
      \expandafter\def\csname LT1\endcsname{\color{black}}%
      \expandafter\def\csname LT2\endcsname{\color{black}}%
      \expandafter\def\csname LT3\endcsname{\color{black}}%
      \expandafter\def\csname LT4\endcsname{\color{black}}%
      \expandafter\def\csname LT5\endcsname{\color{black}}%
      \expandafter\def\csname LT6\endcsname{\color{black}}%
      \expandafter\def\csname LT7\endcsname{\color{black}}%
      \expandafter\def\csname LT8\endcsname{\color{black}}%
    \fi
  \fi
  \setlength{\unitlength}{0.0500bp}%
  \begin{picture}(5328.00,3684.00)%
    \gplgaddtomacro\gplbacktext{%
      \csname LTb\endcsname%
      \put(858,704){\makebox(0,0)[r]{\strut{}\normalsize 0.0}}%
      \csname LTb\endcsname%
      \put(858,1067){\makebox(0,0)[r]{\strut{}\normalsize 0.2}}%
      \csname LTb\endcsname%
      \put(858,1429){\makebox(0,0)[r]{\strut{}\normalsize 0.4}}%
      \csname LTb\endcsname%
      \put(858,1792){\makebox(0,0)[r]{\strut{}\normalsize 0.6}}%
      \csname LTb\endcsname%
      \put(858,2155){\makebox(0,0)[r]{\strut{}\normalsize 0.8}}%
      \csname LTb\endcsname%
      \put(858,2518){\makebox(0,0)[r]{\strut{}\normalsize 1.0}}%
      \csname LTb\endcsname%
      \put(858,2880){\makebox(0,0)[r]{\strut{}\normalsize 1.2}}%
      \csname LTb\endcsname%
      \put(858,3243){\makebox(0,0)[r]{\strut{}\normalsize 1.4}}%
      \csname LTb\endcsname%
      \put(990,484){\makebox(0,0){\strut{}\normalsize 0.0}}%
      \csname LTb\endcsname%
      \put(1647,484){\makebox(0,0){\strut{}\normalsize 0.5}}%
      \csname LTb\endcsname%
      \put(2304,484){\makebox(0,0){\strut{}\normalsize 1.0}}%
      \csname LTb\endcsname%
      \put(2961,484){\makebox(0,0){\strut{}\normalsize 1.5}}%
      \csname LTb\endcsname%
      \put(3617,484){\makebox(0,0){\strut{}\normalsize 2.0}}%
      \csname LTb\endcsname%
      \put(4274,484){\makebox(0,0){\strut{}\normalsize 2.5}}%
      \csname LTb\endcsname%
      \put(4931,484){\makebox(0,0){\strut{}\normalsize 3.0}}%
      \colorrgb{0.75,0.00,0.00}%
      \put(990,3408){\makebox(0,0){\strut{}\color[HTML]{C00000}$\frac{R_{50}}{l_{\mathrm{t}}}$}}%
      \colorrgb{0.75,0.00,0.00}%
      \put(1334,3408){\makebox(0,0){\strut{}\color[HTML]{C00000}0.5}}%
      \colorrgb{0.75,0.00,0.00}%
      \put(1930,3408){\makebox(0,0){\strut{}\color[HTML]{C00000}1.0}}%
      \colorrgb{0.75,0.00,0.00}%
      \put(2865,3408){\makebox(0,0){\strut{}\color[HTML]{C00000}2.0}}%
      \colorrgb{0.75,0.00,0.00}%
      \put(3632,3408){\makebox(0,0){\strut{}\color[HTML]{C00000}3.0}}%
      \colorrgb{0.75,0.00,0.00}%
      \put(4931,3408){\makebox(0,0){\strut{}\color[HTML]{C00000}$\frac{R_{50}}{l_{\mathrm{t}}}$}}%
      \csname LTb\endcsname%
      \put(352,1973){\rotatebox{-270}{\makebox(0,0){\strut{}\large $l_{\mathrm{f}}^2\cdot{\left<\kappa^2 - \left<\kappa\right>_{\mathrm{s},\kappa^-}^2 \right>_{\mathrm{s},\kappa^-}}$ [-]}}}%
      \put(2960,154){\makebox(0,0){\strut{}\large Normalized Time $\frac{t}{\tau_{\mathrm{t}}}$ [-]}}%
    }%
    \gplgaddtomacro\gplfronttext{%
      \csname LTb\endcsname%
      \put(4301,1691){\makebox(0,0)[r]{\strut{}\normalsize Engine Kernel I\,\,}}%
      \csname LTb\endcsname%
      \put(4301,1427){\makebox(0,0)[r]{\strut{}\normalsize Engine Kernel II}}%
      \csname LTb\endcsname%
      \put(4301,1163){\makebox(0,0)[r]{\strut{}\normalsize Large Kernel}}%
      \csname LTb\endcsname%
      \put(4301,899){\makebox(0,0)[r]{\strut{}\normalsize Planar Flame}}%
    }%
    \gplbacktext
    \put(0,0){\includegraphics{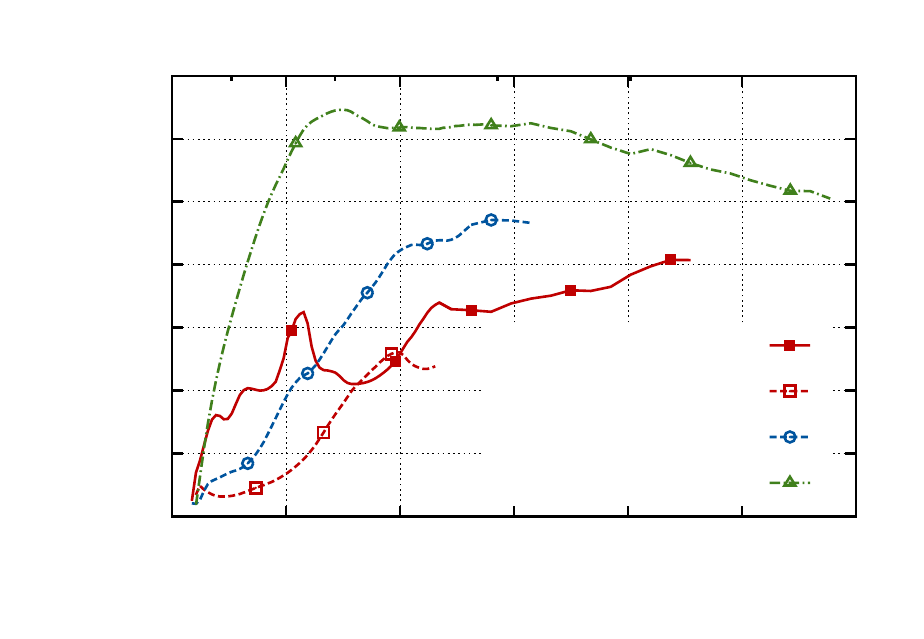}}%
    \gplfronttext
  \end{picture}%
\endgroup

%% file: data/FLAME_KERNEL_DNS_02/181003_curv_strain_kin_restor_diss_divg_skew_190712/flame_kernel_curvVar_plus_02_LOC01_03_P02_JFM_HALF.tex
\begingroup
  \makeatletter
  \providecommand\color[2][]{%
    \GenericError{(gnuplot) \space\space\space\@spaces}{%
      Package color not loaded in conjunction with
      terminal option `colourtext'%
    }{See the gnuplot documentation for explanation.%
    }{Either use 'blacktext' in gnuplot or load the package
      color.sty in LaTeX.}%
    \renewcommand\color[2][]{}%
  }%
  \providecommand\includegraphics[2][]{%
    \GenericError{(gnuplot) \space\space\space\@spaces}{%
      Package graphicx or graphics not loaded%
    }{See the gnuplot documentation for explanation.%
    }{The gnuplot epslatex terminal needs graphicx.sty or graphics.sty.}%
    \renewcommand\includegraphics[2][]{}%
  }%
  \providecommand\rotatebox[2]{#2}%
  \@ifundefined{ifGPcolor}{%
    \newif\ifGPcolor
    \GPcolortrue
  }{}%
  \@ifundefined{ifGPblacktext}{%
    \newif\ifGPblacktext
    \GPblacktexttrue
  }{}%
  \let\gplgaddtomacro\g@addto@macro
  \gdef\gplbacktext{}%
  \gdef\gplfronttext{}%
  \makeatother
  \ifGPblacktext
    \def\colorrgb#1{}%
    \def\colorgray#1{}%
  \else
    \ifGPcolor
      \def\colorrgb#1{\color[rgb]{#1}}%
      \def\colorgray#1{\color[gray]{#1}}%
      \expandafter\def\csname LTw\endcsname{\color{white}}%
      \expandafter\def\csname LTb\endcsname{\color{black}}%
      \expandafter\def\csname LTa\endcsname{\color{black}}%
      \expandafter\def\csname LT0\endcsname{\color[rgb]{1,0,0}}%
      \expandafter\def\csname LT1\endcsname{\color[rgb]{0,1,0}}%
      \expandafter\def\csname LT2\endcsname{\color[rgb]{0,0,1}}%
      \expandafter\def\csname LT3\endcsname{\color[rgb]{1,0,1}}%
      \expandafter\def\csname LT4\endcsname{\color[rgb]{0,1,1}}%
      \expandafter\def\csname LT5\endcsname{\color[rgb]{1,1,0}}%
      \expandafter\def\csname LT6\endcsname{\color[rgb]{0,0,0}}%
      \expandafter\def\csname LT7\endcsname{\color[rgb]{1,0.3,0}}%
      \expandafter\def\csname LT8\endcsname{\color[rgb]{0.5,0.5,0.5}}%
    \else
      \def\colorrgb#1{\color{black}}%
      \def\colorgray#1{\color[gray]{#1}}%
      \expandafter\def\csname LTw\endcsname{\color{white}}%
      \expandafter\def\csname LTb\endcsname{\color{black}}%
      \expandafter\def\csname LTa\endcsname{\color{black}}%
      \expandafter\def\csname LT0\endcsname{\color{black}}%
      \expandafter\def\csname LT1\endcsname{\color{black}}%
      \expandafter\def\csname LT2\endcsname{\color{black}}%
      \expandafter\def\csname LT3\endcsname{\color{black}}%
      \expandafter\def\csname LT4\endcsname{\color{black}}%
      \expandafter\def\csname LT5\endcsname{\color{black}}%
      \expandafter\def\csname LT6\endcsname{\color{black}}%
      \expandafter\def\csname LT7\endcsname{\color{black}}%
      \expandafter\def\csname LT8\endcsname{\color{black}}%
    \fi
  \fi
  \setlength{\unitlength}{0.0500bp}%
  \begin{picture}(5328.00,3684.00)%
    \gplgaddtomacro\gplbacktext{%
      \csname LTb\endcsname%
      \put(858,704){\makebox(0,0)[r]{\strut{}\normalsize 0.0}}%
      \csname LTb\endcsname%
      \put(858,1067){\makebox(0,0)[r]{\strut{}\normalsize 0.2}}%
      \csname LTb\endcsname%
      \put(858,1429){\makebox(0,0)[r]{\strut{}\normalsize 0.4}}%
      \csname LTb\endcsname%
      \put(858,1792){\makebox(0,0)[r]{\strut{}\normalsize 0.6}}%
      \csname LTb\endcsname%
      \put(858,2155){\makebox(0,0)[r]{\strut{}\normalsize 0.8}}%
      \csname LTb\endcsname%
      \put(858,2518){\makebox(0,0)[r]{\strut{}\normalsize 1.0}}%
      \csname LTb\endcsname%
      \put(858,2880){\makebox(0,0)[r]{\strut{}\normalsize 1.2}}%
      \csname LTb\endcsname%
      \put(858,3243){\makebox(0,0)[r]{\strut{}\normalsize 1.4}}%
      \csname LTb\endcsname%
      \put(990,484){\makebox(0,0){\strut{}\normalsize 0.0}}%
      \csname LTb\endcsname%
      \put(1647,484){\makebox(0,0){\strut{}\normalsize 0.5}}%
      \csname LTb\endcsname%
      \put(2304,484){\makebox(0,0){\strut{}\normalsize 1.0}}%
      \csname LTb\endcsname%
      \put(2961,484){\makebox(0,0){\strut{}\normalsize 1.5}}%
      \csname LTb\endcsname%
      \put(3617,484){\makebox(0,0){\strut{}\normalsize 2.0}}%
      \csname LTb\endcsname%
      \put(4274,484){\makebox(0,0){\strut{}\normalsize 2.5}}%
      \csname LTb\endcsname%
      \put(4931,484){\makebox(0,0){\strut{}\normalsize 3.0}}%
      \colorrgb{0.75,0.00,0.00}%
      \put(990,3408){\makebox(0,0){\strut{}\color[HTML]{C00000}$\frac{R_{50}}{l_{\mathrm{t}}}$}}%
      \colorrgb{0.75,0.00,0.00}%
      \put(1334,3408){\makebox(0,0){\strut{}\color[HTML]{C00000}0.5}}%
      \colorrgb{0.75,0.00,0.00}%
      \put(1930,3408){\makebox(0,0){\strut{}\color[HTML]{C00000}1.0}}%
      \colorrgb{0.75,0.00,0.00}%
      \put(2865,3408){\makebox(0,0){\strut{}\color[HTML]{C00000}2.0}}%
      \colorrgb{0.75,0.00,0.00}%
      \put(3632,3408){\makebox(0,0){\strut{}\color[HTML]{C00000}3.0}}%
      \colorrgb{0.75,0.00,0.00}%
      \put(4931,3408){\makebox(0,0){\strut{}\color[HTML]{C00000}$\frac{R_{50}}{l_{\mathrm{t}}}$}}%
      \csname LTb\endcsname%
      \put(352,1973){\rotatebox{-270}{\makebox(0,0){\strut{}\large $l_{\mathrm{f}}^2\cdot{\left<\kappa^2 - \left<\kappa\right>_{\mathrm{s},\kappa^+}^2 \right>_{\mathrm{s},\kappa^+}}$ [-]}}}%
      \put(2960,154){\makebox(0,0){\strut{}\large Normalized Time $\frac{t}{\tau_{\mathrm{t}}}$ [-]}}%
    }%
    \gplgaddtomacro\gplfronttext{%
      \csname LTb\endcsname%
      \put(4301,3048){\makebox(0,0)[r]{\strut{}\normalsize Engine Kernel I\,\,}}%
      \csname LTb\endcsname%
      \put(4301,2784){\makebox(0,0)[r]{\strut{}\normalsize Engine Kernel II}}%
      \csname LTb\endcsname%
      \put(4301,2520){\makebox(0,0)[r]{\strut{}\normalsize Large Kernel}}%
      \csname LTb\endcsname%
      \put(4301,2256){\makebox(0,0)[r]{\strut{}\normalsize Planar Flame}}%
    }%
    \gplbacktext
    \put(0,0){\includegraphics{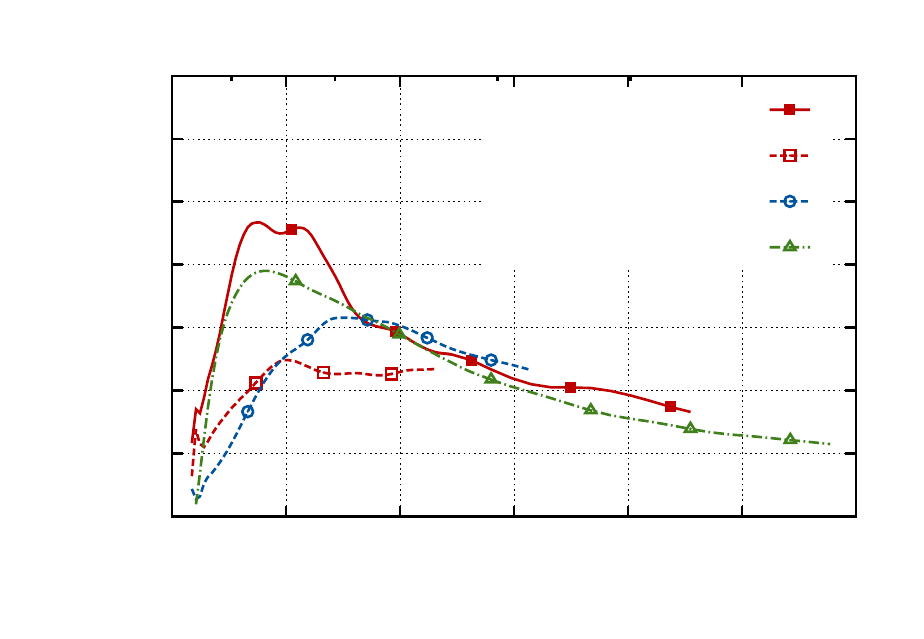}}%
    \gplfronttext
  \end{picture}%
\endgroup

%% file: data/FLAME_KERNEL_DNS_02/181002_curvEq_Ti_diss_skew_mp_curv_mp_progT/flame_kernel_curvEq_flow_flame_net_terms_prodMINUS_02_LOC01_JFM_HALF.tex
\begingroup
  \makeatletter
  \providecommand\color[2][]{%
    \GenericError{(gnuplot) \space\space\space\@spaces}{%
      Package color not loaded in conjunction with
      terminal option `colourtext'%
    }{See the gnuplot documentation for explanation.%
    }{Either use 'blacktext' in gnuplot or load the package
      color.sty in LaTeX.}%
    \renewcommand\color[2][]{}%
  }%
  \providecommand\includegraphics[2][]{%
    \GenericError{(gnuplot) \space\space\space\@spaces}{%
      Package graphicx or graphics not loaded%
    }{See the gnuplot documentation for explanation.%
    }{The gnuplot epslatex terminal needs graphicx.sty or graphics.sty.}%
    \renewcommand\includegraphics[2][]{}%
  }%
  \providecommand\rotatebox[2]{#2}%
  \@ifundefined{ifGPcolor}{%
    \newif\ifGPcolor
    \GPcolortrue
  }{}%
  \@ifundefined{ifGPblacktext}{%
    \newif\ifGPblacktext
    \GPblacktexttrue
  }{}%
  \let\gplgaddtomacro\g@addto@macro
  \gdef\gplbacktext{}%
  \gdef\gplfronttext{}%
  \makeatother
  \ifGPblacktext
    \def\colorrgb#1{}%
    \def\colorgray#1{}%
  \else
    \ifGPcolor
      \def\colorrgb#1{\color[rgb]{#1}}%
      \def\colorgray#1{\color[gray]{#1}}%
      \expandafter\def\csname LTw\endcsname{\color{white}}%
      \expandafter\def\csname LTb\endcsname{\color{black}}%
      \expandafter\def\csname LTa\endcsname{\color{black}}%
      \expandafter\def\csname LT0\endcsname{\color[rgb]{1,0,0}}%
      \expandafter\def\csname LT1\endcsname{\color[rgb]{0,1,0}}%
      \expandafter\def\csname LT2\endcsname{\color[rgb]{0,0,1}}%
      \expandafter\def\csname LT3\endcsname{\color[rgb]{1,0,1}}%
      \expandafter\def\csname LT4\endcsname{\color[rgb]{0,1,1}}%
      \expandafter\def\csname LT5\endcsname{\color[rgb]{1,1,0}}%
      \expandafter\def\csname LT6\endcsname{\color[rgb]{0,0,0}}%
      \expandafter\def\csname LT7\endcsname{\color[rgb]{1,0.3,0}}%
      \expandafter\def\csname LT8\endcsname{\color[rgb]{0.5,0.5,0.5}}%
    \else
      \def\colorrgb#1{\color{black}}%
      \def\colorgray#1{\color[gray]{#1}}%
      \expandafter\def\csname LTw\endcsname{\color{white}}%
      \expandafter\def\csname LTb\endcsname{\color{black}}%
      \expandafter\def\csname LTa\endcsname{\color{black}}%
      \expandafter\def\csname LT0\endcsname{\color{black}}%
      \expandafter\def\csname LT1\endcsname{\color{black}}%
      \expandafter\def\csname LT2\endcsname{\color{black}}%
      \expandafter\def\csname LT3\endcsname{\color{black}}%
      \expandafter\def\csname LT4\endcsname{\color{black}}%
      \expandafter\def\csname LT5\endcsname{\color{black}}%
      \expandafter\def\csname LT6\endcsname{\color{black}}%
      \expandafter\def\csname LT7\endcsname{\color{black}}%
      \expandafter\def\csname LT8\endcsname{\color{black}}%
    \fi
  \fi
  \setlength{\unitlength}{0.0500bp}%
  \begin{picture}(5328.00,3684.00)%
    \gplgaddtomacro\gplbacktext{%
      \csname LTb\endcsname%
      \put(990,704){\makebox(0,0)[r]{\strut{}\normalsize -1.0}}%
      \csname LTb\endcsname%
      \put(990,1212){\makebox(0,0)[r]{\strut{}\normalsize 0.0}}%
      \csname LTb\endcsname%
      \put(990,1720){\makebox(0,0)[r]{\strut{}\normalsize 1.0}}%
      \csname LTb\endcsname%
      \put(990,2227){\makebox(0,0)[r]{\strut{}\normalsize 2.0}}%
      \csname LTb\endcsname%
      \put(990,2735){\makebox(0,0)[r]{\strut{}\normalsize 3.0}}%
      \csname LTb\endcsname%
      \put(990,3243){\makebox(0,0)[r]{\strut{}\normalsize 4.0}}%
      \csname LTb\endcsname%
      \put(1122,484){\makebox(0,0){\strut{}\normalsize 0.0}}%
      \csname LTb\endcsname%
      \put(1757,484){\makebox(0,0){\strut{}\normalsize 0.2}}%
      \csname LTb\endcsname%
      \put(2392,484){\makebox(0,0){\strut{}\normalsize 0.4}}%
      \csname LTb\endcsname%
      \put(3027,484){\makebox(0,0){\strut{}\normalsize 0.6}}%
      \csname LTb\endcsname%
      \put(3661,484){\makebox(0,0){\strut{}\normalsize 0.8}}%
      \csname LTb\endcsname%
      \put(4296,484){\makebox(0,0){\strut{}\normalsize 1.0}}%
      \csname LTb\endcsname%
      \put(4931,484){\makebox(0,0){\strut{}\normalsize 1.2}}%
      \colorrgb{0.75,0.00,0.00}%
      \put(1122,3408){\makebox(0,0){\strut{}\color[HTML]{C00000}$\frac{R_{50}}{l_{\mathrm{t}}}$}}%
      \colorrgb{0.75,0.00,0.00}%
      \put(1953,3408){\makebox(0,0){\strut{}\color[HTML]{C00000}0.5}}%
      \colorrgb{0.75,0.00,0.00}%
      \put(3392,3408){\makebox(0,0){\strut{}\color[HTML]{C00000}1.0}}%
      \colorrgb{0.75,0.00,0.00}%
      \put(4930,3408){\makebox(0,0){\strut{}\color[HTML]{C00000}$\frac{R_{50}}{l_{\mathrm{t}}}$}}%
      \csname LTb\endcsname%
      \put(352,1973){\rotatebox{-270}{\makebox(0,0){\strut{}\large Net Production $\left( \kappa^{-} \right)$ [-]}}}%
      \put(3026,154){\makebox(0,0){\strut{}\large Normalized Time $\frac{t}{\tau_{\mathrm{t}}}$ [-]}}%
    }%
    \gplgaddtomacro\gplfronttext{%
      \csname LTb\endcsname%
      \put(2772,3048){\makebox(0,0)[r]{\strut{}\normalsize Engine Kernel I\,\,}}%
      \csname LTb\endcsname%
      \put(2772,2784){\makebox(0,0)[r]{\strut{}\normalsize Engine Kernel II}}%
      \csname LTb\endcsname%
      \put(2772,2520){\makebox(0,0)[r]{\strut{}\normalsize Large Kernel}}%
    }%
    \gplbacktext
    \put(0,0){\includegraphics{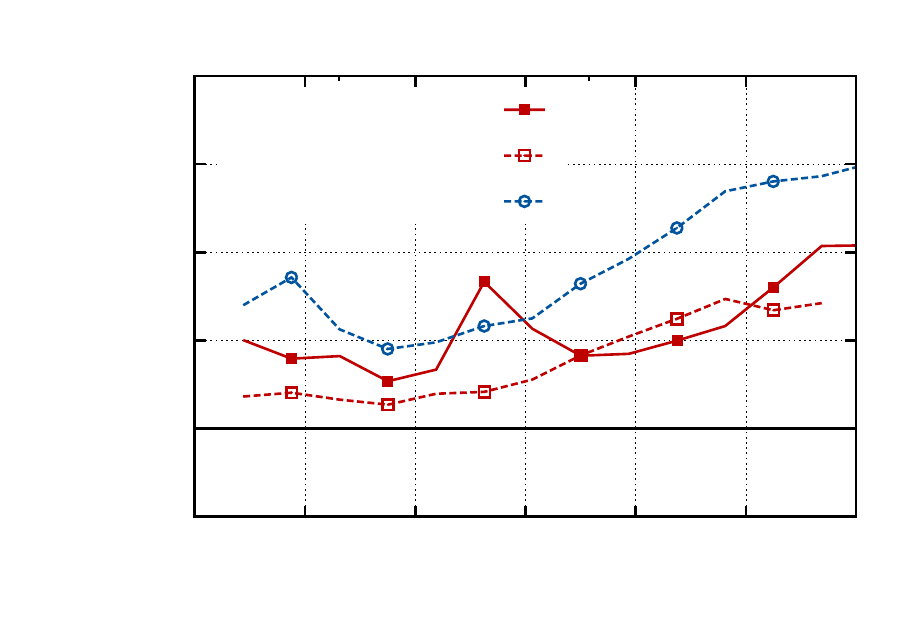}}%
    \gplfronttext
  \end{picture}%
\endgroup

%% file: data/FLAME_KERNEL_DNS_02/181002_curvEq_Ti_diss_skew_mp_curv_mp_progT/flame_kernel_curvEq_flow_flame_net_terms_prodPLUS_02_LOC01_JFM_HALF.tex
\begingroup
  \makeatletter
  \providecommand\color[2][]{%
    \GenericError{(gnuplot) \space\space\space\@spaces}{%
      Package color not loaded in conjunction with
      terminal option `colourtext'%
    }{See the gnuplot documentation for explanation.%
    }{Either use 'blacktext' in gnuplot or load the package
      color.sty in LaTeX.}%
    \renewcommand\color[2][]{}%
  }%
  \providecommand\includegraphics[2][]{%
    \GenericError{(gnuplot) \space\space\space\@spaces}{%
      Package graphicx or graphics not loaded%
    }{See the gnuplot documentation for explanation.%
    }{The gnuplot epslatex terminal needs graphicx.sty or graphics.sty.}%
    \renewcommand\includegraphics[2][]{}%
  }%
  \providecommand\rotatebox[2]{#2}%
  \@ifundefined{ifGPcolor}{%
    \newif\ifGPcolor
    \GPcolortrue
  }{}%
  \@ifundefined{ifGPblacktext}{%
    \newif\ifGPblacktext
    \GPblacktexttrue
  }{}%
  \let\gplgaddtomacro\g@addto@macro
  \gdef\gplbacktext{}%
  \gdef\gplfronttext{}%
  \makeatother
  \ifGPblacktext
    \def\colorrgb#1{}%
    \def\colorgray#1{}%
  \else
    \ifGPcolor
      \def\colorrgb#1{\color[rgb]{#1}}%
      \def\colorgray#1{\color[gray]{#1}}%
      \expandafter\def\csname LTw\endcsname{\color{white}}%
      \expandafter\def\csname LTb\endcsname{\color{black}}%
      \expandafter\def\csname LTa\endcsname{\color{black}}%
      \expandafter\def\csname LT0\endcsname{\color[rgb]{1,0,0}}%
      \expandafter\def\csname LT1\endcsname{\color[rgb]{0,1,0}}%
      \expandafter\def\csname LT2\endcsname{\color[rgb]{0,0,1}}%
      \expandafter\def\csname LT3\endcsname{\color[rgb]{1,0,1}}%
      \expandafter\def\csname LT4\endcsname{\color[rgb]{0,1,1}}%
      \expandafter\def\csname LT5\endcsname{\color[rgb]{1,1,0}}%
      \expandafter\def\csname LT6\endcsname{\color[rgb]{0,0,0}}%
      \expandafter\def\csname LT7\endcsname{\color[rgb]{1,0.3,0}}%
      \expandafter\def\csname LT8\endcsname{\color[rgb]{0.5,0.5,0.5}}%
    \else
      \def\colorrgb#1{\color{black}}%
      \def\colorgray#1{\color[gray]{#1}}%
      \expandafter\def\csname LTw\endcsname{\color{white}}%
      \expandafter\def\csname LTb\endcsname{\color{black}}%
      \expandafter\def\csname LTa\endcsname{\color{black}}%
      \expandafter\def\csname LT0\endcsname{\color{black}}%
      \expandafter\def\csname LT1\endcsname{\color{black}}%
      \expandafter\def\csname LT2\endcsname{\color{black}}%
      \expandafter\def\csname LT3\endcsname{\color{black}}%
      \expandafter\def\csname LT4\endcsname{\color{black}}%
      \expandafter\def\csname LT5\endcsname{\color{black}}%
      \expandafter\def\csname LT6\endcsname{\color{black}}%
      \expandafter\def\csname LT7\endcsname{\color{black}}%
      \expandafter\def\csname LT8\endcsname{\color{black}}%
    \fi
  \fi
  \setlength{\unitlength}{0.0500bp}%
  \begin{picture}(5328.00,3684.00)%
    \gplgaddtomacro\gplbacktext{%
      \csname LTb\endcsname%
      \put(990,704){\makebox(0,0)[r]{\strut{}\normalsize -1.0}}%
      \csname LTb\endcsname%
      \put(990,1212){\makebox(0,0)[r]{\strut{}\normalsize 0.0}}%
      \csname LTb\endcsname%
      \put(990,1720){\makebox(0,0)[r]{\strut{}\normalsize 1.0}}%
      \csname LTb\endcsname%
      \put(990,2227){\makebox(0,0)[r]{\strut{}\normalsize 2.0}}%
      \csname LTb\endcsname%
      \put(990,2735){\makebox(0,0)[r]{\strut{}\normalsize 3.0}}%
      \csname LTb\endcsname%
      \put(990,3243){\makebox(0,0)[r]{\strut{}\normalsize 4.0}}%
      \csname LTb\endcsname%
      \put(1122,484){\makebox(0,0){\strut{}\normalsize 0.0}}%
      \csname LTb\endcsname%
      \put(1757,484){\makebox(0,0){\strut{}\normalsize 0.2}}%
      \csname LTb\endcsname%
      \put(2392,484){\makebox(0,0){\strut{}\normalsize 0.4}}%
      \csname LTb\endcsname%
      \put(3027,484){\makebox(0,0){\strut{}\normalsize 0.6}}%
      \csname LTb\endcsname%
      \put(3661,484){\makebox(0,0){\strut{}\normalsize 0.8}}%
      \csname LTb\endcsname%
      \put(4296,484){\makebox(0,0){\strut{}\normalsize 1.0}}%
      \csname LTb\endcsname%
      \put(4931,484){\makebox(0,0){\strut{}\normalsize 1.2}}%
      \colorrgb{0.75,0.00,0.00}%
      \put(1122,3408){\makebox(0,0){\strut{}\color[HTML]{C00000}$\frac{R_{50}}{l_{\mathrm{t}}}$}}%
      \colorrgb{0.75,0.00,0.00}%
      \put(1953,3408){\makebox(0,0){\strut{}\color[HTML]{C00000}0.5}}%
      \colorrgb{0.75,0.00,0.00}%
      \put(3392,3408){\makebox(0,0){\strut{}\color[HTML]{C00000}1.0}}%
      \colorrgb{0.75,0.00,0.00}%
      \put(4930,3408){\makebox(0,0){\strut{}\color[HTML]{C00000}$\frac{R_{50}}{l_{\mathrm{t}}}$}}%
      \csname LTb\endcsname%
      \put(352,1973){\rotatebox{-270}{\makebox(0,0){\strut{}\large Net Production $\left( \kappa^{+} \right)$ [-]}}}%
      \put(3026,154){\makebox(0,0){\strut{}\large Normalized Time $\frac{t}{\tau_{\mathrm{t}}}$ [-]}}%
    }%
    \gplgaddtomacro\gplfronttext{%
      \csname LTb\endcsname%
      \put(4301,3048){\makebox(0,0)[r]{\strut{}\normalsize Engine Kernel I\,\,}}%
      \csname LTb\endcsname%
      \put(4301,2784){\makebox(0,0)[r]{\strut{}\normalsize Engine Kernel II}}%
      \csname LTb\endcsname%
      \put(4301,2520){\makebox(0,0)[r]{\strut{}\normalsize Large Kernel}}%
    }%
    \gplbacktext
    \put(0,0){\includegraphics{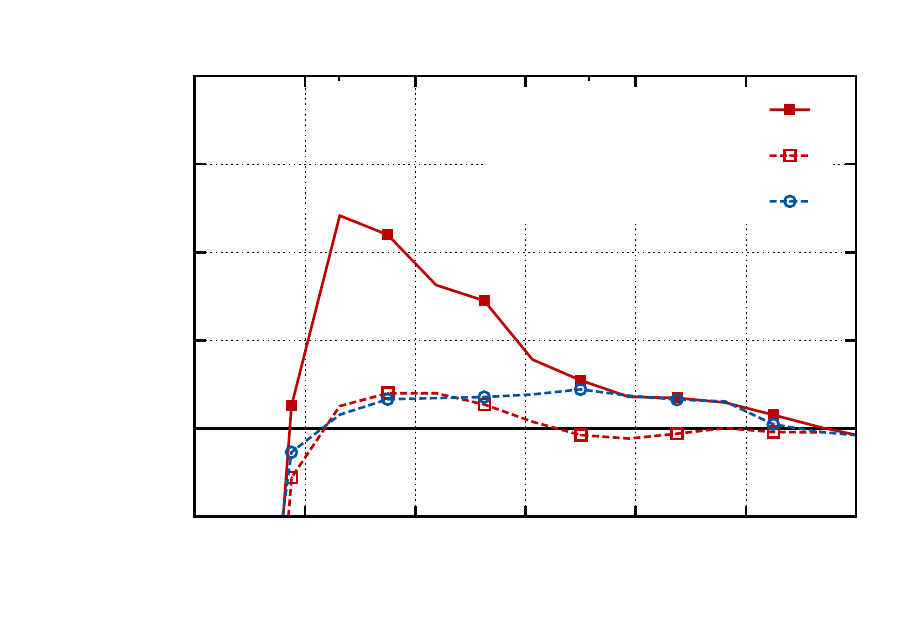}}%
    \gplfronttext
  \end{picture}%
\endgroup

%% file: data/FLAME_KERNEL_DNS_02/181002_curvEq_Ti_diss_skew_mp_curv_mp_progT/flame_kernel_curvEq_flow_terms_prodMINUS_Bending_1p2TAU_02_LOC01_JFM_HALF.tex
\begingroup
  \makeatletter
  \providecommand\color[2][]{%
    \GenericError{(gnuplot) \space\space\space\@spaces}{%
      Package color not loaded in conjunction with
      terminal option `colourtext'%
    }{See the gnuplot documentation for explanation.%
    }{Either use 'blacktext' in gnuplot or load the package
      color.sty in LaTeX.}%
    \renewcommand\color[2][]{}%
  }%
  \providecommand\includegraphics[2][]{%
    \GenericError{(gnuplot) \space\space\space\@spaces}{%
      Package graphicx or graphics not loaded%
    }{See the gnuplot documentation for explanation.%
    }{The gnuplot epslatex terminal needs graphicx.sty or graphics.sty.}%
    \renewcommand\includegraphics[2][]{}%
  }%
  \providecommand\rotatebox[2]{#2}%
  \@ifundefined{ifGPcolor}{%
    \newif\ifGPcolor
    \GPcolortrue
  }{}%
  \@ifundefined{ifGPblacktext}{%
    \newif\ifGPblacktext
    \GPblacktexttrue
  }{}%
  \let\gplgaddtomacro\g@addto@macro
  \gdef\gplbacktext{}%
  \gdef\gplfronttext{}%
  \makeatother
  \ifGPblacktext
    \def\colorrgb#1{}%
    \def\colorgray#1{}%
  \else
    \ifGPcolor
      \def\colorrgb#1{\color[rgb]{#1}}%
      \def\colorgray#1{\color[gray]{#1}}%
      \expandafter\def\csname LTw\endcsname{\color{white}}%
      \expandafter\def\csname LTb\endcsname{\color{black}}%
      \expandafter\def\csname LTa\endcsname{\color{black}}%
      \expandafter\def\csname LT0\endcsname{\color[rgb]{1,0,0}}%
      \expandafter\def\csname LT1\endcsname{\color[rgb]{0,1,0}}%
      \expandafter\def\csname LT2\endcsname{\color[rgb]{0,0,1}}%
      \expandafter\def\csname LT3\endcsname{\color[rgb]{1,0,1}}%
      \expandafter\def\csname LT4\endcsname{\color[rgb]{0,1,1}}%
      \expandafter\def\csname LT5\endcsname{\color[rgb]{1,1,0}}%
      \expandafter\def\csname LT6\endcsname{\color[rgb]{0,0,0}}%
      \expandafter\def\csname LT7\endcsname{\color[rgb]{1,0.3,0}}%
      \expandafter\def\csname LT8\endcsname{\color[rgb]{0.5,0.5,0.5}}%
    \else
      \def\colorrgb#1{\color{black}}%
      \def\colorgray#1{\color[gray]{#1}}%
      \expandafter\def\csname LTw\endcsname{\color{white}}%
      \expandafter\def\csname LTb\endcsname{\color{black}}%
      \expandafter\def\csname LTa\endcsname{\color{black}}%
      \expandafter\def\csname LT0\endcsname{\color{black}}%
      \expandafter\def\csname LT1\endcsname{\color{black}}%
      \expandafter\def\csname LT2\endcsname{\color{black}}%
      \expandafter\def\csname LT3\endcsname{\color{black}}%
      \expandafter\def\csname LT4\endcsname{\color{black}}%
      \expandafter\def\csname LT5\endcsname{\color{black}}%
      \expandafter\def\csname LT6\endcsname{\color{black}}%
      \expandafter\def\csname LT7\endcsname{\color{black}}%
      \expandafter\def\csname LT8\endcsname{\color{black}}%
    \fi
  \fi
  \setlength{\unitlength}{0.0500bp}%
  \begin{picture}(5328.00,3684.00)%
    \gplgaddtomacro\gplbacktext{%
      \csname LTb\endcsname%
      \put(990,704){\makebox(0,0)[r]{\strut{}\normalsize -4.0}}%
      \csname LTb\endcsname%
      \put(990,1247){\makebox(0,0)[r]{\strut{}\normalsize -2.0}}%
      \csname LTb\endcsname%
      \put(990,1790){\makebox(0,0)[r]{\strut{}\normalsize 0.0}}%
      \csname LTb\endcsname%
      \put(990,2333){\makebox(0,0)[r]{\strut{}\normalsize 2.0}}%
      \csname LTb\endcsname%
      \put(990,2876){\makebox(0,0)[r]{\strut{}\normalsize 4.0}}%
      \csname LTb\endcsname%
      \put(990,3419){\makebox(0,0)[r]{\strut{}\normalsize 6.0}}%
      \csname LTb\endcsname%
      \put(1122,484){\makebox(0,0){\strut{}\normalsize 0.0}}%
      \csname LTb\endcsname%
      \put(1757,484){\makebox(0,0){\strut{}\normalsize 0.2}}%
      \csname LTb\endcsname%
      \put(2392,484){\makebox(0,0){\strut{}\normalsize 0.4}}%
      \csname LTb\endcsname%
      \put(3027,484){\makebox(0,0){\strut{}\normalsize 0.6}}%
      \csname LTb\endcsname%
      \put(3661,484){\makebox(0,0){\strut{}\normalsize 0.8}}%
      \csname LTb\endcsname%
      \put(4296,484){\makebox(0,0){\strut{}\normalsize 1.0}}%
      \csname LTb\endcsname%
      \put(4931,484){\makebox(0,0){\strut{}\normalsize 1.2}}%
      \put(352,2061){\rotatebox{-270}{\makebox(0,0){\strut{}\large Production $\left( \kappa^{-} \right)$ [-]}}}%
      \put(3026,154){\makebox(0,0){\strut{}\large Normalized Time $\frac{t}{\tau_{\mathrm{t}}}$ [-]}}%
    }%
    \gplgaddtomacro\gplfronttext{%
      \csname LTb\endcsname%
      \put(1573,3268){\makebox(0,0)[r]{\strut{}\normalsize $-\mathfrak{T}_{\mathrm{n}}^{u}$}}%
      \csname LTb\endcsname%
      \put(1573,2993){\makebox(0,0)[r]{\strut{}\normalsize $-\mathfrak{T}_{\mathrm{t}}^{u}$}}%
      \csname LTb\endcsname%
      \put(2692,3268){\makebox(0,0)[r]{\strut{}\normalsize $-\mathfrak{T}_{\mathrm{b}}^{u}$}}%
      \csname LTb\endcsname%
      \put(2692,2993){\makebox(0,0)[r]{\strut{}\normalsize $-\sum \mathfrak{T}_{\ast}^{u}$}}%
    }%
    \gplbacktext
    \put(0,0){\includegraphics{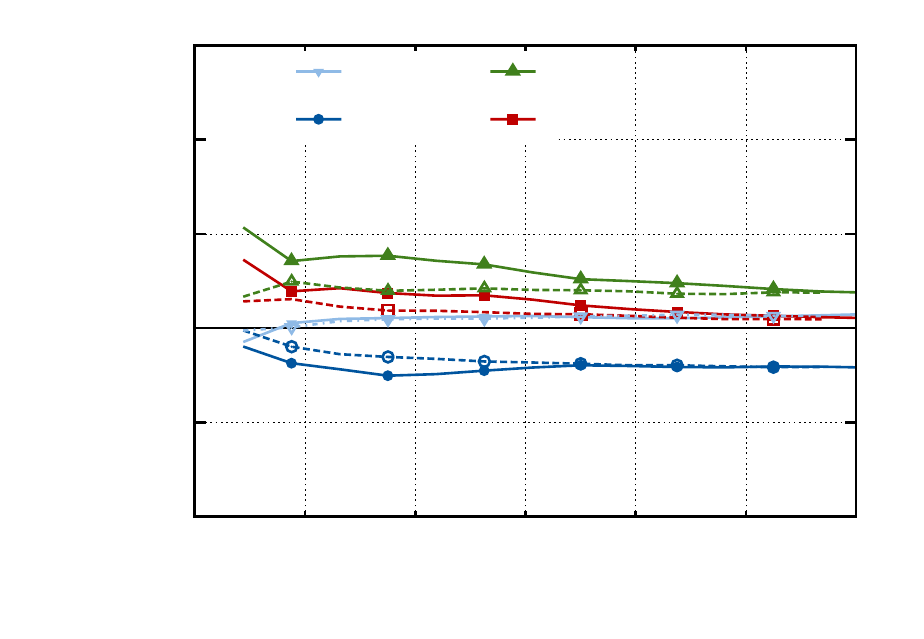}}%
    \gplfronttext
  \end{picture}%
\endgroup

%% file: data/FLAME_KERNEL_DNS_02/181002_curvEq_Ti_diss_skew_mp_curv_mp_progT/flame_kernel_curvEq_flow_terms_prodPLUS_Bending_1p2TAU_02_LOC01_JFM_HALF.tex
\begingroup
  \makeatletter
  \providecommand\color[2][]{%
    \GenericError{(gnuplot) \space\space\space\@spaces}{%
      Package color not loaded in conjunction with
      terminal option `colourtext'%
    }{See the gnuplot documentation for explanation.%
    }{Either use 'blacktext' in gnuplot or load the package
      color.sty in LaTeX.}%
    \renewcommand\color[2][]{}%
  }%
  \providecommand\includegraphics[2][]{%
    \GenericError{(gnuplot) \space\space\space\@spaces}{%
      Package graphicx or graphics not loaded%
    }{See the gnuplot documentation for explanation.%
    }{The gnuplot epslatex terminal needs graphicx.sty or graphics.sty.}%
    \renewcommand\includegraphics[2][]{}%
  }%
  \providecommand\rotatebox[2]{#2}%
  \@ifundefined{ifGPcolor}{%
    \newif\ifGPcolor
    \GPcolortrue
  }{}%
  \@ifundefined{ifGPblacktext}{%
    \newif\ifGPblacktext
    \GPblacktexttrue
  }{}%
  \let\gplgaddtomacro\g@addto@macro
  \gdef\gplbacktext{}%
  \gdef\gplfronttext{}%
  \makeatother
  \ifGPblacktext
    \def\colorrgb#1{}%
    \def\colorgray#1{}%
  \else
    \ifGPcolor
      \def\colorrgb#1{\color[rgb]{#1}}%
      \def\colorgray#1{\color[gray]{#1}}%
      \expandafter\def\csname LTw\endcsname{\color{white}}%
      \expandafter\def\csname LTb\endcsname{\color{black}}%
      \expandafter\def\csname LTa\endcsname{\color{black}}%
      \expandafter\def\csname LT0\endcsname{\color[rgb]{1,0,0}}%
      \expandafter\def\csname LT1\endcsname{\color[rgb]{0,1,0}}%
      \expandafter\def\csname LT2\endcsname{\color[rgb]{0,0,1}}%
      \expandafter\def\csname LT3\endcsname{\color[rgb]{1,0,1}}%
      \expandafter\def\csname LT4\endcsname{\color[rgb]{0,1,1}}%
      \expandafter\def\csname LT5\endcsname{\color[rgb]{1,1,0}}%
      \expandafter\def\csname LT6\endcsname{\color[rgb]{0,0,0}}%
      \expandafter\def\csname LT7\endcsname{\color[rgb]{1,0.3,0}}%
      \expandafter\def\csname LT8\endcsname{\color[rgb]{0.5,0.5,0.5}}%
    \else
      \def\colorrgb#1{\color{black}}%
      \def\colorgray#1{\color[gray]{#1}}%
      \expandafter\def\csname LTw\endcsname{\color{white}}%
      \expandafter\def\csname LTb\endcsname{\color{black}}%
      \expandafter\def\csname LTa\endcsname{\color{black}}%
      \expandafter\def\csname LT0\endcsname{\color{black}}%
      \expandafter\def\csname LT1\endcsname{\color{black}}%
      \expandafter\def\csname LT2\endcsname{\color{black}}%
      \expandafter\def\csname LT3\endcsname{\color{black}}%
      \expandafter\def\csname LT4\endcsname{\color{black}}%
      \expandafter\def\csname LT5\endcsname{\color{black}}%
      \expandafter\def\csname LT6\endcsname{\color{black}}%
      \expandafter\def\csname LT7\endcsname{\color{black}}%
      \expandafter\def\csname LT8\endcsname{\color{black}}%
    \fi
  \fi
  \setlength{\unitlength}{0.0500bp}%
  \begin{picture}(5328.00,3684.00)%
    \gplgaddtomacro\gplbacktext{%
      \csname LTb\endcsname%
      \put(990,704){\makebox(0,0)[r]{\strut{}\normalsize -4.0}}%
      \csname LTb\endcsname%
      \put(990,1247){\makebox(0,0)[r]{\strut{}\normalsize -2.0}}%
      \csname LTb\endcsname%
      \put(990,1790){\makebox(0,0)[r]{\strut{}\normalsize 0.0}}%
      \csname LTb\endcsname%
      \put(990,2333){\makebox(0,0)[r]{\strut{}\normalsize 2.0}}%
      \csname LTb\endcsname%
      \put(990,2876){\makebox(0,0)[r]{\strut{}\normalsize 4.0}}%
      \csname LTb\endcsname%
      \put(990,3419){\makebox(0,0)[r]{\strut{}\normalsize 6.0}}%
      \csname LTb\endcsname%
      \put(1122,484){\makebox(0,0){\strut{}\normalsize 0.0}}%
      \csname LTb\endcsname%
      \put(1757,484){\makebox(0,0){\strut{}\normalsize 0.2}}%
      \csname LTb\endcsname%
      \put(2392,484){\makebox(0,0){\strut{}\normalsize 0.4}}%
      \csname LTb\endcsname%
      \put(3027,484){\makebox(0,0){\strut{}\normalsize 0.6}}%
      \csname LTb\endcsname%
      \put(3661,484){\makebox(0,0){\strut{}\normalsize 0.8}}%
      \csname LTb\endcsname%
      \put(4296,484){\makebox(0,0){\strut{}\normalsize 1.0}}%
      \csname LTb\endcsname%
      \put(4931,484){\makebox(0,0){\strut{}\normalsize 1.2}}%
      \put(352,2061){\rotatebox{-270}{\makebox(0,0){\strut{}\large Production $\left( \kappa^{+} \right)$ [-]}}}%
      \put(3026,154){\makebox(0,0){\strut{}\large Normalized Time $\frac{t}{\tau_{\mathrm{t}}}$ [-]}}%
    }%
    \gplgaddtomacro\gplfronttext{%
      \csname LTb\endcsname%
      \put(3495,3268){\makebox(0,0)[r]{\strut{}\normalsize $\mathfrak{T}_{\mathrm{n}}^{u}$}}%
      \csname LTb\endcsname%
      \put(3495,2993){\makebox(0,0)[r]{\strut{}\normalsize $\mathfrak{T}_{\mathrm{t}}^{u}$}}%
      \csname LTb\endcsname%
      \put(4482,3268){\makebox(0,0)[r]{\strut{}\normalsize $\mathfrak{T}_{\mathrm{b}}^{u}$}}%
      \csname LTb\endcsname%
      \put(4482,2993){\makebox(0,0)[r]{\strut{}\normalsize $\sum \mathfrak{T}_{\ast}^{u}$}}%
    }%
    \gplbacktext
    \put(0,0){\includegraphics{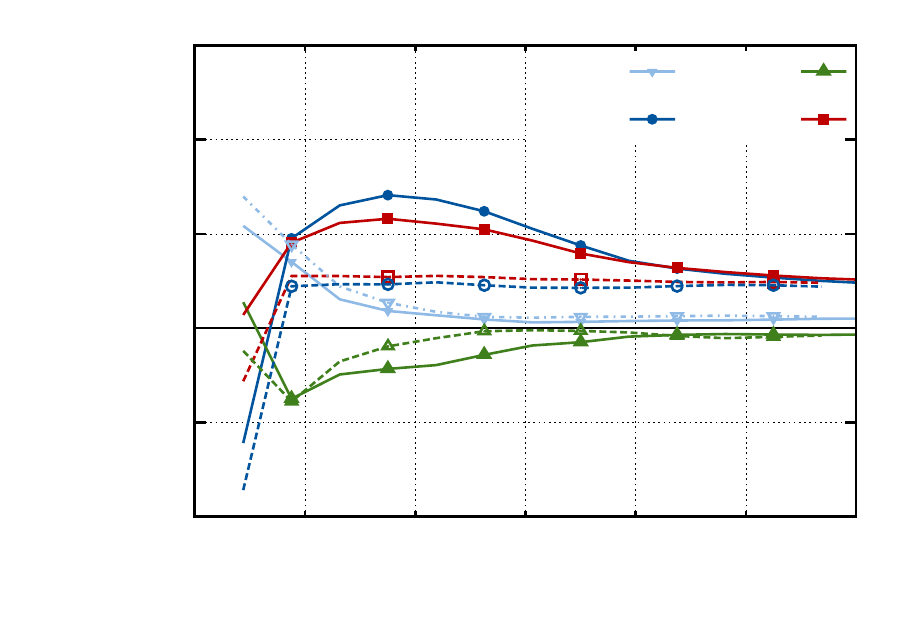}}%
    \gplfronttext
  \end{picture}%
\endgroup

%% file: data/FLAME_KERNEL_DNS_02/181002_curvEq_Ti_a_diss_skew_mp_curv_mp_progT/flame_kernel_curvEq_flame_terms_prodMINUS_Bending_1p2TAU_02_LOC01_JFM_HALF.tex
\begingroup
  \makeatletter
  \providecommand\color[2][]{%
    \GenericError{(gnuplot) \space\space\space\@spaces}{%
      Package color not loaded in conjunction with
      terminal option `colourtext'%
    }{See the gnuplot documentation for explanation.%
    }{Either use 'blacktext' in gnuplot or load the package
      color.sty in LaTeX.}%
    \renewcommand\color[2][]{}%
  }%
  \providecommand\includegraphics[2][]{%
    \GenericError{(gnuplot) \space\space\space\@spaces}{%
      Package graphicx or graphics not loaded%
    }{See the gnuplot documentation for explanation.%
    }{The gnuplot epslatex terminal needs graphicx.sty or graphics.sty.}%
    \renewcommand\includegraphics[2][]{}%
  }%
  \providecommand\rotatebox[2]{#2}%
  \@ifundefined{ifGPcolor}{%
    \newif\ifGPcolor
    \GPcolortrue
  }{}%
  \@ifundefined{ifGPblacktext}{%
    \newif\ifGPblacktext
    \GPblacktexttrue
  }{}%
  \let\gplgaddtomacro\g@addto@macro
  \gdef\gplbacktext{}%
  \gdef\gplfronttext{}%
  \makeatother
  \ifGPblacktext
    \def\colorrgb#1{}%
    \def\colorgray#1{}%
  \else
    \ifGPcolor
      \def\colorrgb#1{\color[rgb]{#1}}%
      \def\colorgray#1{\color[gray]{#1}}%
      \expandafter\def\csname LTw\endcsname{\color{white}}%
      \expandafter\def\csname LTb\endcsname{\color{black}}%
      \expandafter\def\csname LTa\endcsname{\color{black}}%
      \expandafter\def\csname LT0\endcsname{\color[rgb]{1,0,0}}%
      \expandafter\def\csname LT1\endcsname{\color[rgb]{0,1,0}}%
      \expandafter\def\csname LT2\endcsname{\color[rgb]{0,0,1}}%
      \expandafter\def\csname LT3\endcsname{\color[rgb]{1,0,1}}%
      \expandafter\def\csname LT4\endcsname{\color[rgb]{0,1,1}}%
      \expandafter\def\csname LT5\endcsname{\color[rgb]{1,1,0}}%
      \expandafter\def\csname LT6\endcsname{\color[rgb]{0,0,0}}%
      \expandafter\def\csname LT7\endcsname{\color[rgb]{1,0.3,0}}%
      \expandafter\def\csname LT8\endcsname{\color[rgb]{0.5,0.5,0.5}}%
    \else
      \def\colorrgb#1{\color{black}}%
      \def\colorgray#1{\color[gray]{#1}}%
      \expandafter\def\csname LTw\endcsname{\color{white}}%
      \expandafter\def\csname LTb\endcsname{\color{black}}%
      \expandafter\def\csname LTa\endcsname{\color{black}}%
      \expandafter\def\csname LT0\endcsname{\color{black}}%
      \expandafter\def\csname LT1\endcsname{\color{black}}%
      \expandafter\def\csname LT2\endcsname{\color{black}}%
      \expandafter\def\csname LT3\endcsname{\color{black}}%
      \expandafter\def\csname LT4\endcsname{\color{black}}%
      \expandafter\def\csname LT5\endcsname{\color{black}}%
      \expandafter\def\csname LT6\endcsname{\color{black}}%
      \expandafter\def\csname LT7\endcsname{\color{black}}%
      \expandafter\def\csname LT8\endcsname{\color{black}}%
    \fi
  \fi
  \setlength{\unitlength}{0.0500bp}%
  \begin{picture}(5328.00,3684.00)%
    \gplgaddtomacro\gplbacktext{%
      \csname LTb\endcsname%
      \put(990,704){\makebox(0,0)[r]{\strut{}\normalsize -4.0}}%
      \csname LTb\endcsname%
      \put(990,1247){\makebox(0,0)[r]{\strut{}\normalsize -2.0}}%
      \csname LTb\endcsname%
      \put(990,1790){\makebox(0,0)[r]{\strut{}\normalsize 0.0}}%
      \csname LTb\endcsname%
      \put(990,2333){\makebox(0,0)[r]{\strut{}\normalsize 2.0}}%
      \csname LTb\endcsname%
      \put(990,2876){\makebox(0,0)[r]{\strut{}\normalsize 4.0}}%
      \csname LTb\endcsname%
      \put(990,3419){\makebox(0,0)[r]{\strut{}\normalsize 6.0}}%
      \csname LTb\endcsname%
      \put(1122,484){\makebox(0,0){\strut{}\normalsize 0.0}}%
      \csname LTb\endcsname%
      \put(1757,484){\makebox(0,0){\strut{}\normalsize 0.2}}%
      \csname LTb\endcsname%
      \put(2392,484){\makebox(0,0){\strut{}\normalsize 0.4}}%
      \csname LTb\endcsname%
      \put(3027,484){\makebox(0,0){\strut{}\normalsize 0.6}}%
      \csname LTb\endcsname%
      \put(3661,484){\makebox(0,0){\strut{}\normalsize 0.8}}%
      \csname LTb\endcsname%
      \put(4296,484){\makebox(0,0){\strut{}\normalsize 1.0}}%
      \csname LTb\endcsname%
      \put(4931,484){\makebox(0,0){\strut{}\normalsize 1.2}}%
      \put(352,2061){\rotatebox{-270}{\makebox(0,0){\strut{}\large Production $\left( \kappa^{-} \right)$ [-]}}}%
      \put(3026,154){\makebox(0,0){\strut{}\large Normalized Time $\frac{t}{\tau_{\mathrm{t}}}$ [-]}}%
    }%
    \gplgaddtomacro\gplfronttext{%
      \csname LTb\endcsname%
      \put(1573,3268){\makebox(0,0)[r]{\strut{}\normalsize $-\mathfrak{T}_{\mathrm{n}}^{\mathrm{p}}$}}%
      \csname LTb\endcsname%
      \put(1573,2993){\makebox(0,0)[r]{\strut{}\normalsize $-\mathfrak{T}_{\mathrm{t}}^{\mathrm{p}}$}}%
      \csname LTb\endcsname%
      \put(2692,3268){\makebox(0,0)[r]{\strut{}\normalsize $-\mathfrak{T}_{\mathrm{b}}^{\mathrm{p}}$}}%
      \csname LTb\endcsname%
      \put(2692,2993){\makebox(0,0)[r]{\strut{}\normalsize $-\sum \mathfrak{T}_{\ast}^{\mathrm{p}}$}}%
    }%
    \gplbacktext
    \put(0,0){\includegraphics{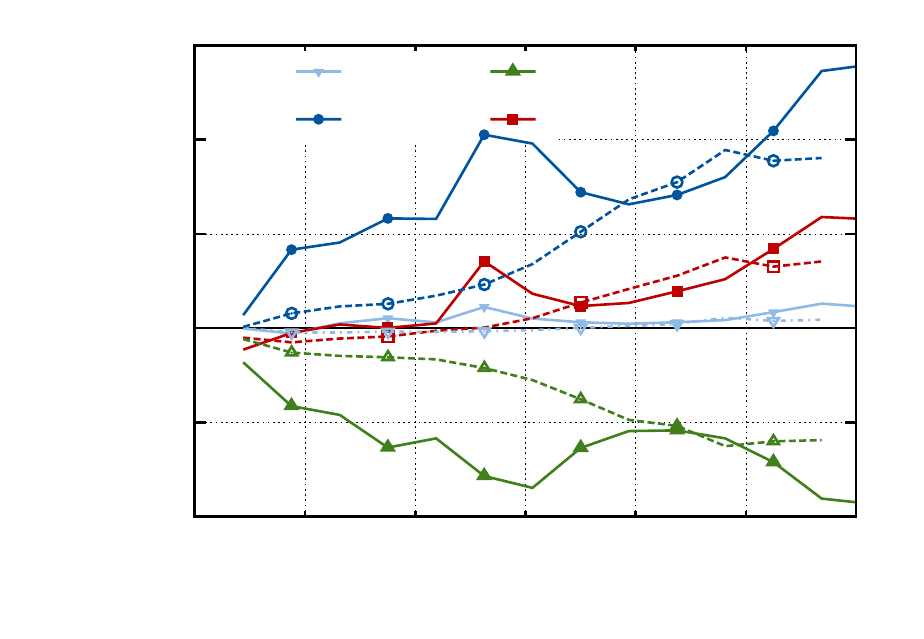}}%
    \gplfronttext
  \end{picture}%
\endgroup

%% file: data/FLAME_KERNEL_DNS_02/181002_curvEq_Ti_a_diss_skew_mp_curv_mp_progT/flame_kernel_curvEq_flame_terms_prodPLUS_Bending_1p2TAU_02_LOC01_JFM_HALF.tex
\begingroup
  \makeatletter
  \providecommand\color[2][]{%
    \GenericError{(gnuplot) \space\space\space\@spaces}{%
      Package color not loaded in conjunction with
      terminal option `colourtext'%
    }{See the gnuplot documentation for explanation.%
    }{Either use 'blacktext' in gnuplot or load the package
      color.sty in LaTeX.}%
    \renewcommand\color[2][]{}%
  }%
  \providecommand\includegraphics[2][]{%
    \GenericError{(gnuplot) \space\space\space\@spaces}{%
      Package graphicx or graphics not loaded%
    }{See the gnuplot documentation for explanation.%
    }{The gnuplot epslatex terminal needs graphicx.sty or graphics.sty.}%
    \renewcommand\includegraphics[2][]{}%
  }%
  \providecommand\rotatebox[2]{#2}%
  \@ifundefined{ifGPcolor}{%
    \newif\ifGPcolor
    \GPcolortrue
  }{}%
  \@ifundefined{ifGPblacktext}{%
    \newif\ifGPblacktext
    \GPblacktexttrue
  }{}%
  \let\gplgaddtomacro\g@addto@macro
  \gdef\gplbacktext{}%
  \gdef\gplfronttext{}%
  \makeatother
  \ifGPblacktext
    \def\colorrgb#1{}%
    \def\colorgray#1{}%
  \else
    \ifGPcolor
      \def\colorrgb#1{\color[rgb]{#1}}%
      \def\colorgray#1{\color[gray]{#1}}%
      \expandafter\def\csname LTw\endcsname{\color{white}}%
      \expandafter\def\csname LTb\endcsname{\color{black}}%
      \expandafter\def\csname LTa\endcsname{\color{black}}%
      \expandafter\def\csname LT0\endcsname{\color[rgb]{1,0,0}}%
      \expandafter\def\csname LT1\endcsname{\color[rgb]{0,1,0}}%
      \expandafter\def\csname LT2\endcsname{\color[rgb]{0,0,1}}%
      \expandafter\def\csname LT3\endcsname{\color[rgb]{1,0,1}}%
      \expandafter\def\csname LT4\endcsname{\color[rgb]{0,1,1}}%
      \expandafter\def\csname LT5\endcsname{\color[rgb]{1,1,0}}%
      \expandafter\def\csname LT6\endcsname{\color[rgb]{0,0,0}}%
      \expandafter\def\csname LT7\endcsname{\color[rgb]{1,0.3,0}}%
      \expandafter\def\csname LT8\endcsname{\color[rgb]{0.5,0.5,0.5}}%
    \else
      \def\colorrgb#1{\color{black}}%
      \def\colorgray#1{\color[gray]{#1}}%
      \expandafter\def\csname LTw\endcsname{\color{white}}%
      \expandafter\def\csname LTb\endcsname{\color{black}}%
      \expandafter\def\csname LTa\endcsname{\color{black}}%
      \expandafter\def\csname LT0\endcsname{\color{black}}%
      \expandafter\def\csname LT1\endcsname{\color{black}}%
      \expandafter\def\csname LT2\endcsname{\color{black}}%
      \expandafter\def\csname LT3\endcsname{\color{black}}%
      \expandafter\def\csname LT4\endcsname{\color{black}}%
      \expandafter\def\csname LT5\endcsname{\color{black}}%
      \expandafter\def\csname LT6\endcsname{\color{black}}%
      \expandafter\def\csname LT7\endcsname{\color{black}}%
      \expandafter\def\csname LT8\endcsname{\color{black}}%
    \fi
  \fi
  \setlength{\unitlength}{0.0500bp}%
  \begin{picture}(5328.00,3684.00)%
    \gplgaddtomacro\gplbacktext{%
      \csname LTb\endcsname%
      \put(990,704){\makebox(0,0)[r]{\strut{}\normalsize -4.0}}%
      \csname LTb\endcsname%
      \put(990,1247){\makebox(0,0)[r]{\strut{}\normalsize -2.0}}%
      \csname LTb\endcsname%
      \put(990,1790){\makebox(0,0)[r]{\strut{}\normalsize 0.0}}%
      \csname LTb\endcsname%
      \put(990,2333){\makebox(0,0)[r]{\strut{}\normalsize 2.0}}%
      \csname LTb\endcsname%
      \put(990,2876){\makebox(0,0)[r]{\strut{}\normalsize 4.0}}%
      \csname LTb\endcsname%
      \put(990,3419){\makebox(0,0)[r]{\strut{}\normalsize 6.0}}%
      \csname LTb\endcsname%
      \put(1122,484){\makebox(0,0){\strut{}\normalsize 0.0}}%
      \csname LTb\endcsname%
      \put(1757,484){\makebox(0,0){\strut{}\normalsize 0.2}}%
      \csname LTb\endcsname%
      \put(2392,484){\makebox(0,0){\strut{}\normalsize 0.4}}%
      \csname LTb\endcsname%
      \put(3027,484){\makebox(0,0){\strut{}\normalsize 0.6}}%
      \csname LTb\endcsname%
      \put(3661,484){\makebox(0,0){\strut{}\normalsize 0.8}}%
      \csname LTb\endcsname%
      \put(4296,484){\makebox(0,0){\strut{}\normalsize 1.0}}%
      \csname LTb\endcsname%
      \put(4931,484){\makebox(0,0){\strut{}\normalsize 1.2}}%
      \put(352,2061){\rotatebox{-270}{\makebox(0,0){\strut{}\large Production $\left( \kappa^{+} \right)$ [-]}}}%
      \put(3026,154){\makebox(0,0){\strut{}\large Normalized Time $\frac{t}{\tau_{\mathrm{t}}}$ [-]}}%
    }%
    \gplgaddtomacro\gplfronttext{%
      \csname LTb\endcsname%
      \put(3561,3268){\makebox(0,0)[r]{\strut{}\normalsize $\mathfrak{T}_{\mathrm{n}}^{\mathrm{p}}$}}%
      \csname LTb\endcsname%
      \put(3561,2993){\makebox(0,0)[r]{\strut{}\normalsize $\mathfrak{T}_{\mathrm{t}}^{\mathrm{p}}$}}%
      \csname LTb\endcsname%
      \put(4482,3268){\makebox(0,0)[r]{\strut{}\normalsize $\mathfrak{T}_{\mathrm{b}}^{\mathrm{p}}$}}%
      \csname LTb\endcsname%
      \put(4482,2993){\makebox(0,0)[r]{\strut{}\normalsize $\sum \mathfrak{T}_{\ast}^{\mathrm{p}}$}}%
    }%
    \gplbacktext
    \put(0,0){\includegraphics{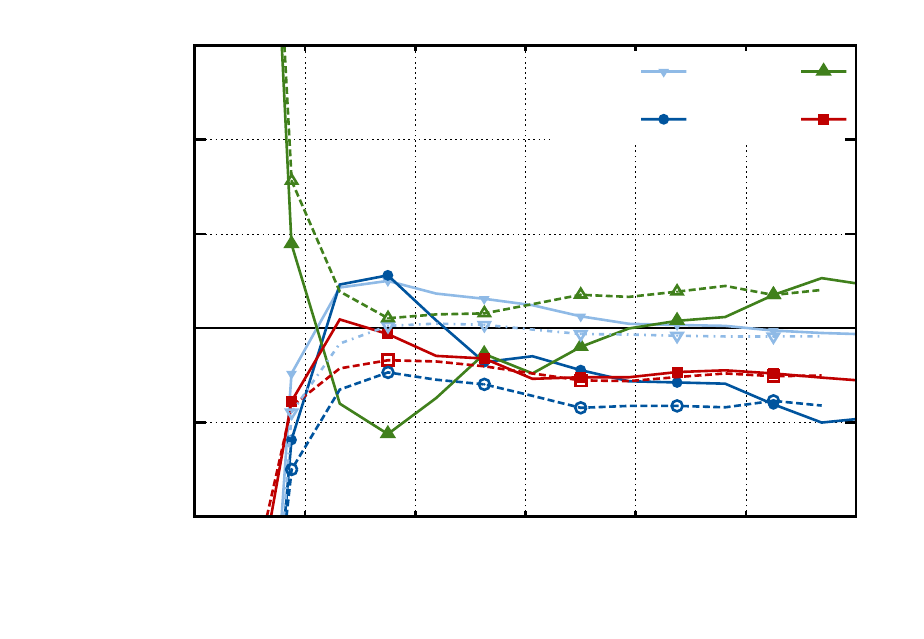}}%
    \gplfronttext
  \end{picture}%
\endgroup

%% file: data/FLAME_KERNEL_DNS_02/181003_curv_strain_kin_restor_diss_divg_skew/flame_kernel_curvSkew_02_LOC01_03_P02_JFM_HALF.tex
\begingroup
  \makeatletter
  \providecommand\color[2][]{%
    \GenericError{(gnuplot) \space\space\space\@spaces}{%
      Package color not loaded in conjunction with
      terminal option `colourtext'%
    }{See the gnuplot documentation for explanation.%
    }{Either use 'blacktext' in gnuplot or load the package
      color.sty in LaTeX.}%
    \renewcommand\color[2][]{}%
  }%
  \providecommand\includegraphics[2][]{%
    \GenericError{(gnuplot) \space\space\space\@spaces}{%
      Package graphicx or graphics not loaded%
    }{See the gnuplot documentation for explanation.%
    }{The gnuplot epslatex terminal needs graphicx.sty or graphics.sty.}%
    \renewcommand\includegraphics[2][]{}%
  }%
  \providecommand\rotatebox[2]{#2}%
  \@ifundefined{ifGPcolor}{%
    \newif\ifGPcolor
    \GPcolortrue
  }{}%
  \@ifundefined{ifGPblacktext}{%
    \newif\ifGPblacktext
    \GPblacktexttrue
  }{}%
  \let\gplgaddtomacro\g@addto@macro
  \gdef\gplbacktext{}%
  \gdef\gplfronttext{}%
  \makeatother
  \ifGPblacktext
    \def\colorrgb#1{}%
    \def\colorgray#1{}%
  \else
    \ifGPcolor
      \def\colorrgb#1{\color[rgb]{#1}}%
      \def\colorgray#1{\color[gray]{#1}}%
      \expandafter\def\csname LTw\endcsname{\color{white}}%
      \expandafter\def\csname LTb\endcsname{\color{black}}%
      \expandafter\def\csname LTa\endcsname{\color{black}}%
      \expandafter\def\csname LT0\endcsname{\color[rgb]{1,0,0}}%
      \expandafter\def\csname LT1\endcsname{\color[rgb]{0,1,0}}%
      \expandafter\def\csname LT2\endcsname{\color[rgb]{0,0,1}}%
      \expandafter\def\csname LT3\endcsname{\color[rgb]{1,0,1}}%
      \expandafter\def\csname LT4\endcsname{\color[rgb]{0,1,1}}%
      \expandafter\def\csname LT5\endcsname{\color[rgb]{1,1,0}}%
      \expandafter\def\csname LT6\endcsname{\color[rgb]{0,0,0}}%
      \expandafter\def\csname LT7\endcsname{\color[rgb]{1,0.3,0}}%
      \expandafter\def\csname LT8\endcsname{\color[rgb]{0.5,0.5,0.5}}%
    \else
      \def\colorrgb#1{\color{black}}%
      \def\colorgray#1{\color[gray]{#1}}%
      \expandafter\def\csname LTw\endcsname{\color{white}}%
      \expandafter\def\csname LTb\endcsname{\color{black}}%
      \expandafter\def\csname LTa\endcsname{\color{black}}%
      \expandafter\def\csname LT0\endcsname{\color{black}}%
      \expandafter\def\csname LT1\endcsname{\color{black}}%
      \expandafter\def\csname LT2\endcsname{\color{black}}%
      \expandafter\def\csname LT3\endcsname{\color{black}}%
      \expandafter\def\csname LT4\endcsname{\color{black}}%
      \expandafter\def\csname LT5\endcsname{\color{black}}%
      \expandafter\def\csname LT6\endcsname{\color{black}}%
      \expandafter\def\csname LT7\endcsname{\color{black}}%
      \expandafter\def\csname LT8\endcsname{\color{black}}%
    \fi
  \fi
  \setlength{\unitlength}{0.0500bp}%
  \begin{picture}(5328.00,3684.00)%
    \gplgaddtomacro\gplbacktext{%
      \csname LTb\endcsname%
      \put(990,704){\makebox(0,0)[r]{\strut{}\normalsize -4.0}}%
      \csname LTb\endcsname%
      \put(990,1067){\makebox(0,0)[r]{\strut{}\normalsize -3.0}}%
      \csname LTb\endcsname%
      \put(990,1429){\makebox(0,0)[r]{\strut{}\normalsize -2.0}}%
      \csname LTb\endcsname%
      \put(990,1792){\makebox(0,0)[r]{\strut{}\normalsize -1.0}}%
      \csname LTb\endcsname%
      \put(990,2155){\makebox(0,0)[r]{\strut{}\normalsize 0.0}}%
      \csname LTb\endcsname%
      \put(990,2518){\makebox(0,0)[r]{\strut{}\normalsize 1.0}}%
      \csname LTb\endcsname%
      \put(990,2880){\makebox(0,0)[r]{\strut{}\normalsize 2.0}}%
      \csname LTb\endcsname%
      \put(990,3243){\makebox(0,0)[r]{\strut{}\normalsize 3.0}}%
      \csname LTb\endcsname%
      \put(1122,484){\makebox(0,0){\strut{}\normalsize 0.0}}%
      \csname LTb\endcsname%
      \put(1757,484){\makebox(0,0){\strut{}\normalsize 0.5}}%
      \csname LTb\endcsname%
      \put(2392,484){\makebox(0,0){\strut{}\normalsize 1.0}}%
      \csname LTb\endcsname%
      \put(3027,484){\makebox(0,0){\strut{}\normalsize 1.5}}%
      \csname LTb\endcsname%
      \put(3661,484){\makebox(0,0){\strut{}\normalsize 2.0}}%
      \csname LTb\endcsname%
      \put(4296,484){\makebox(0,0){\strut{}\normalsize 2.5}}%
      \csname LTb\endcsname%
      \put(4931,484){\makebox(0,0){\strut{}\normalsize 3.0}}%
      \colorrgb{0.75,0.00,0.00}%
      \put(1122,3408){\makebox(0,0){\strut{}\color[HTML]{C00000}$\frac{R_{50}}{l_{\mathrm{t}}}$}}%
      \colorrgb{0.75,0.00,0.00}%
      \put(1454,3408){\makebox(0,0){\strut{}\color[HTML]{C00000}0.5}}%
      \colorrgb{0.75,0.00,0.00}%
      \put(2030,3408){\makebox(0,0){\strut{}\color[HTML]{C00000}1.0}}%
      \colorrgb{0.75,0.00,0.00}%
      \put(2934,3408){\makebox(0,0){\strut{}\color[HTML]{C00000}2.0}}%
      \colorrgb{0.75,0.00,0.00}%
      \put(3676,3408){\makebox(0,0){\strut{}\color[HTML]{C00000}3.0}}%
      \colorrgb{0.75,0.00,0.00}%
      \put(4931,3408){\makebox(0,0){\strut{}\color[HTML]{C00000}$\frac{R_{50}}{l_{\mathrm{t}}}$}}%
      \csname LTb\endcsname%
      \put(352,1973){\rotatebox{-270}{\makebox(0,0){\strut{}\large Skewness of $\left(\kappa \cdot l_{\mathrm{f}}\right)$ [-]}}}%
      \put(3026,154){\makebox(0,0){\strut{}\large Normalized Time $\frac{t}{\tau_{\mathrm{t}}}$ [-]}}%
    }%
    \gplgaddtomacro\gplfronttext{%
      \csname LTb\endcsname%
      \put(4301,3059){\makebox(0,0)[r]{\strut{}\normalsize Engine Kernel I\,\,}}%
      \csname LTb\endcsname%
      \put(4301,2817){\makebox(0,0)[r]{\strut{}\normalsize Engine Kernel II}}%
      \csname LTb\endcsname%
      \put(4301,2575){\makebox(0,0)[r]{\strut{}\normalsize Large Kernel}}%
      \csname LTb\endcsname%
      \put(4301,2333){\makebox(0,0)[r]{\strut{}\normalsize Planar Flame}}%
    }%
    \gplbacktext
    \put(0,0){\includegraphics{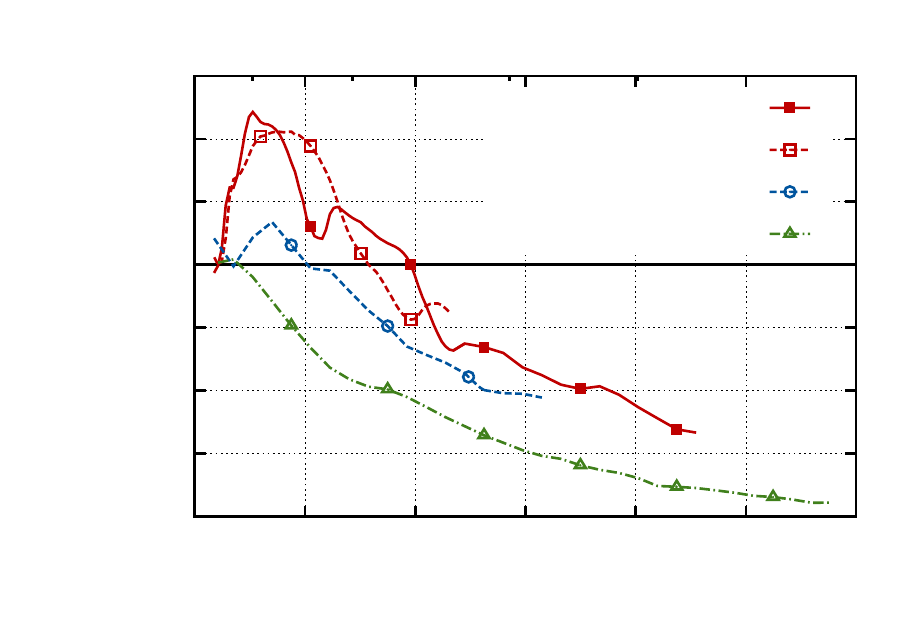}}%
    \gplfronttext
  \end{picture}%
\endgroup

%% file: data/FLAME_KERNEL_DNS_02/181021_alignment_curv_strain_progT_filt33/flame_kernel_nvec_ComprStrain_align_filt33_0p25TAU_02_JFM_HALF.tex
\begingroup
  \makeatletter
  \providecommand\color[2][]{%
    \GenericError{(gnuplot) \space\space\space\@spaces}{%
      Package color not loaded in conjunction with
      terminal option `colourtext'%
    }{See the gnuplot documentation for explanation.%
    }{Either use 'blacktext' in gnuplot or load the package
      color.sty in LaTeX.}%
    \renewcommand\color[2][]{}%
  }%
  \providecommand\includegraphics[2][]{%
    \GenericError{(gnuplot) \space\space\space\@spaces}{%
      Package graphicx or graphics not loaded%
    }{See the gnuplot documentation for explanation.%
    }{The gnuplot epslatex terminal needs graphicx.sty or graphics.sty.}%
    \renewcommand\includegraphics[2][]{}%
  }%
  \providecommand\rotatebox[2]{#2}%
  \@ifundefined{ifGPcolor}{%
    \newif\ifGPcolor
    \GPcolortrue
  }{}%
  \@ifundefined{ifGPblacktext}{%
    \newif\ifGPblacktext
    \GPblacktexttrue
  }{}%
  \let\gplgaddtomacro\g@addto@macro
  \gdef\gplbacktext{}%
  \gdef\gplfronttext{}%
  \makeatother
  \ifGPblacktext
    \def\colorrgb#1{}%
    \def\colorgray#1{}%
  \else
    \ifGPcolor
      \def\colorrgb#1{\color[rgb]{#1}}%
      \def\colorgray#1{\color[gray]{#1}}%
      \expandafter\def\csname LTw\endcsname{\color{white}}%
      \expandafter\def\csname LTb\endcsname{\color{black}}%
      \expandafter\def\csname LTa\endcsname{\color{black}}%
      \expandafter\def\csname LT0\endcsname{\color[rgb]{1,0,0}}%
      \expandafter\def\csname LT1\endcsname{\color[rgb]{0,1,0}}%
      \expandafter\def\csname LT2\endcsname{\color[rgb]{0,0,1}}%
      \expandafter\def\csname LT3\endcsname{\color[rgb]{1,0,1}}%
      \expandafter\def\csname LT4\endcsname{\color[rgb]{0,1,1}}%
      \expandafter\def\csname LT5\endcsname{\color[rgb]{1,1,0}}%
      \expandafter\def\csname LT6\endcsname{\color[rgb]{0,0,0}}%
      \expandafter\def\csname LT7\endcsname{\color[rgb]{1,0.3,0}}%
      \expandafter\def\csname LT8\endcsname{\color[rgb]{0.5,0.5,0.5}}%
    \else
      \def\colorrgb#1{\color{black}}%
      \def\colorgray#1{\color[gray]{#1}}%
      \expandafter\def\csname LTw\endcsname{\color{white}}%
      \expandafter\def\csname LTb\endcsname{\color{black}}%
      \expandafter\def\csname LTa\endcsname{\color{black}}%
      \expandafter\def\csname LT0\endcsname{\color{black}}%
      \expandafter\def\csname LT1\endcsname{\color{black}}%
      \expandafter\def\csname LT2\endcsname{\color{black}}%
      \expandafter\def\csname LT3\endcsname{\color{black}}%
      \expandafter\def\csname LT4\endcsname{\color{black}}%
      \expandafter\def\csname LT5\endcsname{\color{black}}%
      \expandafter\def\csname LT6\endcsname{\color{black}}%
      \expandafter\def\csname LT7\endcsname{\color{black}}%
      \expandafter\def\csname LT8\endcsname{\color{black}}%
    \fi
  \fi
  \setlength{\unitlength}{0.0500bp}%
  \begin{picture}(5328.00,3684.00)%
    \gplgaddtomacro\gplbacktext{%
      \csname LTb\endcsname%
      \put(990,704){\makebox(0,0)[r]{\strut{}\normalsize 0.00}}%
      \csname LTb\endcsname%
      \put(990,1109){\makebox(0,0)[r]{\strut{}\normalsize 0.75}}%
      \csname LTb\endcsname%
      \put(990,1514){\makebox(0,0)[r]{\strut{}\normalsize 1.50}}%
      \csname LTb\endcsname%
      \put(990,1919){\makebox(0,0)[r]{\strut{}\normalsize 2.25}}%
      \csname LTb\endcsname%
      \put(990,2323){\makebox(0,0)[r]{\strut{}\normalsize 3.00}}%
      \csname LTb\endcsname%
      \put(990,2728){\makebox(0,0)[r]{\strut{}\normalsize 3.75}}%
      \csname LTb\endcsname%
      \put(990,3133){\makebox(0,0)[r]{\strut{}\normalsize 4.50}}%
      \csname LTb\endcsname%
      \put(1122,484){\makebox(0,0){\strut{}\normalsize 0.0}}%
      \csname LTb\endcsname%
      \put(1884,484){\makebox(0,0){\strut{}\normalsize 0.2}}%
      \csname LTb\endcsname%
      \put(2646,484){\makebox(0,0){\strut{}\normalsize 0.4}}%
      \csname LTb\endcsname%
      \put(3407,484){\makebox(0,0){\strut{}\normalsize 0.6}}%
      \csname LTb\endcsname%
      \put(4169,484){\makebox(0,0){\strut{}\normalsize 0.8}}%
      \csname LTb\endcsname%
      \put(4931,484){\makebox(0,0){\strut{}\normalsize 1.0}}%
      \put(352,1918){\rotatebox{-270}{\makebox(0,0){\strut{}\large PDF}}}%
      \put(3026,154){\makebox(0,0){\strut{}\large $cos\left(\varphi \right)$}}%
      \put(3026,3353){\makebox(0,0){\strut{}\normalsize Engine Kernel I, $t=0.25\cdot \tau_{\mathrm{t}}$, $R_{50}=0.5 \cdot l_{\mathrm{t}}$}}%
    }%
    \gplgaddtomacro\gplfronttext{%
      \csname LTb\endcsname%
      \put(2121,2919){\makebox(0,0)[r]{\strut{}\normalsize no Filter}}%
      \csname LTb\endcsname%
      \put(2121,2589){\makebox(0,0)[r]{\strut{}\normalsize $\Delta = 0.5\; l_{\mathrm{t}}$}}%
    }%
    \gplbacktext
    \put(0,0){\includegraphics{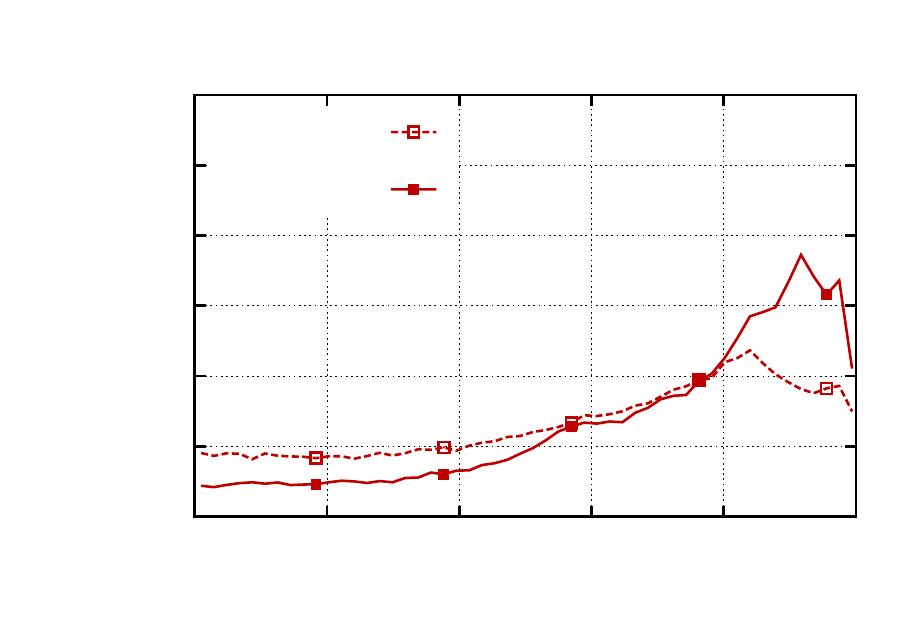}}%
    \gplfronttext
  \end{picture}%
\endgroup

%% file: data/FLAME_KERNEL_DNS_02/181021_alignment_curv_strain_progT_filt33/flame_kernel_nvec_ComprStrain_align_filt33_1p00TAU_02_JFM_HALF.tex
\begingroup
  \makeatletter
  \providecommand\color[2][]{%
    \GenericError{(gnuplot) \space\space\space\@spaces}{%
      Package color not loaded in conjunction with
      terminal option `colourtext'%
    }{See the gnuplot documentation for explanation.%
    }{Either use 'blacktext' in gnuplot or load the package
      color.sty in LaTeX.}%
    \renewcommand\color[2][]{}%
  }%
  \providecommand\includegraphics[2][]{%
    \GenericError{(gnuplot) \space\space\space\@spaces}{%
      Package graphicx or graphics not loaded%
    }{See the gnuplot documentation for explanation.%
    }{The gnuplot epslatex terminal needs graphicx.sty or graphics.sty.}%
    \renewcommand\includegraphics[2][]{}%
  }%
  \providecommand\rotatebox[2]{#2}%
  \@ifundefined{ifGPcolor}{%
    \newif\ifGPcolor
    \GPcolortrue
  }{}%
  \@ifundefined{ifGPblacktext}{%
    \newif\ifGPblacktext
    \GPblacktexttrue
  }{}%
  \let\gplgaddtomacro\g@addto@macro
  \gdef\gplbacktext{}%
  \gdef\gplfronttext{}%
  \makeatother
  \ifGPblacktext
    \def\colorrgb#1{}%
    \def\colorgray#1{}%
  \else
    \ifGPcolor
      \def\colorrgb#1{\color[rgb]{#1}}%
      \def\colorgray#1{\color[gray]{#1}}%
      \expandafter\def\csname LTw\endcsname{\color{white}}%
      \expandafter\def\csname LTb\endcsname{\color{black}}%
      \expandafter\def\csname LTa\endcsname{\color{black}}%
      \expandafter\def\csname LT0\endcsname{\color[rgb]{1,0,0}}%
      \expandafter\def\csname LT1\endcsname{\color[rgb]{0,1,0}}%
      \expandafter\def\csname LT2\endcsname{\color[rgb]{0,0,1}}%
      \expandafter\def\csname LT3\endcsname{\color[rgb]{1,0,1}}%
      \expandafter\def\csname LT4\endcsname{\color[rgb]{0,1,1}}%
      \expandafter\def\csname LT5\endcsname{\color[rgb]{1,1,0}}%
      \expandafter\def\csname LT6\endcsname{\color[rgb]{0,0,0}}%
      \expandafter\def\csname LT7\endcsname{\color[rgb]{1,0.3,0}}%
      \expandafter\def\csname LT8\endcsname{\color[rgb]{0.5,0.5,0.5}}%
    \else
      \def\colorrgb#1{\color{black}}%
      \def\colorgray#1{\color[gray]{#1}}%
      \expandafter\def\csname LTw\endcsname{\color{white}}%
      \expandafter\def\csname LTb\endcsname{\color{black}}%
      \expandafter\def\csname LTa\endcsname{\color{black}}%
      \expandafter\def\csname LT0\endcsname{\color{black}}%
      \expandafter\def\csname LT1\endcsname{\color{black}}%
      \expandafter\def\csname LT2\endcsname{\color{black}}%
      \expandafter\def\csname LT3\endcsname{\color{black}}%
      \expandafter\def\csname LT4\endcsname{\color{black}}%
      \expandafter\def\csname LT5\endcsname{\color{black}}%
      \expandafter\def\csname LT6\endcsname{\color{black}}%
      \expandafter\def\csname LT7\endcsname{\color{black}}%
      \expandafter\def\csname LT8\endcsname{\color{black}}%
    \fi
  \fi
  \setlength{\unitlength}{0.0500bp}%
  \begin{picture}(5328.00,3684.00)%
    \gplgaddtomacro\gplbacktext{%
      \csname LTb\endcsname%
      \put(990,704){\makebox(0,0)[r]{\strut{}\normalsize 0.00}}%
      \csname LTb\endcsname%
      \put(990,1109){\makebox(0,0)[r]{\strut{}\normalsize 0.75}}%
      \csname LTb\endcsname%
      \put(990,1514){\makebox(0,0)[r]{\strut{}\normalsize 1.50}}%
      \csname LTb\endcsname%
      \put(990,1919){\makebox(0,0)[r]{\strut{}\normalsize 2.25}}%
      \csname LTb\endcsname%
      \put(990,2323){\makebox(0,0)[r]{\strut{}\normalsize 3.00}}%
      \csname LTb\endcsname%
      \put(990,2728){\makebox(0,0)[r]{\strut{}\normalsize 3.75}}%
      \csname LTb\endcsname%
      \put(990,3133){\makebox(0,0)[r]{\strut{}\normalsize 4.50}}%
      \csname LTb\endcsname%
      \put(1122,484){\makebox(0,0){\strut{}\normalsize 0.0}}%
      \csname LTb\endcsname%
      \put(1884,484){\makebox(0,0){\strut{}\normalsize 0.2}}%
      \csname LTb\endcsname%
      \put(2646,484){\makebox(0,0){\strut{}\normalsize 0.4}}%
      \csname LTb\endcsname%
      \put(3407,484){\makebox(0,0){\strut{}\normalsize 0.6}}%
      \csname LTb\endcsname%
      \put(4169,484){\makebox(0,0){\strut{}\normalsize 0.8}}%
      \csname LTb\endcsname%
      \put(4931,484){\makebox(0,0){\strut{}\normalsize 1.0}}%
      \put(352,1918){\rotatebox{-270}{\makebox(0,0){\strut{}\large PDF}}}%
      \put(3026,154){\makebox(0,0){\strut{}\large $cos\left(\varphi \right)$}}%
      \put(3026,3353){\makebox(0,0){\strut{}\normalsize Engine Kernel I, $t=1.0\cdot \tau_{\mathrm{t}}$, $R_{50}=1.4 \cdot l_{\mathrm{t}}$}}%
    }%
    \gplgaddtomacro\gplfronttext{%
      \csname LTb\endcsname%
      \put(2121,2919){\makebox(0,0)[r]{\strut{}\normalsize no Filter}}%
      \csname LTb\endcsname%
      \put(2121,2589){\makebox(0,0)[r]{\strut{}\normalsize $\Delta = 0.5\; l_{\mathrm{t}}$}}%
    }%
    \gplbacktext
    \put(0,0){\includegraphics{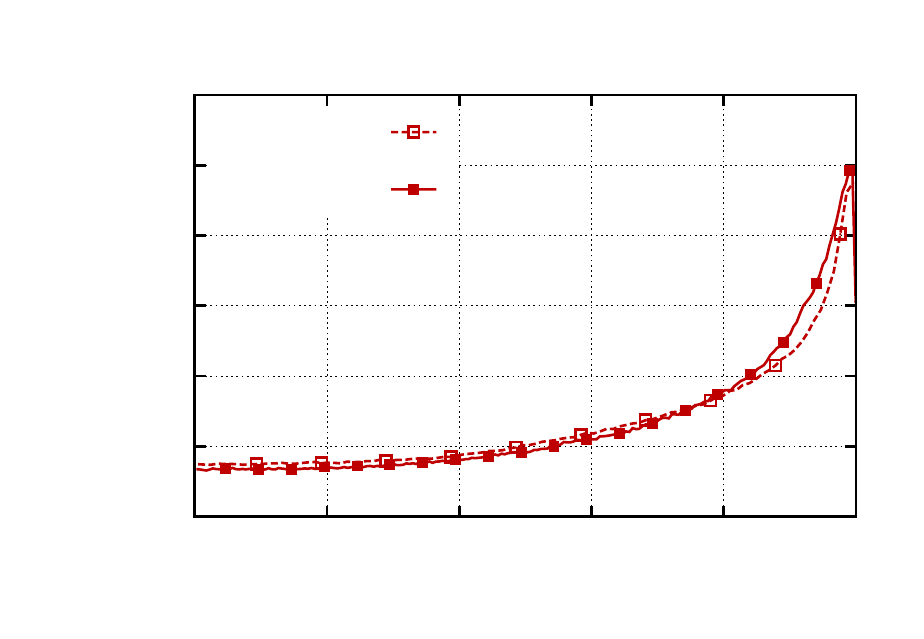}}%
    \gplfronttext
  \end{picture}%
\endgroup

%% file: data/FLAME_KERNEL_DNS_02_LOC1/181021_alignment_curv_strain_progT_filt33/flame_kernel_nvec_ComprStrain_align_filt33_0p25TAU_02_LOC1_JFM_HALF.tex
\begingroup
  \makeatletter
  \providecommand\color[2][]{%
    \GenericError{(gnuplot) \space\space\space\@spaces}{%
      Package color not loaded in conjunction with
      terminal option `colourtext'%
    }{See the gnuplot documentation for explanation.%
    }{Either use 'blacktext' in gnuplot or load the package
      color.sty in LaTeX.}%
    \renewcommand\color[2][]{}%
  }%
  \providecommand\includegraphics[2][]{%
    \GenericError{(gnuplot) \space\space\space\@spaces}{%
      Package graphicx or graphics not loaded%
    }{See the gnuplot documentation for explanation.%
    }{The gnuplot epslatex terminal needs graphicx.sty or graphics.sty.}%
    \renewcommand\includegraphics[2][]{}%
  }%
  \providecommand\rotatebox[2]{#2}%
  \@ifundefined{ifGPcolor}{%
    \newif\ifGPcolor
    \GPcolortrue
  }{}%
  \@ifundefined{ifGPblacktext}{%
    \newif\ifGPblacktext
    \GPblacktexttrue
  }{}%
  \let\gplgaddtomacro\g@addto@macro
  \gdef\gplbacktext{}%
  \gdef\gplfronttext{}%
  \makeatother
  \ifGPblacktext
    \def\colorrgb#1{}%
    \def\colorgray#1{}%
  \else
    \ifGPcolor
      \def\colorrgb#1{\color[rgb]{#1}}%
      \def\colorgray#1{\color[gray]{#1}}%
      \expandafter\def\csname LTw\endcsname{\color{white}}%
      \expandafter\def\csname LTb\endcsname{\color{black}}%
      \expandafter\def\csname LTa\endcsname{\color{black}}%
      \expandafter\def\csname LT0\endcsname{\color[rgb]{1,0,0}}%
      \expandafter\def\csname LT1\endcsname{\color[rgb]{0,1,0}}%
      \expandafter\def\csname LT2\endcsname{\color[rgb]{0,0,1}}%
      \expandafter\def\csname LT3\endcsname{\color[rgb]{1,0,1}}%
      \expandafter\def\csname LT4\endcsname{\color[rgb]{0,1,1}}%
      \expandafter\def\csname LT5\endcsname{\color[rgb]{1,1,0}}%
      \expandafter\def\csname LT6\endcsname{\color[rgb]{0,0,0}}%
      \expandafter\def\csname LT7\endcsname{\color[rgb]{1,0.3,0}}%
      \expandafter\def\csname LT8\endcsname{\color[rgb]{0.5,0.5,0.5}}%
    \else
      \def\colorrgb#1{\color{black}}%
      \def\colorgray#1{\color[gray]{#1}}%
      \expandafter\def\csname LTw\endcsname{\color{white}}%
      \expandafter\def\csname LTb\endcsname{\color{black}}%
      \expandafter\def\csname LTa\endcsname{\color{black}}%
      \expandafter\def\csname LT0\endcsname{\color{black}}%
      \expandafter\def\csname LT1\endcsname{\color{black}}%
      \expandafter\def\csname LT2\endcsname{\color{black}}%
      \expandafter\def\csname LT3\endcsname{\color{black}}%
      \expandafter\def\csname LT4\endcsname{\color{black}}%
      \expandafter\def\csname LT5\endcsname{\color{black}}%
      \expandafter\def\csname LT6\endcsname{\color{black}}%
      \expandafter\def\csname LT7\endcsname{\color{black}}%
      \expandafter\def\csname LT8\endcsname{\color{black}}%
    \fi
  \fi
  \setlength{\unitlength}{0.0500bp}%
  \begin{picture}(5328.00,3684.00)%
    \gplgaddtomacro\gplbacktext{%
      \csname LTb\endcsname%
      \put(990,704){\makebox(0,0)[r]{\strut{}\normalsize 0.00}}%
      \csname LTb\endcsname%
      \put(990,1109){\makebox(0,0)[r]{\strut{}\normalsize 0.75}}%
      \csname LTb\endcsname%
      \put(990,1514){\makebox(0,0)[r]{\strut{}\normalsize 1.50}}%
      \csname LTb\endcsname%
      \put(990,1919){\makebox(0,0)[r]{\strut{}\normalsize 2.25}}%
      \csname LTb\endcsname%
      \put(990,2323){\makebox(0,0)[r]{\strut{}\normalsize 3.00}}%
      \csname LTb\endcsname%
      \put(990,2728){\makebox(0,0)[r]{\strut{}\normalsize 3.75}}%
      \csname LTb\endcsname%
      \put(990,3133){\makebox(0,0)[r]{\strut{}\normalsize 4.50}}%
      \csname LTb\endcsname%
      \put(1122,484){\makebox(0,0){\strut{}\normalsize 0.0}}%
      \csname LTb\endcsname%
      \put(1884,484){\makebox(0,0){\strut{}\normalsize 0.2}}%
      \csname LTb\endcsname%
      \put(2646,484){\makebox(0,0){\strut{}\normalsize 0.4}}%
      \csname LTb\endcsname%
      \put(3407,484){\makebox(0,0){\strut{}\normalsize 0.6}}%
      \csname LTb\endcsname%
      \put(4169,484){\makebox(0,0){\strut{}\normalsize 0.8}}%
      \csname LTb\endcsname%
      \put(4931,484){\makebox(0,0){\strut{}\normalsize 1.0}}%
      \put(352,1918){\rotatebox{-270}{\makebox(0,0){\strut{}\large PDF}}}%
      \put(3026,154){\makebox(0,0){\strut{}\large $cos\left(\varphi \right)$}}%
      \put(3026,3353){\makebox(0,0){\strut{}\normalsize Engine Kernel II, $t=0.25\cdot \tau_{\mathrm{t}}$, $R_{50}=0.5 \cdot l_{\mathrm{t}}$}}%
    }%
    \gplgaddtomacro\gplfronttext{%
      \csname LTb\endcsname%
      \put(2121,2919){\makebox(0,0)[r]{\strut{}\normalsize no Filter}}%
      \csname LTb\endcsname%
      \put(2121,2589){\makebox(0,0)[r]{\strut{}\normalsize $\Delta = 0.5\; l_{\mathrm{t}}$}}%
    }%
    \gplbacktext
    \put(0,0){\includegraphics{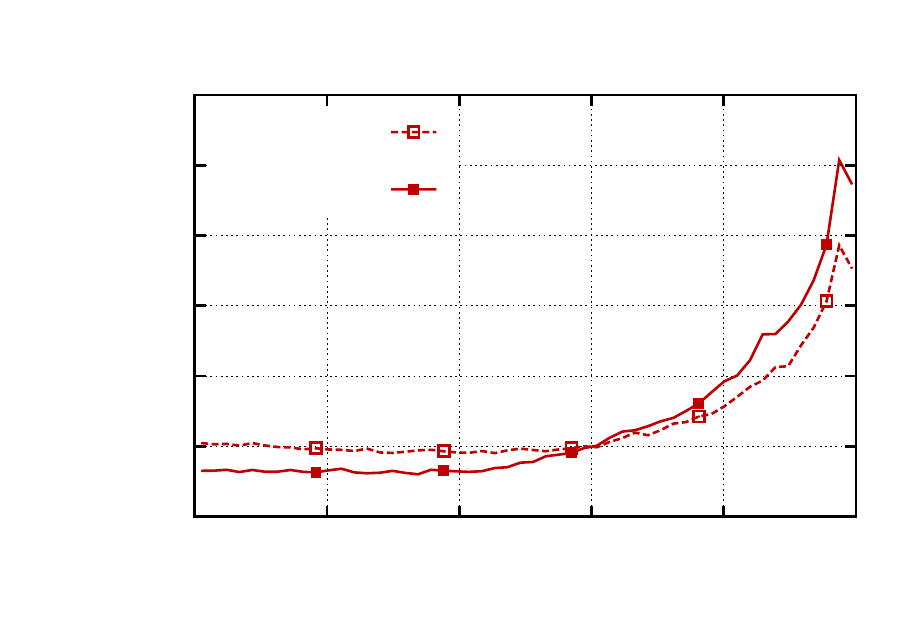}}%
    \gplfronttext
  \end{picture}%
\endgroup

%% file: data/FLAME_KERNEL_DNS_02_LOC1/181021_alignment_curv_strain_progT_filt33/flame_kernel_nvec_ComprStrain_align_filt33_1p00TAU_02_LOC1_JFM_HALF.tex
\begingroup
  \makeatletter
  \providecommand\color[2][]{%
    \GenericError{(gnuplot) \space\space\space\@spaces}{%
      Package color not loaded in conjunction with
      terminal option `colourtext'%
    }{See the gnuplot documentation for explanation.%
    }{Either use 'blacktext' in gnuplot or load the package
      color.sty in LaTeX.}%
    \renewcommand\color[2][]{}%
  }%
  \providecommand\includegraphics[2][]{%
    \GenericError{(gnuplot) \space\space\space\@spaces}{%
      Package graphicx or graphics not loaded%
    }{See the gnuplot documentation for explanation.%
    }{The gnuplot epslatex terminal needs graphicx.sty or graphics.sty.}%
    \renewcommand\includegraphics[2][]{}%
  }%
  \providecommand\rotatebox[2]{#2}%
  \@ifundefined{ifGPcolor}{%
    \newif\ifGPcolor
    \GPcolortrue
  }{}%
  \@ifundefined{ifGPblacktext}{%
    \newif\ifGPblacktext
    \GPblacktexttrue
  }{}%
  \let\gplgaddtomacro\g@addto@macro
  \gdef\gplbacktext{}%
  \gdef\gplfronttext{}%
  \makeatother
  \ifGPblacktext
    \def\colorrgb#1{}%
    \def\colorgray#1{}%
  \else
    \ifGPcolor
      \def\colorrgb#1{\color[rgb]{#1}}%
      \def\colorgray#1{\color[gray]{#1}}%
      \expandafter\def\csname LTw\endcsname{\color{white}}%
      \expandafter\def\csname LTb\endcsname{\color{black}}%
      \expandafter\def\csname LTa\endcsname{\color{black}}%
      \expandafter\def\csname LT0\endcsname{\color[rgb]{1,0,0}}%
      \expandafter\def\csname LT1\endcsname{\color[rgb]{0,1,0}}%
      \expandafter\def\csname LT2\endcsname{\color[rgb]{0,0,1}}%
      \expandafter\def\csname LT3\endcsname{\color[rgb]{1,0,1}}%
      \expandafter\def\csname LT4\endcsname{\color[rgb]{0,1,1}}%
      \expandafter\def\csname LT5\endcsname{\color[rgb]{1,1,0}}%
      \expandafter\def\csname LT6\endcsname{\color[rgb]{0,0,0}}%
      \expandafter\def\csname LT7\endcsname{\color[rgb]{1,0.3,0}}%
      \expandafter\def\csname LT8\endcsname{\color[rgb]{0.5,0.5,0.5}}%
    \else
      \def\colorrgb#1{\color{black}}%
      \def\colorgray#1{\color[gray]{#1}}%
      \expandafter\def\csname LTw\endcsname{\color{white}}%
      \expandafter\def\csname LTb\endcsname{\color{black}}%
      \expandafter\def\csname LTa\endcsname{\color{black}}%
      \expandafter\def\csname LT0\endcsname{\color{black}}%
      \expandafter\def\csname LT1\endcsname{\color{black}}%
      \expandafter\def\csname LT2\endcsname{\color{black}}%
      \expandafter\def\csname LT3\endcsname{\color{black}}%
      \expandafter\def\csname LT4\endcsname{\color{black}}%
      \expandafter\def\csname LT5\endcsname{\color{black}}%
      \expandafter\def\csname LT6\endcsname{\color{black}}%
      \expandafter\def\csname LT7\endcsname{\color{black}}%
      \expandafter\def\csname LT8\endcsname{\color{black}}%
    \fi
  \fi
  \setlength{\unitlength}{0.0500bp}%
  \begin{picture}(5328.00,3684.00)%
    \gplgaddtomacro\gplbacktext{%
      \csname LTb\endcsname%
      \put(990,704){\makebox(0,0)[r]{\strut{}\normalsize 0.00}}%
      \csname LTb\endcsname%
      \put(990,1109){\makebox(0,0)[r]{\strut{}\normalsize 0.75}}%
      \csname LTb\endcsname%
      \put(990,1514){\makebox(0,0)[r]{\strut{}\normalsize 1.50}}%
      \csname LTb\endcsname%
      \put(990,1919){\makebox(0,0)[r]{\strut{}\normalsize 2.25}}%
      \csname LTb\endcsname%
      \put(990,2323){\makebox(0,0)[r]{\strut{}\normalsize 3.00}}%
      \csname LTb\endcsname%
      \put(990,2728){\makebox(0,0)[r]{\strut{}\normalsize 3.75}}%
      \csname LTb\endcsname%
      \put(990,3133){\makebox(0,0)[r]{\strut{}\normalsize 4.50}}%
      \csname LTb\endcsname%
      \put(1122,484){\makebox(0,0){\strut{}\normalsize 0.0}}%
      \csname LTb\endcsname%
      \put(1884,484){\makebox(0,0){\strut{}\normalsize 0.2}}%
      \csname LTb\endcsname%
      \put(2646,484){\makebox(0,0){\strut{}\normalsize 0.4}}%
      \csname LTb\endcsname%
      \put(3407,484){\makebox(0,0){\strut{}\normalsize 0.6}}%
      \csname LTb\endcsname%
      \put(4169,484){\makebox(0,0){\strut{}\normalsize 0.8}}%
      \csname LTb\endcsname%
      \put(4931,484){\makebox(0,0){\strut{}\normalsize 1.0}}%
      \put(352,1918){\rotatebox{-270}{\makebox(0,0){\strut{}\large PDF}}}%
      \put(3026,154){\makebox(0,0){\strut{}\large $cos\left(\varphi \right)$}}%
      \put(3026,3353){\makebox(0,0){\strut{}\normalsize Engine Kernel II, $t=1.0\cdot \tau_{\mathrm{t}}$, $R_{50}=1.5 \cdot l_{\mathrm{t}}$}}%
    }%
    \gplgaddtomacro\gplfronttext{%
      \csname LTb\endcsname%
      \put(2121,2919){\makebox(0,0)[r]{\strut{}\normalsize no Filter}}%
      \csname LTb\endcsname%
      \put(2121,2589){\makebox(0,0)[r]{\strut{}\normalsize $\Delta = 0.5\; l_{\mathrm{t}}$}}%
    }%
    \gplbacktext
    \put(0,0){\includegraphics{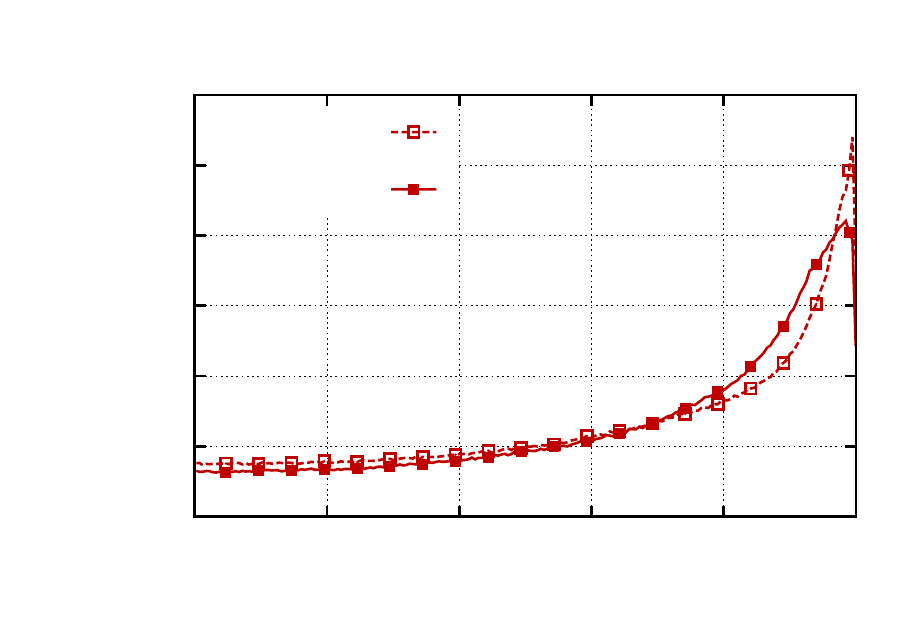}}%
    \gplfronttext
  \end{picture}%
\endgroup

%% file: data/FLAME_KERNEL_DNS_03/181021_alignment_curv_strain_progT_filt33/flame_kernel_nvec_ComprStrain_align_filt33_0p25TAU_03_JFM_HALF.tex
\begingroup
  \makeatletter
  \providecommand\color[2][]{%
    \GenericError{(gnuplot) \space\space\space\@spaces}{%
      Package color not loaded in conjunction with
      terminal option `colourtext'%
    }{See the gnuplot documentation for explanation.%
    }{Either use 'blacktext' in gnuplot or load the package
      color.sty in LaTeX.}%
    \renewcommand\color[2][]{}%
  }%
  \providecommand\includegraphics[2][]{%
    \GenericError{(gnuplot) \space\space\space\@spaces}{%
      Package graphicx or graphics not loaded%
    }{See the gnuplot documentation for explanation.%
    }{The gnuplot epslatex terminal needs graphicx.sty or graphics.sty.}%
    \renewcommand\includegraphics[2][]{}%
  }%
  \providecommand\rotatebox[2]{#2}%
  \@ifundefined{ifGPcolor}{%
    \newif\ifGPcolor
    \GPcolortrue
  }{}%
  \@ifundefined{ifGPblacktext}{%
    \newif\ifGPblacktext
    \GPblacktexttrue
  }{}%
  \let\gplgaddtomacro\g@addto@macro
  \gdef\gplbacktext{}%
  \gdef\gplfronttext{}%
  \makeatother
  \ifGPblacktext
    \def\colorrgb#1{}%
    \def\colorgray#1{}%
  \else
    \ifGPcolor
      \def\colorrgb#1{\color[rgb]{#1}}%
      \def\colorgray#1{\color[gray]{#1}}%
      \expandafter\def\csname LTw\endcsname{\color{white}}%
      \expandafter\def\csname LTb\endcsname{\color{black}}%
      \expandafter\def\csname LTa\endcsname{\color{black}}%
      \expandafter\def\csname LT0\endcsname{\color[rgb]{1,0,0}}%
      \expandafter\def\csname LT1\endcsname{\color[rgb]{0,1,0}}%
      \expandafter\def\csname LT2\endcsname{\color[rgb]{0,0,1}}%
      \expandafter\def\csname LT3\endcsname{\color[rgb]{1,0,1}}%
      \expandafter\def\csname LT4\endcsname{\color[rgb]{0,1,1}}%
      \expandafter\def\csname LT5\endcsname{\color[rgb]{1,1,0}}%
      \expandafter\def\csname LT6\endcsname{\color[rgb]{0,0,0}}%
      \expandafter\def\csname LT7\endcsname{\color[rgb]{1,0.3,0}}%
      \expandafter\def\csname LT8\endcsname{\color[rgb]{0.5,0.5,0.5}}%
    \else
      \def\colorrgb#1{\color{black}}%
      \def\colorgray#1{\color[gray]{#1}}%
      \expandafter\def\csname LTw\endcsname{\color{white}}%
      \expandafter\def\csname LTb\endcsname{\color{black}}%
      \expandafter\def\csname LTa\endcsname{\color{black}}%
      \expandafter\def\csname LT0\endcsname{\color{black}}%
      \expandafter\def\csname LT1\endcsname{\color{black}}%
      \expandafter\def\csname LT2\endcsname{\color{black}}%
      \expandafter\def\csname LT3\endcsname{\color{black}}%
      \expandafter\def\csname LT4\endcsname{\color{black}}%
      \expandafter\def\csname LT5\endcsname{\color{black}}%
      \expandafter\def\csname LT6\endcsname{\color{black}}%
      \expandafter\def\csname LT7\endcsname{\color{black}}%
      \expandafter\def\csname LT8\endcsname{\color{black}}%
    \fi
  \fi
  \setlength{\unitlength}{0.0500bp}%
  \begin{picture}(5328.00,3684.00)%
    \gplgaddtomacro\gplbacktext{%
      \csname LTb\endcsname%
      \put(990,704){\makebox(0,0)[r]{\strut{}\normalsize 0.00}}%
      \csname LTb\endcsname%
      \put(990,1109){\makebox(0,0)[r]{\strut{}\normalsize 0.75}}%
      \csname LTb\endcsname%
      \put(990,1514){\makebox(0,0)[r]{\strut{}\normalsize 1.50}}%
      \csname LTb\endcsname%
      \put(990,1919){\makebox(0,0)[r]{\strut{}\normalsize 2.25}}%
      \csname LTb\endcsname%
      \put(990,2323){\makebox(0,0)[r]{\strut{}\normalsize 3.00}}%
      \csname LTb\endcsname%
      \put(990,2728){\makebox(0,0)[r]{\strut{}\normalsize 3.75}}%
      \csname LTb\endcsname%
      \put(990,3133){\makebox(0,0)[r]{\strut{}\normalsize 4.50}}%
      \csname LTb\endcsname%
      \put(1122,484){\makebox(0,0){\strut{}\normalsize 0.0}}%
      \csname LTb\endcsname%
      \put(1884,484){\makebox(0,0){\strut{}\normalsize 0.2}}%
      \csname LTb\endcsname%
      \put(2646,484){\makebox(0,0){\strut{}\normalsize 0.4}}%
      \csname LTb\endcsname%
      \put(3407,484){\makebox(0,0){\strut{}\normalsize 0.6}}%
      \csname LTb\endcsname%
      \put(4169,484){\makebox(0,0){\strut{}\normalsize 0.8}}%
      \csname LTb\endcsname%
      \put(4931,484){\makebox(0,0){\strut{}\normalsize 1.0}}%
      \put(352,1918){\rotatebox{-270}{\makebox(0,0){\strut{}\large PDF}}}%
      \put(3026,154){\makebox(0,0){\strut{}\large $cos\left(\varphi \right)$}}%
      \put(3026,3353){\makebox(0,0){\strut{}\normalsize Large Kernel, $t=0.25\cdot \tau_{\mathrm{t}}$, $R_{50}=1.1 \cdot l_{\mathrm{t}}$}}%
    }%
    \gplgaddtomacro\gplfronttext{%
      \csname LTb\endcsname%
      \put(2121,2919){\makebox(0,0)[r]{\strut{}\normalsize no Filter}}%
      \csname LTb\endcsname%
      \put(2121,2589){\makebox(0,0)[r]{\strut{}\normalsize $\Delta = 0.5\; l_{\mathrm{t}}$}}%
    }%
    \gplbacktext
    \put(0,0){\includegraphics{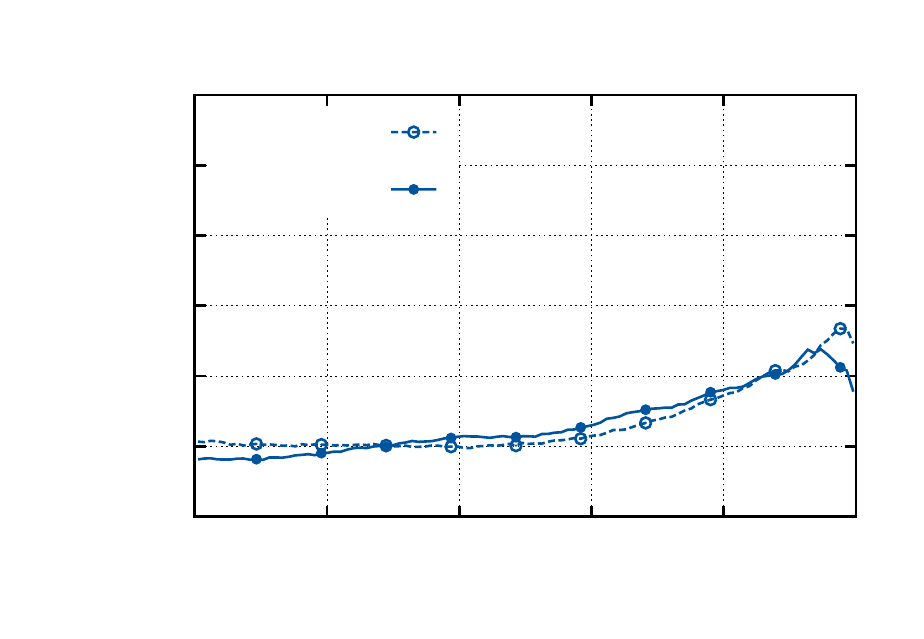}}%
    \gplfronttext
  \end{picture}%
\endgroup

%% file: data/FLAME_KERNEL_DNS_03/181021_alignment_curv_strain_progT_filt33/flame_kernel_nvec_ComprStrain_align_filt33_1p00TAU_03_JFM_HALF.tex
\begingroup
  \makeatletter
  \providecommand\color[2][]{%
    \GenericError{(gnuplot) \space\space\space\@spaces}{%
      Package color not loaded in conjunction with
      terminal option `colourtext'%
    }{See the gnuplot documentation for explanation.%
    }{Either use 'blacktext' in gnuplot or load the package
      color.sty in LaTeX.}%
    \renewcommand\color[2][]{}%
  }%
  \providecommand\includegraphics[2][]{%
    \GenericError{(gnuplot) \space\space\space\@spaces}{%
      Package graphicx or graphics not loaded%
    }{See the gnuplot documentation for explanation.%
    }{The gnuplot epslatex terminal needs graphicx.sty or graphics.sty.}%
    \renewcommand\includegraphics[2][]{}%
  }%
  \providecommand\rotatebox[2]{#2}%
  \@ifundefined{ifGPcolor}{%
    \newif\ifGPcolor
    \GPcolortrue
  }{}%
  \@ifundefined{ifGPblacktext}{%
    \newif\ifGPblacktext
    \GPblacktexttrue
  }{}%
  \let\gplgaddtomacro\g@addto@macro
  \gdef\gplbacktext{}%
  \gdef\gplfronttext{}%
  \makeatother
  \ifGPblacktext
    \def\colorrgb#1{}%
    \def\colorgray#1{}%
  \else
    \ifGPcolor
      \def\colorrgb#1{\color[rgb]{#1}}%
      \def\colorgray#1{\color[gray]{#1}}%
      \expandafter\def\csname LTw\endcsname{\color{white}}%
      \expandafter\def\csname LTb\endcsname{\color{black}}%
      \expandafter\def\csname LTa\endcsname{\color{black}}%
      \expandafter\def\csname LT0\endcsname{\color[rgb]{1,0,0}}%
      \expandafter\def\csname LT1\endcsname{\color[rgb]{0,1,0}}%
      \expandafter\def\csname LT2\endcsname{\color[rgb]{0,0,1}}%
      \expandafter\def\csname LT3\endcsname{\color[rgb]{1,0,1}}%
      \expandafter\def\csname LT4\endcsname{\color[rgb]{0,1,1}}%
      \expandafter\def\csname LT5\endcsname{\color[rgb]{1,1,0}}%
      \expandafter\def\csname LT6\endcsname{\color[rgb]{0,0,0}}%
      \expandafter\def\csname LT7\endcsname{\color[rgb]{1,0.3,0}}%
      \expandafter\def\csname LT8\endcsname{\color[rgb]{0.5,0.5,0.5}}%
    \else
      \def\colorrgb#1{\color{black}}%
      \def\colorgray#1{\color[gray]{#1}}%
      \expandafter\def\csname LTw\endcsname{\color{white}}%
      \expandafter\def\csname LTb\endcsname{\color{black}}%
      \expandafter\def\csname LTa\endcsname{\color{black}}%
      \expandafter\def\csname LT0\endcsname{\color{black}}%
      \expandafter\def\csname LT1\endcsname{\color{black}}%
      \expandafter\def\csname LT2\endcsname{\color{black}}%
      \expandafter\def\csname LT3\endcsname{\color{black}}%
      \expandafter\def\csname LT4\endcsname{\color{black}}%
      \expandafter\def\csname LT5\endcsname{\color{black}}%
      \expandafter\def\csname LT6\endcsname{\color{black}}%
      \expandafter\def\csname LT7\endcsname{\color{black}}%
      \expandafter\def\csname LT8\endcsname{\color{black}}%
    \fi
  \fi
  \setlength{\unitlength}{0.0500bp}%
  \begin{picture}(5328.00,3684.00)%
    \gplgaddtomacro\gplbacktext{%
      \csname LTb\endcsname%
      \put(990,704){\makebox(0,0)[r]{\strut{}\normalsize 0.00}}%
      \csname LTb\endcsname%
      \put(990,1109){\makebox(0,0)[r]{\strut{}\normalsize 0.75}}%
      \csname LTb\endcsname%
      \put(990,1514){\makebox(0,0)[r]{\strut{}\normalsize 1.50}}%
      \csname LTb\endcsname%
      \put(990,1919){\makebox(0,0)[r]{\strut{}\normalsize 2.25}}%
      \csname LTb\endcsname%
      \put(990,2323){\makebox(0,0)[r]{\strut{}\normalsize 3.00}}%
      \csname LTb\endcsname%
      \put(990,2728){\makebox(0,0)[r]{\strut{}\normalsize 3.75}}%
      \csname LTb\endcsname%
      \put(990,3133){\makebox(0,0)[r]{\strut{}\normalsize 4.50}}%
      \csname LTb\endcsname%
      \put(1122,484){\makebox(0,0){\strut{}\normalsize 0.0}}%
      \csname LTb\endcsname%
      \put(1884,484){\makebox(0,0){\strut{}\normalsize 0.2}}%
      \csname LTb\endcsname%
      \put(2646,484){\makebox(0,0){\strut{}\normalsize 0.4}}%
      \csname LTb\endcsname%
      \put(3407,484){\makebox(0,0){\strut{}\normalsize 0.6}}%
      \csname LTb\endcsname%
      \put(4169,484){\makebox(0,0){\strut{}\normalsize 0.8}}%
      \csname LTb\endcsname%
      \put(4931,484){\makebox(0,0){\strut{}\normalsize 1.0}}%
      \put(352,1918){\rotatebox{-270}{\makebox(0,0){\strut{}\large PDF}}}%
      \put(3026,154){\makebox(0,0){\strut{}\large $cos\left(\varphi \right)$}}%
      \put(3026,3353){\makebox(0,0){\strut{}\normalsize Large Kernel, $t=1.0\cdot \tau_{\mathrm{t}}$, $R_{50}=2.3 \cdot l_{\mathrm{t}}$}}%
    }%
    \gplgaddtomacro\gplfronttext{%
      \csname LTb\endcsname%
      \put(2121,2919){\makebox(0,0)[r]{\strut{}\normalsize no Filter}}%
      \csname LTb\endcsname%
      \put(2121,2589){\makebox(0,0)[r]{\strut{}\normalsize $\Delta = 0.5\; l_{\mathrm{t}}$}}%
    }%
    \gplbacktext
    \put(0,0){\includegraphics{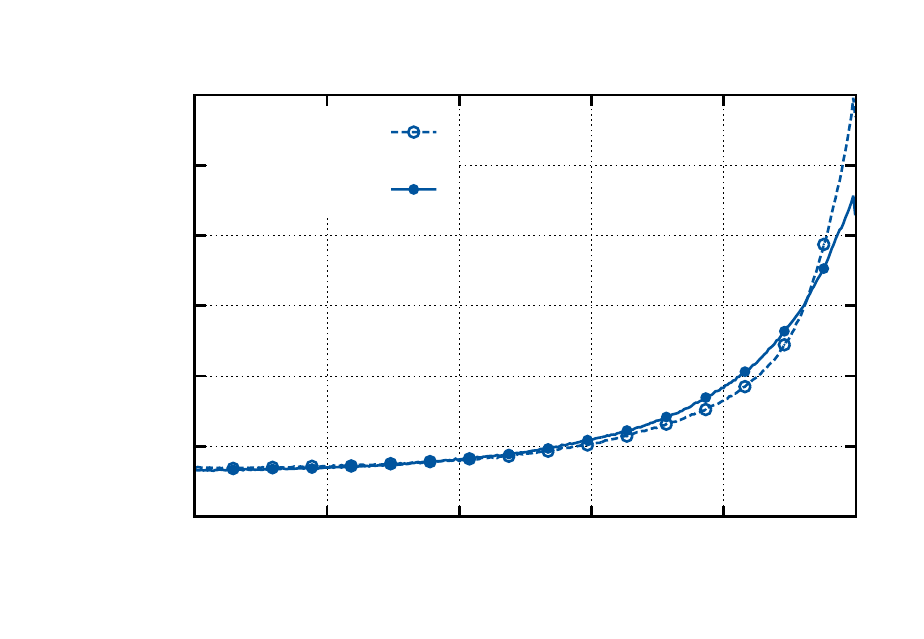}}%
    \gplfronttext
  \end{picture}%
\endgroup

%% file: data/FLAME_KERNEL_DNS_02/181003_curv_strain_kin_restor_diss_divg_skew/kernel_dns_mean_I0_Sd_1d_sizeT_02_LOC01_JFM_HALF.tex
\begingroup
  \makeatletter
  \providecommand\color[2][]{%
    \GenericError{(gnuplot) \space\space\space\@spaces}{%
      Package color not loaded in conjunction with
      terminal option `colourtext'%
    }{See the gnuplot documentation for explanation.%
    }{Either use 'blacktext' in gnuplot or load the package
      color.sty in LaTeX.}%
    \renewcommand\color[2][]{}%
  }%
  \providecommand\includegraphics[2][]{%
    \GenericError{(gnuplot) \space\space\space\@spaces}{%
      Package graphicx or graphics not loaded%
    }{See the gnuplot documentation for explanation.%
    }{The gnuplot epslatex terminal needs graphicx.sty or graphics.sty.}%
    \renewcommand\includegraphics[2][]{}%
  }%
  \providecommand\rotatebox[2]{#2}%
  \@ifundefined{ifGPcolor}{%
    \newif\ifGPcolor
    \GPcolortrue
  }{}%
  \@ifundefined{ifGPblacktext}{%
    \newif\ifGPblacktext
    \GPblacktexttrue
  }{}%
  \let\gplgaddtomacro\g@addto@macro
  \gdef\gplbacktext{}%
  \gdef\gplfronttext{}%
  \makeatother
  \ifGPblacktext
    \def\colorrgb#1{}%
    \def\colorgray#1{}%
  \else
    \ifGPcolor
      \def\colorrgb#1{\color[rgb]{#1}}%
      \def\colorgray#1{\color[gray]{#1}}%
      \expandafter\def\csname LTw\endcsname{\color{white}}%
      \expandafter\def\csname LTb\endcsname{\color{black}}%
      \expandafter\def\csname LTa\endcsname{\color{black}}%
      \expandafter\def\csname LT0\endcsname{\color[rgb]{1,0,0}}%
      \expandafter\def\csname LT1\endcsname{\color[rgb]{0,1,0}}%
      \expandafter\def\csname LT2\endcsname{\color[rgb]{0,0,1}}%
      \expandafter\def\csname LT3\endcsname{\color[rgb]{1,0,1}}%
      \expandafter\def\csname LT4\endcsname{\color[rgb]{0,1,1}}%
      \expandafter\def\csname LT5\endcsname{\color[rgb]{1,1,0}}%
      \expandafter\def\csname LT6\endcsname{\color[rgb]{0,0,0}}%
      \expandafter\def\csname LT7\endcsname{\color[rgb]{1,0.3,0}}%
      \expandafter\def\csname LT8\endcsname{\color[rgb]{0.5,0.5,0.5}}%
    \else
      \def\colorrgb#1{\color{black}}%
      \def\colorgray#1{\color[gray]{#1}}%
      \expandafter\def\csname LTw\endcsname{\color{white}}%
      \expandafter\def\csname LTb\endcsname{\color{black}}%
      \expandafter\def\csname LTa\endcsname{\color{black}}%
      \expandafter\def\csname LT0\endcsname{\color{black}}%
      \expandafter\def\csname LT1\endcsname{\color{black}}%
      \expandafter\def\csname LT2\endcsname{\color{black}}%
      \expandafter\def\csname LT3\endcsname{\color{black}}%
      \expandafter\def\csname LT4\endcsname{\color{black}}%
      \expandafter\def\csname LT5\endcsname{\color{black}}%
      \expandafter\def\csname LT6\endcsname{\color{black}}%
      \expandafter\def\csname LT7\endcsname{\color{black}}%
      \expandafter\def\csname LT8\endcsname{\color{black}}%
    \fi
  \fi
  \setlength{\unitlength}{0.0500bp}%
  \begin{picture}(5328.00,3684.00)%
    \gplgaddtomacro\gplbacktext{%
      \csname LTb\endcsname%
      \put(858,873){\makebox(0,0)[r]{\strut{}\normalsize 0.6}}%
      \csname LTb\endcsname%
      \put(858,1212){\makebox(0,0)[r]{\strut{}\normalsize 0.8}}%
      \csname LTb\endcsname%
      \put(858,1550){\makebox(0,0)[r]{\strut{}\normalsize 1.0}}%
      \csname LTb\endcsname%
      \put(858,1889){\makebox(0,0)[r]{\strut{}\normalsize 1.2}}%
      \csname LTb\endcsname%
      \put(858,2227){\makebox(0,0)[r]{\strut{}\normalsize 1.4}}%
      \csname LTb\endcsname%
      \put(858,2566){\makebox(0,0)[r]{\strut{}\normalsize 1.6}}%
      \csname LTb\endcsname%
      \put(858,2904){\makebox(0,0)[r]{\strut{}\normalsize 1.8}}%
      \csname LTb\endcsname%
      \put(858,3243){\makebox(0,0)[r]{\strut{}\normalsize 2.0}}%
      \csname LTb\endcsname%
      \put(990,484){\makebox(0,0){\strut{}\normalsize 0.0}}%
      \csname LTb\endcsname%
      \put(1553,484){\makebox(0,0){\strut{}\normalsize 0.5}}%
      \csname LTb\endcsname%
      \put(2116,484){\makebox(0,0){\strut{}\normalsize 1.0}}%
      \csname LTb\endcsname%
      \put(2679,484){\makebox(0,0){\strut{}\normalsize 1.5}}%
      \csname LTb\endcsname%
      \put(3242,484){\makebox(0,0){\strut{}\normalsize 2.0}}%
      \csname LTb\endcsname%
      \put(3805,484){\makebox(0,0){\strut{}\normalsize 2.5}}%
      \csname LTb\endcsname%
      \put(4368,484){\makebox(0,0){\strut{}\normalsize 3.0}}%
      \csname LTb\endcsname%
      \put(4931,484){\makebox(0,0){\strut{}\normalsize 3.5}}%
      \colorrgb{0.47,0.47,0.47}%
      \put(990,3408){\makebox(0,0){\strut{}\color[HTML]{777777}$\frac{t}{\tau_{\mathrm{t}}}$}}%
      \colorrgb{0.47,0.47,0.47}%
      \put(1818,3408){\makebox(0,0){\strut{}\color[HTML]{777777}0.5}}%
      \colorrgb{0.47,0.47,0.47}%
      \put(2542,3408){\makebox(0,0){\strut{}\color[HTML]{777777}1.0}}%
      \colorrgb{0.47,0.47,0.47}%
      \put(3353,3408){\makebox(0,0){\strut{}\color[HTML]{777777}1.5}}%
      \colorrgb{0.47,0.47,0.47}%
      \put(4346,3408){\makebox(0,0){\strut{}\color[HTML]{777777}2.0}}%
      \colorrgb{0.47,0.47,0.47}%
      \put(4931,3408){\makebox(0,0){\strut{}\color[HTML]{777777}$\frac{t}{\tau_{\mathrm{t}}}$}}%
      \csname LTb\endcsname%
      \put(352,1973){\rotatebox{-270}{\makebox(0,0){\strut{}\large Stretch Factor [-]}}}%
      \put(2960,154){\makebox(0,0){\strut{}\large Normalized Median Radius $\frac{R_{50}}{l_{\mathrm{t}}}$ [-]}}%
    }%
    \gplgaddtomacro\gplfronttext{%
      \csname LTb\endcsname%
      \put(4274,3015){\makebox(0,0)[r]{\strut{}\normalsize $\left(\mathrm{I}_{0\;\;\;}\right)$}}%
      \csname LTb\endcsname%
      \put(4274,2685){\makebox(0,0)[r]{\strut{}\normalsize $\left(\mathrm{I}_{0,\mathrm{rn}}\right)$}}%
      \csname LTb\endcsname%
      \put(4274,2355){\makebox(0,0)[r]{\strut{}\normalsize $\left(1+\mathrm{I}_{0,\kappa}\right)$}}%
    }%
    \gplbacktext
    \put(0,0){\includegraphics{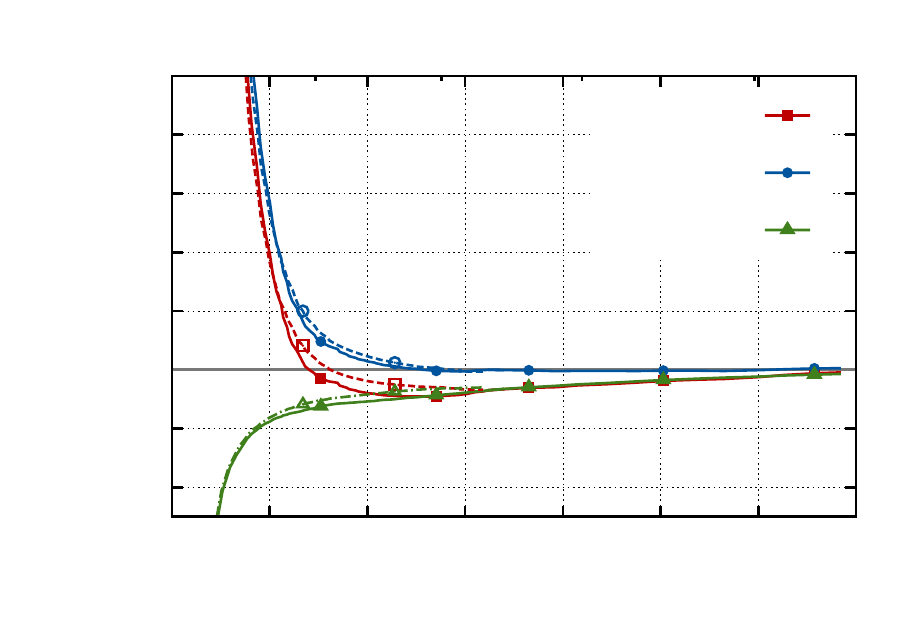}}%
    \gplfronttext
  \end{picture}%
\endgroup

%% file: main.bbl
\begin{thebibliography}{65}
\expandafter\ifx\csname natexlab\endcsname\relax\def\natexlab#1{#1}\fi
\def\au#1{#1} \def\ed#1{#1} \def\yr#1{#1}\def\at#1{#1}\def\jt#1{\textit{#1}}
  \def\bt#1{#1}\def\bvol#1{\textbf{#1}} \def\vol#1{#1} \def\pg#1{#1}
  \def\publ#1{#1}\def\arxiv#1{#1}\def\org#1{#1}\def\st#1{\textit{#1}}

\bibitem[Abdel-Gayed {\em et~al.\/}(1984)Abdel-Gayed, Al-Khishali \&
  Bradley]{AbdelGayed84_kernel_flow_interact}
{\sc \au{Abdel-Gayed, R.~G.}, \au{Al-Khishali, K.~J.} \& \au{Bradley, D.}}
  \yr{1984}  \at{Turbulent burning velocities and flame straining in
  explosions}.  \jt{Proceedings of the Royal Society of London. Series A,
  Mathematical and Physical Sciences}  \bvol{391}~(1801),  \pg{393--414}.

\bibitem[Akindele {\em et~al.\/}(1982)Akindele, Bradley, Mak \&
  McMahon]{Akindele82_kernel_flow_interact}
{\sc \au{Akindele, O.}, \au{Bradley, D.}, \au{Mak, P.} \& \au{McMahon, M.}}
  \yr{1982}  \at{Spark ignition of turbulent gases}.  \jt{Combustion and Flame}
   \bvol{47},  \pg{129{--}155}.

\bibitem[Aleiferis {\em et~al.\/}(2004)Aleiferis, Taylor, Ishii \&
  Urata]{Aleiferis04_lean_ccv}
{\sc \au{Aleiferis, P.}, \au{Taylor, A.}, \au{Ishii, K.} \& \au{Urata, Y.}}
  \yr{2004}  \at{The nature of early flame development in a lean-burn
  stratified-charge spark-ignition engine}.  \jt{Combustion and Flame}
  \bvol{136}~(3),  \pg{283{--}302}.

\bibitem[Alqallaf {\em et~al.\/}(2019)Alqallaf, Klein \&
  Chakraborty]{Alqallaf19_kernel_curv_eq}
{\sc \au{Alqallaf, A.}, \au{Klein, M.} \& \au{Chakraborty, N.}} \yr{2019}
  \at{Effects of {L}ewis number on the evolution of curvature in spherically
  expanding turbulent premixed flames}.  \jt{Fluids}  \bvol{4}~(1),  \pg{12}.

\bibitem[Ashurst {\em et~al.\/}(1987)Ashurst, Kerstein, Kerr \&
  Gibson]{Ashurst87_sc_align}
{\sc \au{Ashurst, W.~T.}, \au{Kerstein, A.~R.}, \au{Kerr, R.~M.} \& \au{Gibson,
  C.~H.}} \yr{1987}  \at{Alignment of vorticity and scalar gradient with strain
  rate in simulated {N}avier-{S}tokes turbulence}.  \jt{The Physics of Fluids}
  \bvol{30}~(8),  \pg{2343--2353}.

\bibitem[Batchelor(1959)]{Batchelor59}
{\sc \au{Batchelor, G.~K.}} \yr{1959}  \at{Small-scale variation of convected
  quantities like temperature in turbulent fluid part 1. general discussion and
  the case of small conductivity}.  \jt{Journal of Fluid Mechanics}
  \bvol{5}~(1),  \pg{113--133}.

\bibitem[Bode {\em et~al.\/}(2017)Bode, Schorr, Kr{\"u}ger, Dreizler \&
  B{\"o}hm]{Bode17_piv_early_kernel}
{\sc \au{Bode, J.}, \au{Schorr, J.}, \au{Kr{\"u}ger, C.}, \au{Dreizler, A.} \&
  \au{B{\"o}hm, B.}} \yr{2017}  \at{Influence of three-dimensional in-cylinder
  flows on cycle-to-cycle variations in a fired stratified {DISI} engine
  measured by time-resolved dual-plane {PIV}}.  \jt{Proceedings of the
  Combustion Institute}  \bvol{36}~(3),  \pg{3477{--}3485}.

\bibitem[Boger {\em et~al.\/}(1998)Boger, Veynante, Boughanem \&
  Trouv{\'e}]{Boger98_gen_fsd}
{\sc \au{Boger, M.}, \au{Veynante, D.}, \au{Boughanem, H.} \& \au{Trouv{\'e},
  A.}} \yr{1998}  \at{Direct numerical simulation analysis of flame surface
  density concept for large eddy simulation of turbulent premixed combustion}.
  \jt{Symposium (International) on Combustion}  \bvol{27}~(1),
  \pg{917{--}925}.

\bibitem[Bradley {\em et~al.\/}(1994)Bradley, Lawes, Scott \&
  Mushi]{Bradley94_kernel_stretching}
{\sc \au{Bradley, D.}, \au{Lawes, M.}, \au{Scott, M.} \& \au{Mushi, E.}}
  \yr{1994}  \at{Afterburning in spherical premixed turbulent explosions}.
  \jt{Combustion and Flame}  \bvol{99}~(3),  \pg{581{--}590}.

\bibitem[Bradley \& Lung(1987)]{Bradley87_kernel_flow_interact}
{\sc \au{Bradley, D.} \& \au{Lung, F.-K.}} \yr{1987}  \at{Spark ignition and
  the early stages of turbulent flame propagation}.  \jt{Combustion and Flame}
  \bvol{69}~(1),  \pg{71{--}93}.

\bibitem[Buschbeck {\em et~al.\/}(2012)Buschbeck, Bittner, Halfmann \&
  Arndt]{Buschbeck12_piv_kernel}
{\sc \au{Buschbeck, M.}, \au{Bittner, N.}, \au{Halfmann, T.} \& \au{Arndt, S.}}
  \yr{2012}  \at{Dependence of combustion dynamics in a gasoline engine upon
  the in-cylinder flow field, determined by high-speed {PIV}}.  \jt{Experiments
  in Fluids}  \bvol{53}~(6),  \pg{1701--1712}.

\bibitem[Buxton {\em et~al.\/}(2011)Buxton, Laizet \&
  Ganapathisubramani]{Buxton11_enthrophy_prod}
{\sc \au{Buxton, O. R.~H.}, \au{Laizet, S.} \& \au{Ganapathisubramani, B.}}
  \yr{2011}  \at{The interaction between strain-rate and rotation in shear flow
  turbulence from inertial range to dissipative length scales}.  \jt{Physics of
  Fluids}  \bvol{23}~(6),  \pg{061704}.

\bibitem[Candel \& Poinsot(1990)]{Candel90_fsd_eq}
{\sc \au{Candel, S.~M.} \& \au{Poinsot, T.~J.}} \yr{1990}  \at{Flame stretch
  and the balance equation for the flame area}.  \jt{Combustion Science and
  Technology}  \bvol{70}~(1--3),  \pg{1--15}.

\bibitem[Castela {\em et~al.\/}(2016)Castela, Fiorina, Coussement, Gicquel,
  Darabiha \& Laux]{Castela16_spark_dns}
{\sc \au{Castela, M.}, \au{Fiorina, B.}, \au{Coussement, A.}, \au{Gicquel, O.},
  \au{Darabiha, N.} \& \au{Laux, C.~O.}} \yr{2016}  \at{Modelling the impact of
  non-equilibrium discharges on reactive mixtures for simulations of
  plasma-assisted ignition in turbulent flows}.  \jt{Combustion and Flame}
  \bvol{166},  \pg{133{--}147}.

\bibitem[Chakraborty \& Swaminathan(2007)]{Chakraborty07_sc_align}
{\sc \au{Chakraborty, N.} \& \au{Swaminathan, N.}} \yr{2007}  \at{Influence of
  the {D}amk{\"o}hler number on turbulence-scalar interaction in premixed
  flames. {I}. {P}hysical insight}.  \jt{Physics of Fluids}  \bvol{19}~(4),
  \pg{045103}.

\bibitem[Cifuentes {\em et~al.\/}(2018)Cifuentes, Dopazo, Sandeep, Chakraborty
  \& Kempf]{Cifuentes18_curv_eq_kempf}
{\sc \au{Cifuentes, L.}, \au{Dopazo, C.}, \au{Sandeep, A.}, \au{Chakraborty,
  N.} \& \au{Kempf, A.}} \yr{2018}  \at{Analysis of flame curvature evolution
  in a turbulent premixed bluff body burner}.  \jt{Physics of Fluids}
  \bvol{30}~(9),  \pg{095101}.

\bibitem[Colin \& Truffin(2011)]{Colin11_spark_model}
{\sc \au{Colin, O.} \& \au{Truffin, K.}} \yr{2011}  \at{A spark ignition model
  for large eddy simulation based on an {FSD} transport equation
  ({ISSIM}-{LES})}.  \jt{Proceedings of the Combustion Institute}
  \bvol{33}~(2),  \pg{3097{--}3104}.

\bibitem[Creta {\em et~al.\/}(2016)Creta, Lamioni, Lapenna \&
  Troiani]{Creta16_lam_turb_skew}
{\sc \au{Creta, F.}, \au{Lamioni, R.}, \au{Lapenna, P.~E.} \& \au{Troiani, G.}}
  \yr{2016}  \at{Interplay of {D}arrieus-{L}andau instability and weak
  turbulence in premixed flame propagation}.  \jt{Physical Review~E}
  \bvol{94},  \pg{053102}.

\bibitem[Damk{\"o}hler(1940)]{Damkoehler40}
{\sc \au{Damk{\"o}hler, G.}} \yr{1940}  \at{Der {E}influss der {T}urbulenz auf
  die {F}lammengeschwindigkeit in {G}asgemischen}.  \jt{Zeitschrift für
  Elektrochemie und angewandte physikalische Chemie}  \bvol{46}~(11),
  \pg{601--626}.

\bibitem[Dopazo {\em et~al.\/}(2018)Dopazo, Martin, Cifuentes \&
  Hierro]{Dopazo18_curvEq}
{\sc \au{Dopazo, C.}, \au{Martin, J.}, \au{Cifuentes, L.} \& \au{Hierro, J.}}
  \yr{2018}  \at{Strain, rotation and curvature of non-material propagating
  iso-scalar surfaces in homogeneous turbulence}.  \jt{Flow, Turbulence and
  Combustion}  \bvol{101}~(1),  \pg{1--32}.

\bibitem[Echekki \& Chen(1999)]{Echekki99_diff_term_split}
{\sc \au{Echekki, T.} \& \au{Chen, J.}} \yr{1999}  \at{Analysis of the
  contribution of curvature to premixed flame propagation}.  \jt{Combustion and
  Flame}  \bvol{118},  \pg{308--311}.

\bibitem[Echekki \& Kolera-Gokula(2007)]{Echekki07_kernel_vortex_map_01}
{\sc \au{Echekki, T.} \& \au{Kolera-Gokula, H.}} \yr{2007}  \at{A regime
  diagram for premixed flame kernel-vortex interactions}.  \jt{Physics of
  Fluids}  \bvol{19}~(4),  \pg{043604}.

\bibitem[Echekki {\em et~al.\/}(1994)Echekki, Poinsot, Baritaud \&
  Trouv{\'e}]{Echekki94_kernel_dns_Le_effect}
{\sc \au{Echekki, T.}, \au{Poinsot, T.}, \au{Baritaud, T.} \& \au{Trouv{\'e},
  A.}} \yr{1994}  \bt{Modeling and simulation of turbulent flame kernel
  evolution}. {\em Tech. Rep.\/} 41525.  \org{Institut Fran{\c c}ais du
  P{\'e}trole}.

\bibitem[Falkenstein {\em et~al.\/}(2019)Falkenstein, Kang, Cai, Bode \&
  Pitsch]{Falkenstein19_kernel_Le1_cnf}
{\sc \au{Falkenstein, T.}, \au{Kang, S.}, \au{Cai, L.}, \au{Bode, M.} \&
  \au{Pitsch, H.}} \yr{2019}  \at{{DNS} study of the global heat release rate
  during early flame kernel development under engine conditions}.
  \jt{Submitted to Combustion and Flame} ,  \arxiv{arXiv:
  http://arxiv.org/abs/1908.07556}.

\bibitem[Fansler \& Wagner(2015)]{Fansler15_ccv}
{\sc \au{Fansler, T.~D.} \& \au{Wagner, R.~M.}} \yr{2015}  \at{Cyclic
  dispersion in engine combustion {--} introduction by the special issue
  editors}.  \jt{International Journal of Engine Research}  \bvol{16}~(3),
  \pg{255--259}.

\bibitem[Gashi {\em et~al.\/}(2005)Gashi, Hult, Jenkins, Chakraborty, Cant \&
  Kaminski]{Gashi05_curv_pdf_turb}
{\sc \au{Gashi, S.}, \au{Hult, J.}, \au{Jenkins, K.~W.}, \au{Chakraborty, N.},
  \au{Cant, S.} \& \au{Kaminski, C.~F.}} \yr{2005}  \at{Curvature and wrinkling
  of premixed flame kernelsâ€”comparisons of oh plif and dns data}.
  \jt{Proceedings of the Combustion Institute}  \bvol{30}~(1),
  \pg{809{--}817}.

\bibitem[Haq {\em et~al.\/}(2002)Haq, Sheppard, Woolley, Greenhalgh \&
  Lockett]{Haq02_curv_pdf_turb}
{\sc \au{Haq, M.}, \au{Sheppard, C.}, \au{Woolley, R.}, \au{Greenhalgh, D.} \&
  \au{Lockett, R.}} \yr{2002}  \at{Wrinkling and curvature of laminar and
  turbulent premixed flames}.  \jt{Combustion and Flame}  \bvol{131}~(1),
  \pg{1{--}15}.

\bibitem[Hasse(2016)]{Hasse16_LES}
{\sc \au{Hasse, C.}} \yr{2016}  \at{Scale-resolving simulations in engine
  combustion process design based on a systematic approach for model
  development}.  \jt{International Journal of Engine Research}  \bvol{17}~(1),
  \pg{44--62}.

\bibitem[Heim \& Ghandhi(2011)]{Heim11_engine_turb}
{\sc \au{Heim, D.} \& \au{Ghandhi, J.}} \yr{2011}  \at{A detailed study of
  in-cylinder flow and turbulence using {PIV}}.  \jt{SAE International Journal
  of Engines}  \bvol{4}~(1),  \pg{1642--1668}.

\bibitem[Herweg \& Maly(1992)]{Herweg92_model_valid}
{\sc \au{Herweg, R.} \& \au{Maly, R.~R.}} \yr{1992} A fundamental model for
  flame kernel formation in {S.~I.} engines.  \bt{In {\em International Fuels
  \& Lubricants Meeting \& Exposition\/}}.  \publ{SAE International}.

\bibitem[Hesse {\em et~al.\/}(2009)Hesse, Chakraborty \&
  Mastorakos]{Hesse09_spark_duration}
{\sc \au{Hesse, H.}, \au{Chakraborty, N.} \& \au{Mastorakos, E.}} \yr{2009}
  \at{The effects of the lewis number of the fuel on the displacement speed of
  edge flames in igniting turbulent mixing layers}.  \jt{Proceedings of the
  Combustion Institute}  \bvol{32}~(1),  \pg{1399{--}1407}.

\bibitem[Heywood(1994)]{Heywood94_COMODIA}
{\sc \au{Heywood, J.}} \yr{1994} Combustion and its modeling in spark-ignition
  engines.  \bt{In {\em Proceedings of the Third International Symposium on
  Diagnostics and Modeling of Combustion in Internal Combustion Engines
  (COMODIA), Yokohama, Japan, July 11{--}14\/}},  \pg{pp. 1--15}.

\bibitem[Hirschfelder {\em et~al.\/}(1954)Hirschfelder, Curtiss \&
  Bird]{Hirschfelder54_diff_model}
{\sc \au{Hirschfelder, J.~O.}, \au{Curtiss, C.~F.} \& \au{Bird, R.~B.}}
  \yr{1954} {\em Molecular theory of gases and liquids\/}.  \publ{John Wiley
  and Sons, New York}.

\bibitem[Jung {\em et~al.\/}(2017)Jung, Sasaki \& Iida]{Jung17_lean_si_ccv}
{\sc \au{Jung, D.}, \au{Sasaki, K.} \& \au{Iida, N.}} \yr{2017}  \at{Effects of
  increased spark discharge energy and enhanced in-cylinder turbulence level on
  lean limits and cycle-to-cycle variations of combustion for {SI} engine
  operation}.  \jt{Applied Energy}  \bvol{205},  \pg{1467{--}1477}.

\bibitem[Karlovitz {\em et~al.\/}(1951)Karlovitz, Denniston \&
  Wells]{Karlovitz51_turb}
{\sc \au{Karlovitz, B.}, \au{Denniston, D.~W.} \& \au{Wells, F.~E.}} \yr{1951}
  \at{Investigation of turbulent flames}.  \jt{The Journal of Chemical Physics}
   \bvol{19}~(5),  \pg{541--547}.

\bibitem[Kim \& Pitsch(2007)]{Kim07_sc_align}
{\sc \au{Kim, S.~H.} \& \au{Pitsch, H.}} \yr{2007}  \at{Scalar gradient and
  small-scale structure in turbulent premixed combustion}.  \jt{Physics of
  Fluids}  \bvol{19}~(11),  \pg{115104}.

\bibitem[Luca {\em et~al.\/}(2019)Luca, Attili, Schiavo, Creta \&
  Bisetti]{Luca19_dns_resolution}
{\sc \au{Luca, S.}, \au{Attili, A.}, \au{Schiavo, E.~L.}, \au{Creta, F.} \&
  \au{Bisetti, F.}} \yr{2019}  \at{On the statistics of flame stretch in
  turbulent premixed jet flames in the thin reaction zone regime at varying
  reynolds number}.  \jt{Proceedings of the Combustion Institute}
  \bvol{37}~(2),  \pg{2451{--}2459}.

\bibitem[Meneveau \& Poinsot(1991)]{Meneveau91_ITNFS_model}
{\sc \au{Meneveau, C.} \& \au{Poinsot, T.}} \yr{1991}  \at{Stretching and
  quenching of flamelets in premixed turbulent combustion}.  \jt{Combustion and
  Flame}  \bvol{86}~(4),  \pg{311{--}332}.

\bibitem[M{\"u}ller(1998)]{Mueller98_lowMach}
{\sc \au{M{\"u}ller, B.}} \yr{1998}  \at{Low-{M}ach-number asymptotics of the
  {N}avier-{S}tokes equations}.  \jt{Journal of Engineering Mathematics}
  \bvol{34}~(1),  \pg{97--109}.

\bibitem[Peters(1992)]{Peters92_jfm}
{\sc \au{Peters, N.}} \yr{1992}  \at{A spectral closure for premixed turbulent
  combustion in the flamelet regime}.  \jt{Journal of Fluid Mechanics}
  \bvol{242},  \pg{611--629}.

\bibitem[Peters(1999)]{Peters99_regime_diagr}
{\sc \au{Peters, N.}} \yr{1999}  \at{The turbulent burning velocity for
  large-scale and small-scale turbulence}.  \jt{Journal of Fluid Mechanics}
  \bvol{384},  \pg{107--132}.

\bibitem[Peterson {\em et~al.\/}(2011)Peterson, Reuss \&
  Sick]{Peterson11_exp_misfire}
{\sc \au{Peterson, B.}, \au{Reuss, D.~L.} \& \au{Sick, V.}} \yr{2011}
  \at{High-speed imaging analysis of misfires in a spray-guided direct
  injection engine}.  \jt{Proceedings of the Combustion Institute}
  \bvol{33}~(2),  \pg{3089{--}3096}.

\bibitem[Peterson {\em et~al.\/}(2014)Peterson, Reuss \&
  Sick]{Peterson14_early_flame_exp}
{\sc \au{Peterson, B.}, \au{Reuss, D.~L.} \& \au{Sick, V.}} \yr{2014}  \at{On
  the ignition and flame development in a spray-guided direct-injection
  spark-ignition engine}.  \jt{Combustion and Flame}  \bvol{161}~(1),
  \pg{240{--}255}.

\bibitem[Pitsch \& Peters(1996)]{Pitsch96_iso_octane_mech}
{\sc \au{Pitsch, H.} \& \au{Peters, N.}} \yr{1996}  \at{Numerical and asymtotic
  studies of the structure of premixed iso-octane flames}.  \jt{Symposium
  (International) on Combustion}  \bvol{26}~(1),  \pg{763{--}771}.

\bibitem[Poludnenko \& Oran(2011)]{Poludnenko11_cusp_turb}
{\sc \au{Poludnenko, A.} \& \au{Oran, E.}} \yr{2011}  \at{The interaction of
  high-speed turbulence with flames: Turbulent flame speed}.  \jt{Combustion
  and Flame}  \bvol{158}~(2),  \pg{301{--}326}.

\bibitem[{Pope}(1988)]{Pope88_surf_turb}
{\sc \au{{Pope}, S.~B.}} \yr{1988}  \at{{The evolution of surfaces in
  turbulence}}.  \jt{International Journal of Engineering Science}  \bvol{26},
  \pg{445--469}.

\bibitem[Reddy \& Abraham(2013)]{Reddy11_regime_map}
{\sc \au{Reddy, H.} \& \au{Abraham, J.}} \yr{2013}  \at{Influence of
  turbulence-kernel interactions on flame development in lean methane/air
  mixtures under natural gas-fueled engine conditions}.  \jt{Fuel}  \bvol{103},
   \pg{1090--1105}.

\bibitem[Richard {\em et~al.\/}(2007)Richard, Colin, Vermorel, Benkenida,
  Angelberger \& Veynante]{Richard07_spark_model}
{\sc \au{Richard, S.}, \au{Colin, O.}, \au{Vermorel, O.}, \au{Benkenida, A.},
  \au{Angelberger, C.} \& \au{Veynante, D.}} \yr{2007}  \at{Towards large eddy
  simulation of combustion in spark ignition engines}.  \jt{Proceedings of the
  Combustion Institute}  \bvol{31}~(2),  \pg{3059{--}3066}.

\bibitem[Rutland(2011)]{Rutland11_LES_review}
{\sc \au{Rutland, C.~J.}} \yr{2011}  \at{Large-eddy simulations for internal
  combustion engines -- a review}.  \jt{International Journal of Engine
  Research}  \bvol{12}~(5),  \pg{421--451}.

\bibitem[Schiffmann {\em et~al.\/}(2018)Schiffmann, Reuss \&
  Sick]{Schiffmann17_exp}
{\sc \au{Schiffmann, P.}, \au{Reuss, D.~L.} \& \au{Sick, V.}} \yr{2018}
  \at{Empirical investigation of spark-ignited flame-initiation cycle-to-cycle
  variability in a homogeneous charge reciprocating engine}.  \jt{International
  Journal of Engine Research}  \bvol{19}~(5),  \pg{491--508}.

\bibitem[Scurlock \& Grover(1953)]{Scurlock53_planar_flame_devel_theory}
{\sc \au{Scurlock, A.} \& \au{Grover, J.}} \yr{1953}  \at{Propagation of
  turbulent flames}.  \jt{Symposium (International) on Combustion}
  \bvol{4}~(1),  \pg{645{--}658}.

\bibitem[Shepherd \& Ashurst(1992)]{Shephered92_curv_symm_turb}
{\sc \au{Shepherd, I.} \& \au{Ashurst, W.}} \yr{1992}  \at{Flame front geometry
  in premixed turbulent flames}.  \jt{Symposium (International) on Combustion}
  \bvol{24}~(1),  \pg{485{--}491}.

\bibitem[Shepherd {\em et~al.\/}(2002{\natexlab{{\em a\/}}})Shepherd, Cheng,
  Plessing, Kortschik \& Peters]{Shepherd02_curv_pdf_turb}
{\sc \au{Shepherd, I.}, \au{Cheng, R.}, \au{Plessing, T.}, \au{Kortschik, C.}
  \& \au{Peters, N.}} \yr{2002{\natexlab{{\em a\/}}}}  \at{Premixed flame front
  structure in intense turbulence}.  \jt{Proceedings of the Combustion
  Institute}  \bvol{29}~(2),  \pg{1833{--}1840}.

\bibitem[Shepherd {\em et~al.\/}(2002{\natexlab{{\em b\/}}})Shepherd, Cheng,
  Plessing, Kortschik \& Peters]{Shephered02_curv_turb}
{\sc \au{Shepherd, I.}, \au{Cheng, R.}, \au{Plessing, T.}, \au{Kortschik, C.}
  \& \au{Peters, N.}} \yr{2002{\natexlab{{\em b\/}}}}  \at{Premixed flame front
  structure in intense turbulence}.  \jt{Proceedings of the Combustion
  Institute}  \bvol{29}~(2),  \pg{1833{--}1840}.

\bibitem[Subramanian {\em et~al.\/}(2009)Subramanian, Domingo \&
  Vervisch]{Subramanian09_spark_src}
{\sc \au{Subramanian, V.}, \au{Domingo, P.} \& \au{Vervisch, L.}} \yr{2009}
  \at{Turbulent flame spreading mechanisms after spark ignition}.  \jt{AIP
  Conference Proceedings}  \bvol{1190}~(1),  \pg{68--89}.

\bibitem[Th{\'e}venin(2005)]{Thevenin05_kernel_dns_3d}
{\sc \au{Th{\'e}venin, D.}} \yr{2005}  \at{Three-dimensional direct simulations
  and structure of expanding turbulent methane flames}.  \jt{Proceedings of the
  Combustion Institute}  \bvol{30}~(1),  \pg{629{--}637}.

\bibitem[Th{\'e}venin {\em et~al.\/}(2002)Th{\'e}venin, Gicquel, Charentenay,
  Hilbert \& Veynante]{Thevenin02_dns_2d_3d}
{\sc \au{Th{\'e}venin, D.}, \au{Gicquel, O.}, \au{Charentenay, J.~D.},
  \au{Hilbert, R.} \& \au{Veynante, D.}} \yr{2002}  \at{Two- versus
  three-dimensional direct simulations of turbulent methane flame kernels using
  realistic chemistry}.  \jt{Proceedings of the Combustion Institute}
  \bvol{29}~(2),  \pg{2031{--}2039}.

\bibitem[Uranakara {\em et~al.\/}(2017)Uranakara, Chaudhuri \&
  Lakshmisha]{Uranakara17_ign_kernel_3d}
{\sc \au{Uranakara, H.~A.}, \au{Chaudhuri, S.} \& \au{Lakshmisha, K.}}
  \yr{2017}  \at{On the extinction of igniting kernels in near-isotropic
  turbulence}.  \jt{Proceedings of the Combustion Institute}  \bvol{36}~(2),
  \pg{1793{--}1800}.

\bibitem[Vasudeo {\em et~al.\/}(2010)Vasudeo, Echekki, Day \&
  Bell]{Vasudeo10_kernel_vortex_map_01}
{\sc \au{Vasudeo, N.}, \au{Echekki, T.}, \au{Day, M.~S.} \& \au{Bell, J.~B.}}
  \yr{2010}  \at{The regime diagram for premixed flame kernel-vortex
  interactions -- revisited}.  \jt{Physics of Fluids}  \bvol{22}~(4),
  \pg{043602}.

\bibitem[Wang {\em et~al.\/}(2017{\natexlab{{\em a\/}}})Wang, Hawkes, Chen,
  Zhou, Li \& Ald{\'e}n]{Wang17_jet_dns}
{\sc \au{Wang, H.}, \au{Hawkes, E.~R.}, \au{Chen, J.~H.}, \au{Zhou, B.},
  \au{Li, Z.} \& \au{Ald{\'e}n, M.}} \yr{2017{\natexlab{{\em a\/}}}}
  \at{Direct numerical simulations of a high {K}arlovitz number laboratory
  premixed jet flame {--} an analysis of flame stretch and flame thickening}.
  \jt{Journal of Fluid Mechanics}  \bvol{815},  \pg{511--536}.

\bibitem[Wang {\em et~al.\/}(2017{\natexlab{{\em b\/}}})Wang, Liu \&
  Reitz]{Wang17_knock_precs}
{\sc \au{Wang, Z.}, \au{Liu, H.} \& \au{Reitz, R.~D.}} \yr{2017{\natexlab{{\em
  b\/}}}}  \at{Knocking combustion in spark-ignition engines}.  \jt{Progress in
  Energy and Combustion Science}  \bvol{61},  \pg{78{--}112}.

\bibitem[Wenzel \& Peters(2000)]{Wenzel00_dns_flame_area}
{\sc \au{Wenzel, H.} \& \au{Peters, N.}} \yr{2000}  \at{Direct numerical
  simulation and modeling of kinematic restoration, dissipation and gas
  expansion effects of premixed flames in homogeneous turbulence}.
  \jt{Combustion Science and Technology}  \bvol{158}~(1),  \pg{273--297}.

\bibitem[Young(1981)]{Young81_ccv}
{\sc \au{Young, M.~B.}} \yr{1981}  \at{Cyclic dispersion in the
  homogeneous-charge spark-ignition engine -- a literature survey}.  \jt{SAE
  Transactions}  \bvol{90},  \pg{49--73}.

\bibitem[Zel'dovich(1966)]{Zeldovich66_lam_cusp}
{\sc \au{Zel'dovich, Y.~B.}} \yr{1966}  \at{An effect which stabilizes the
  curved front of a laminar flame}.  \jt{Journal of Applied Mechanics and
  Technical Physics}  \bvol{7}~(1),  \pg{68--69}.

\bibitem[Zeng {\em et~al.\/}(2019)Zeng, Keum, Kuo \& Sick]{Zeng18_exp}
{\sc \au{Zeng, W.}, \au{Keum, S.}, \au{Kuo, T.-W.} \& \au{Sick, V.}} \yr{2019}
  \at{Role of large scale flow features on cycle-to-cycle variations of
  spark-ignited flame-initiation and its transition to turbulent combustion}.
  \jt{Proceedings of the Combustion Institute}  \bvol{37}~(4),
  \pg{4945{--}4953}.

\end{thebibliography}
